\documentclass[journal]{IEEEtran}
\usepackage{}
\usepackage{amsfonts}
\usepackage{amssymb}
\usepackage{amsmath}
\usepackage{mathrsfs}
\usepackage{amsmath,amssymb}
\usepackage{amsmath}
\usepackage{graphicx}
\usepackage{amsmath,amssymb,amsthm}
\usepackage{leftidx}
\usepackage{extarrows}
\usepackage{caption}
\usepackage{enumerate}
\usepackage{graphics}
\usepackage{subfigure}
\usepackage{float}
\usepackage{pict2e}
\usepackage{tikz}
\usepackage{float}
\usepackage{pict2e}
\usepackage{tikz}
\usepackage{cases}
\ifCLASSINFOpdf
\else
\fi
\begin{document}
%
\title{A Stochastic Hybrid Approach to Decentralized Networked Control: Stochastic Network Delays and Poisson Pulsing Attacks}




%


%
%


\author{Dandan Zhang, Xin Jin, Hongye Su
\thanks{This work was partially supported by National Key Research and Development Program of
China (Grant No. 2022YFB3305903) and by the National Natural Science Foundation of China (Grant No. 62403141). 
}
\thanks{D. Zhang and H. Su are with the
State Key Laboratory of Industrial Control Technology, the Institute of
Cyber-Systems and Control, the College of Control Science and Engineering, Zhejiang University, Hangzhou 310027,
China (e-mail: zdandan3@zju.edu.cn; hysu@iipc.zju.edu.cn).

X. Jin is with the Research Institute of Intelligent Complex Systems, Fudan University,
Shanghai 200433, China (e-mail: xinjin@fudan.edu.cn)}
}

\maketitle

\begin{abstract}
By designing the decentralized time-regularized (Zeno-free) event-triggered strategies for the state-feedback control law, this paper considers the stochastic stabilization of a class of networked control systems, where two sources of randomness exist in multiple decentralized networks that operate asynchronously and independently: the communication channels are constrained by the stochastic network delays and also by Poisson pulsing denial-of-service (Pp-DoS) attacks. The time delay in the network denotes the length from a transmission instant to the corresponding update instant, and is supposed to be a continuous random variable subject to certain continuous probability distribution; while the attacks' cardinal number is a discrete random variable supposed to be subject to Poisson distribution, so the inter-attack time, i.e., the time between two consecutive attack instants, is subject to exponential distribution. The considered system is modeled as a stochastic hybrid formalism, where the randomness enters through the jump map into the reset value (the inter-attack time directly related) of each triggered strategy. By only sampling/transmitting state measurements when needed and simultaneously by taking the specific medium access protocols into account, the designed event-triggered strategies are synthesized in a state-based and decentralized form, which are robust (tolerable well) to stochastic network delays, under different tradeoff-conditions between the minimum inter-event times, maximum allowable delays (i.e., potentially tolerable delays) and the frequencies of attacks. Using stochastic hybrid tools to combine attack-active parts with attack-over parts, the designed triggered strategies, if designed well according to the actual system needs, can tolerate (be resilient to) the Pp-DoS attacks and stochastic network delays without jeopardizing the stability and Zeno-freeness, which is verified by the robust global attitude stabilization of flexible combined rotorcraft-like aerial vehicles.

\end{abstract}

\begin{IEEEkeywords}
Event-triggered strategy, stochastic network delay, denial-of-service attack, Poisson process.
\end{IEEEkeywords}

%
\IEEEpeerreviewmaketitle

\section{Introduction}

\IEEEPARstart{S}{tochastic} hybrid systems have attracted considerable attention
in recent years \cite{3.1shs}, \cite{5.2shs}, i.e., systems with states
that sometimes change continuously (flow) and sometimes change instantaneously (jump), with each type of evolution
possibly driven by a random process. With immense applications in the engineering field, see \cite{4.1shs} for a review, stochastic hybrid system models are expected to be very effective especially for dealing with networked control systems based on certain event-triggered control strategy \cite{romain2022}, which is generally resilient to attacks and varying communication delays \cite{pan2022}, \cite{q2010hemeels}, \cite{q15}. If designed without considering the delay, either constant (up to jitter) or random time-varying \cite{walsh2001}, the network delays can potentially degrade the performance of control systems and can even destabilize the system. Unlike other system-specific attacks, denial-of-service (DoS) attacks are more generic in plaguing the network communication (causing packet losses), resulting in very significant impact in scope and damage. See \cite{key2009} for polymorphic DoS attacks. The conventional flood-based DoS attacks (zero attack-over interval) can be easily detected at the victim's side \cite{2002flood}, \cite{Golait2016}, so this paper considers the stochastic pulsing DoS (p-DoS) attacks, i.e., each attack pulse lasts for a negligible short period of time, which occurs in multiple decentralized (asynchronous and independent) communication channels constrained by the stochastic network delays: The time delay with the length from a transmission instant to the corresponding update instant is a continuous random variable subject to certain continuous probability distribution; while the attacks' cardinal number is a discrete random variable supposed to be subject to Poisson distribution, so the inter-attack time (the time between two consecutive attack instants) is subject to an exponential distribution.




The first contribution of this paper is to establish a novel stochastic hybrid model for multiple decentralized networks
under the reasonable assumptions and constraints imposed on the stochastic network delays and Poisson pulsing DoS (Pp-DoS) attacks, in view of the stochastic hybrid formalism of \cite{3.1shs}. The primary objective of this study is to develop a sophisticated model that can effectively accommodate decentralized event-triggered control strategies and stochastic stability analysis. The primary challenge lies in accurately representing data transmissions by integrating network delays with Pp-DoS attacks, particularly in capturing the transitions between transmission and update segments due to delays. To address this, we begin by characterizing Pp-DoS attacks, assuming that the cardinal number of attacks follows a Poisson distribution, and the inter-attack time follows an exponential distribution. Within each network, we introduce a hybrid model featuring two boolean variables. These variables respectively track whether the next event is a transmission or an update event, and whether the next update is successful or not. Additionally, we introduce a time variable and an integer variable, representing the time elapsed since the latest transmission and the total number of transmissions over time. Furthermore, we introduce a memory variable to store certain values related to the medium access protocol at the moment of a transmission, also used to model the update event at an update instant. This approach represents a significant advancement over existing methods, particularly in its ability to handle the intricacies of decentralized event-triggered control strategies and the nuanced dynamics of Pp-DoS attacks. By integrating these variables into a hybrid model, we are able to provide a more comprehensive understanding of the system's behavior, enabling more effective control strategies and improved resilience against cyber-attacks.




The second contribution of this paper is an emulation result for the multiple decentralized networks constrained by the stochastic network delays and Pp-DoS attacks. The main idea in the emulation method is to first design the decentralized controllers that should asymptotically stabilize the closed-loop systems in the absence of networks and attacks, where the closed-loop system can be $\mathcal {L}_{2}$-gain stable (robust; tolerable well) ensured by the designed controllers. To do so, for each network, the designed controller enables the subsystem to tolerate attacks before undergoing instability, which requires limiting the frequency of attacks; otherwise, stability can be lost no matter what control law and triggered strategy are adopted. Hence, the premise of designing a triggered strategy is the frequency of attacks under an appropriate controller with the two-step design approach: designing control law using the nonlinear continuous-time design control methodologies to achieve stabilities, then choosing the network strategy with the appropriate channel bandwidth under appropriate frequency of attacks to maintain the closed-loop asymptotic stability when an attacked network is inserted into the feedback loop. The triggered strategy is designed for the attack-over parts by treating the network induced-error terms as vanishing perturbations, to maintain the asymptotic stability as opposed to the role of the controller (to achieve stabilities); while another condition is required to compensate for the attack-active partes where the network induced-error terms cannot be treated as vanishing perturbations. See \cite{c2007}, \cite{2004a} and \cite{walsh2001} for the techniques applicable to a wide class of networked control systems and network protocols. However, none
of the available results are directly applicable to a class of multiple networks subject to stochastic network delays and Pp-DoS attacks, especially when designing the specific triggered strategies to generate transmission instants. Indeed, the coexistence and randomness of delays and attacks require a more complicated model and novel triggered strategies for transmissions, which remains an open problem.


In particular, we assume that the specific medium access protocol works for each delayed network, for instance, static round-robin (RR) protocol or dynamic try-once-discard (TOD) protocol \cite{2004a}: The access protocol can prioritize nodes (groups of sensor signals) when many (sensor) nodes compete for access to a network in which only a single node can transmit at each instant. That is, the
scheduling protocol decides which components (instead of the entire vector) of the single-network-induced error are set to zero at the update instants \cite{walsh2001}. By building upon \cite{c2007} and \cite{2004a}, a framework can be obtained that allows us to study next to the standard sampled-data setup, also other access protocols including the dynamic TOD protocol or the static RR protocol in one framework without additional burden. In this work, we show that probabilistic asymptotic stability of the networked control system can be maintained when the decentralized controllers are implemented over the delayed and attacked networks, under the tradeoff-conditions between the minimum inter-transmission times, maximum allowable delays (i.e., potentially tolerable delays), and the frequency of attacks.

The third significant contribution of this paper lies in the development of a novel class of decentralized time-regularized (Zeno-free) event-triggered strategies for state-feedback networked control laws under Pp-DoS attacks. In the presence of pulsing attacks occurring at specific update instants, our proposed strategies demonstrate resilience by promptly scheduling new sampled data after failed transmission attempts, thereby ensuring the timely update of the controller. To do so, for each network, we employ a decomposition technique by decomposing the time axis into two mutually independent attack-active and attack-over parts, then prolonging the duration of almost each pulsing attack using a fixed time period, so that the total time intervals of the attack-active parts are related to the attacks' cardinal number which is subject to a Poisson distribution. From here, combining the attack-active and attack-over parts derives constraints to maintain stochastic stability, after the designed triggered strategy works on treating the network induced-error terms as non-vanishing (attack-active) and vanishing (attack-over) perturbations--during which the randomness of the inter-attack time enters into the reset value of the triggered strategy. The memoryless property of exponential distribution and the inherent property (involving the cardinal number) of Poisson distribution justify the rationality and feasibility of the decomposition technique. Furthermore, we address Zeno behavior (infinitely many impulses over a finite period of time) by time-regularizing the triggered strategy, introducing minimum inter-transmission times respectively for the two modes of each network (i.e., attack-active and attack-over). This ensures that an event is triggered only after one minimum inter-transmission time has elapsed since the latest update instant. We note that the two modes of each network will not appear at the same time and have no overlap on the time scale, so Zeno solution will not exhibit for each network if no Zeno behavior exists in each mode. The decentralized nature of the multiple networks (operating asynchronously and independently) further ensures that Zeno solutions do not manifest for the entire system, provided that no Zeno behavior exists within each network. In summary, our approach offers a novel and sophisticated solution, effectively excluding Zeno behavior and demonstrating the system's resilience to Pp-DoS attacks.



The remainder of this paper is organized as follows. Section \ref{subsection2.1} presents the notations. Sections \ref{subsection2.2}, \ref{subsection2.3}, \ref{subsection2.4} and \ref{subsection2.5} present the stochastic hybrid formalism, networked control configuration, Pp-DoS attacks and event-based communication, respectively. Section \ref{section3} establishes mathematical formulation. Sections \ref{subsection3.1} and \ref{subsection3.2} show the constraints on the designed event-triggered strategy and stability guarantees, respectively. Section \ref{exam} shows a cluster formation form of combining multiple small rotorcraft-like aerial vehicles, and conclusions are derived in Section \ref{conclusion}.

\section{Preliminaries}\label{section2}


\subsection{Notations}\label{subsection2.1}

Define $\mathbb{R}:= (-\infty,+\infty)$, $\mathbb{R}_{\geq a}:= [a,+\infty)$, $\mathbb{Z}_{\geq0}:=\{0,1,2,\cdots\}$ and $\mathbb{Z}_{>0}:=\mathbb{Z}_{\geq0}-\{0\}$. For $N \in \mathbb{Z}_{>0}$, define $\bar{N}:=\{1,2,\ldots,N\}$. For $N$ vectors
$x_i \in \mathbb{R}^{n_i}$ , $i \in \bar{N}$, we denote the vector obtained by stacking all
vectors in one (column) vector $x \in\mathbb{R}^{n}$ with $n :=\sum_{i=1}^{N} n_i$ by
$(x_1, x_2,\ldots,x_N )$, i.e., $(x_1, x_2,\ldots,x_N )=[x^{T}_1 , x^{T}_2 ,\ldots,x^{T}_N]^{T}$.
The vectors in $\mathbb{R}^{N}$ consisting of all ones and zeros are denoted
by $\textbf{1}_N$ and $\textbf{0}_N$, respectively. By $|\cdot|$ and $\langle \cdot, \cdot \rangle$ we denote the
Euclidean norm and the usual inner product of real vectors,
respectively. Denote $|A|$ as the determinant of $A\in \mathbb{R}^{n\times n}$, $\lambda_{\max}(A)$ and $\lambda_{\min}(A)$ as the maximum and minimum eigenvalues of $A$. $\mathbb{B}$ (resp., $\mathbb{B}^{\circ}$) is the
closed (resp., open) unit ball in $\mathbb{R}^{n}$. Notations $\vee$ and $\wedge$ denote logical words ``or'' and ``and'', respectively. For a closed set $\mathcal {A}\subset\mathbb{R}^{n}$ and vector $x\in\mathbb{R}^{n}$, define the Euclidean distance as $|x|_{\mathcal {A}}:= \mbox{inf}_{y\in \mathcal {A}} |x-y|$, and the $\varepsilon$-neighborhood for $\varepsilon>0$ as $\mathcal {A}+\varepsilon\mathbb{B}:=\{x\in \mathbb{R}^{n}: |x|_{\mathcal {A}}\leq\varepsilon\}$ (resp., $\mathcal {A}+\varepsilon\mathbb{B}^{\circ}:=\{x\in \mathbb{R}^{n}: |x|_{\mathcal {A}}<\varepsilon\}$). Function $\gamma: \mathbb{R}_{\geq0}\rightarrow \mathbb{R}_{\geq0}$ belongs to the class $\mathcal {K}_{\infty}$ if it is continuous and unbounded, zero at zero and strictly increasing. Function $f: \mathbb{R}^{n}\rightarrow\mathbb{R}^{n}$ is locally Lipschitz if for each $x_{0}\in\mathbb{R}^{n}$, there exist constants $\delta, L>0$ such that for all $x\in\mathbb{R}^{n}$, it holds that $|x-x_{0}|\leq\delta \Rightarrow |f(x)-f(x_{0})|\leq L|x-x_{0}|$ \cite[pg. 9]{clarke1990}. Function $\phi: \mathbb{R}^{n}\rightarrow \mathbb{R}$ is upper semicontinuous if $\limsup_{i\rightarrow\infty}\phi(x_{i})\leq\phi(x)$ holds whenever $\lim_{i\rightarrow\infty}x_{i}=x$ \cite[pg. 13]{Rock1998}. Set-valued mapping $M: \mathbb{R}^{p} \rightrightarrows \mathbb{R}^{n}$ is outer semicontinuous if, for each $(x_{i}, y_{i})\rightarrow(x, y) \in \mathbb{R}^{p}\times\mathbb{R}^{n}$ satisfying $y_{i}\in M(x_{i})$ for $i\in \mathbb{Z}_{\geq0}$, one has $y\in M(x)$ \cite[Def. 5.4]{Rock1998}. Mapping $M$ is locally bounded if, for each bounded set $K\subset\mathbb{R}^{p}$, $M(K):=\bigcup_{x\in K}M(x)$ is bounded.
Given a closed set $\mathcal {A}\subset\mathbb{R}^{n}$, $\rho: \mathbb{R}^{n}\rightarrow\mathbb{R}_{\geq0}$ is \emph{of class} $\mathcal {P}\mathcal {D}(\mathcal {A})$ if it is continuous and $\rho(x)=0\Leftrightarrow x\in \mathcal {A}$.

Let $(\Omega,\mathcal {F})$ be a measure space, e.g., $(\Omega,\mathcal {F}):=(\mathbb{R}^{m}, \textbf{B}(\mathbb{R}^{m}))$, where $\textbf{B}(\mathbb{R}^{m})$ denotes the Borel field over $\mathbb{R}^{m}$, consisting of the subsets of $\mathbb{R}^{m}$ generated from open subsets of $\mathbb{R}^{m}$ through complements and finite and countable unions. A set $A\subset \mathbb{R}^{m}$ is measurable if $A\in \textbf{B}(\mathbb{R}^{m})$, and a set $F\subseteq \Omega$ is $\mathcal {F}$-measurable if $F\in \mathcal {F}$. A mapping $M: \mathbb{R}^{r} \rightrightarrows \mathbb{R}^{m}$ is $\mathcal {F}$-measurable if, for each
open set $\mathcal {O}\subseteq\mathbb{R}^{m}$, $M^{-1}(\mathcal {O}):=\{v\in\mathbb{R}^{r}: M(v)\cap \mathcal {O}\neq\emptyset\}\in \mathcal {F}$ \cite[Def. 14.1]{Rock1998}; if the values of $M$ are closed, the measurability of $M$ is equivalent to the measurability of $M^{-1}(\mathcal {C})$ for any closed set $\mathcal {C}\subseteq \mathbb{R}^{m}$ \cite[Thm. 14.3]{Rock1998}. Denote $(\cdot)^{+}$ as the next jump value, and $\hat{\cdot}$ as the networked value (available at the controller side). For the sake of brevity, we sometimes omit the arguments of a function for the sake of simplicity without causing confusion, especially when the function is with respect to more than one arguments.

\subsection{Stochastic Hybrid Formalism}\label{subsection2.2}

In this paper, the considered network control systems exhibiting two sources of randomness will be modeled as a stochastic hybrid system, so we first report from \cite{3.1shs} the following stochastic hybrid formalism to let the paper be as self-contained as possible,
\begin{subequations}\label{4.5shs:1}
\begin{align}
&\dot{x}\in F(x),x\in C, \label{4.5shs:1A}\\
&x^{+}\in G(x, v^{+}),x\in D,\label{4.5shs:1B}
\end{align}
\end{subequations}
where $x\in \mathbb{R}^{n}$ denotes the system state, $F$ and $G$ denote flow and jump set-valued mappings, respectively, the flow set $C$ and the jump set $D$ respectively allow the flowing and jumping states. Notation $x^{+}$ denotes the instantaneous change depending on jumping states. Placeholder $v^{+}$ holds the place for a sequence of independent, identically distributed (i.i.d.) input random variables $\{\textbf{v}_{n}\}_{n=1}^{\infty}$, which is defined on a probability space $(\Omega,\mathcal {F},\mathbb{P})$, i.e., $\textbf{v}_{n}: \Omega \rightarrow \mathbb{R}^{m}$ for $n\in \mathbb{Z}_{\geq1}$. Given the Borel $\sigma$-field  $\textbf{B}(\mathbb{R}^{m})$ over $\mathbb{R}^{m}$, it holds that $\textbf{v}^{-1}_{n}(A) := \{\omega\in \Omega: \textbf{v}_{n}(\omega)\in A\}\in\mathcal {F}$, and $\mathbb{P}\{\omega\in\Omega: \textbf{v}_{n}(\omega)\in A\}$ is well-defined and independent of $n$, for $A\in \textbf{B}(\mathbb{R}^{m})$. Notation $\mu$ denotes the distribution function derived from $(\Omega,\mathcal {F},\mathbb{P})$, then $\mu: \textbf{B}(\mathbb{R}^{m})\rightarrow[0,1]$ holds with $\mu(A):=\mathbb{P}\{\omega\in\Omega: \textbf{v}_{n}(\omega)\in A\}$, see \cite[Sec. 2.1 and 11.1]{Fristedt1997}. Denote the formalism of the stochastic hybrid system (\ref{4.5shs:1}) as $\mathcal {H}:= (C,F,D,G, \mu)$, where the solution' uniqueness is not assumed due to set-valued flow/jump mappings and the potentially non-trivial overlap of flow and jump sets, see \cite{3.1shs}.

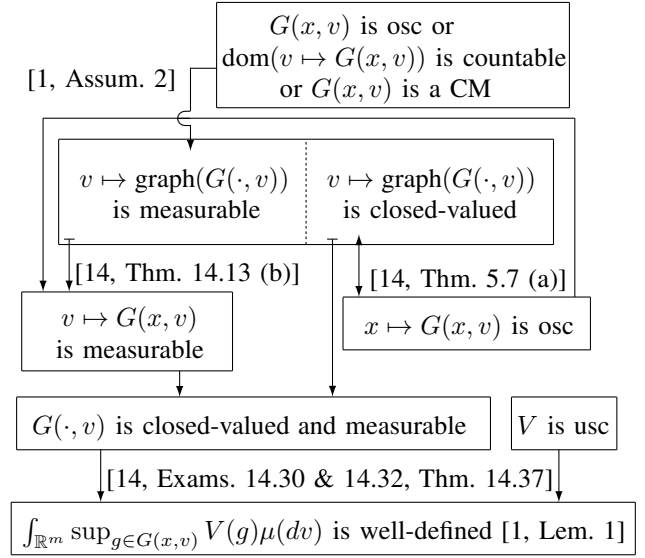
\begin{figure}[!htb]
\setlength{\unitlength}{0.7mm}
$$\begin{picture}(116,100)

\put(12,70){$v\mapsto \mbox{graph}(G(\cdot, v))$~~~~$v\mapsto \mbox{graph}(G(\cdot, v))$}
\put(13,64){~~~is measurable~~~~~~~~~is closed-valued}
\multiput(55,59)(0,1){20}{\line(0,1){0.5}}
\put(8,59){\line(1,0){94}}
\put(8,79){\line(1,0){94}}
\put(8,59){\line(0,1){20}}
\put(102,59){\line(0,1){20}}

\put(65,59){\vector(0,1){2}}
\put(65,59){\vector(0,-1){10}}
\put(67,51){\cite[Thm. 5.7 (a)]{Rock1998}}

\put(66,42){$x\mapsto G(x, v)$ is osc}
\put(62,49){\line(1,0){47}}
\put(62,39){\line(1,0){47}}
\put(62,39){\line(0,1){10}}
\put(109,39){\line(0,1){10}}

\put(106,49){\line(0,1){34}}
\put(106,83){\line(-1,0){101}}
\put(5,83){\vector(0,-1){33}}

\put(59,60){\line(1,0){2}}
\put(60,60){\vector(0,-1){30}}

\put(9,60){\line(1,0){2}}
\put(10,60){\vector(0,-1){10}}
\put(11,52){\cite[Thm. 14.13 (b)]{Rock1998}}

\put(8.5,43.5){$v\mapsto G(x, v)$}
\put(7.5,37.5){is measurable}
\put(1,35){\line(1,0){40}}
\put(1,50){\line(1,0){40}}
\put(1,35){\line(0,1){15}}
\put(41,35){\line(0,1){15}}

\put(31,35){\vector(0,-1){5}}
\put(3,23){$G(\cdot, v)$ is closed-valued and measurable}
\put(0,20){\line(1,0){90}}
\put(0,30){\line(1,0){90}}
\put(0,20){\line(0,1){10}}
\put(90,20){\line(0,1){10}}

\put(95,23){$V$ is usc}
\put(94,20){\line(1,0){20}}
\put(94,30){\line(1,0){20}}
\put(94,20){\line(0,1){10}}
\put(114,20){\line(0,1){10}}

\put(103,20){\vector(0,-1){10}}
\put(16,20){\vector(0,-1){10}}
\put(17,13){\cite[Exams. 14.30 \& 14.32, Thm. 14.37]{Rock1998}}
\put(1,3){$\int_{\mathbb{R}^{m}}\sup_{g\in G(x,v)}V(g)\mu(dv)$ is well-defined \cite[Lem. 1]{3.1shs}}
\put(-1,0){\line(1,0){118.5}}
\put(-1,10){\line(1,0){118.5}}
\put(-1,0){\line(0,1){10}}
\put(117.5,0){\line(0,1){10}}

\put(48,99){$G(x, v)$ is osc or}
\put(39.1,93){$\mbox{dom}(v\mapsto G(x, v))$ is countable}
\put(50,87){or $G(x, v)$ is a CM}
\put(38,105){\line(1,0){67}}
\put(38,85){\line(1,0){67}}
\put(38,85){\line(0,1){20}}
\put(105,85){\line(0,1){20}}

\put(2,89){\cite[Assum. 2]{3.1shs}}
\put(38,92.5){\line(-1,0){5}}
\put(33,92.6){\line(0,-1){8.24}}
\put(33,81.74){\vector(0,-1){5}}
\put(30.1,81.75){$\in$}
\end{picture}$$
\caption{A roadmap of the relationship for Assumption 1-3b), where ``$\mbox{dom}$'' denotes domain (i.e., the domain of a mapping), ``CM'' denotes Carath\'{e}odory mapping, ``osc'' denotes ``outer semicontinuous'' and ``usc'' denotes ``upper semicontinuous''.}\label{item3b}
\end{figure}


The following mild regularity conditions are adopted for the stochastic hybrid system (\ref{4.5shs:1}) from \cite[Assum. 1-2]{3.1shs}.

\textbf{Assumption 1} (Stochastic hybrid conditions).
\begin{itemize}
  \item [A1)] The sets $C,D \subset \mathbb{R}^{n}$ are closed;
  \item [A2)] The mapping $F: \mathbb{R}^{n}\rightrightarrows\mathbb{R}^{n}$ is outer-semicontinuous, locally
bounded with nonempty convex values on $C$;
  \item [A3)] \begin{itemize}
             \item [3a)] The mapping $G: \mathbb{R}^{n}\times\mathbb{R}^{m}\rightrightarrows\mathbb{R}^{n}$ is locally bounded;
             \item [3b)] The mapping $v\mapsto \mbox{graph}(G(\cdot, v)):= \{(x, y)\in \mathbb{R}^{2n}: y\in G(x, v)\}$ is measurable (with respect to the Borel $\sigma$-algebra on $\mathbb{R}^{m}$) with closed values;
           \end{itemize}
\end{itemize}

Explanations on these regularity conditions in Assumption 1: Random solutions to (\ref{4.5shs:1}) are ensured simultaneously by the local boundedness (see items A2) and 3a)) and by the measurability (see item 3b)) \cite[Sec. IV]{4.5shs}; and additionally, the nominal robustness is guaranteed by the local boundedness and the outer semicontinuity, which enables the equivalence between non-uniform and uniform versions of stochastic stability properties, see \cite[Claim 1, Thms. 2-3]{4.6shs} and \cite[Thms. 4-5]{3.2shs} for the difference inclusions, as well as \cite{7.1shs} for (\ref{4.5shs:1}). The measurability in item 3b) implies that $v\mapsto G(x, v)$ is measurable for any $x\in \mathbb{R}^{n}$, see \cite[Prop. 2]{3.6shs}, \cite[Remark 1]{4.7shs} and \cite[Thm. 14.13(b)]{Rock1998}, used to establish certain well-posed integrals used later; and additionally, the condition of closed values in item 3b) is equivalent to the outer semicontinuity of $x\mapsto G(x, v)$ for each $v\in \mathbb{R}^{n}$, see \cite[Thm. 5.7(a)]{Rock1998}. As analyzed in \cite{3.1shs} and \cite{3.3shs}, item 3b) holds if $G(x, v)$ is outer semicontinuous; holds if the domain of $v\mapsto G(x, v)$ is countable; also holds if a single-valued mapping $G$ is a Carath\'{e}odory mapping \cite[Exam. 14.15]{Rock1998}, i.e., $x\mapsto G(x, v)$ is continuous and $v\mapsto G(x, v)$ is measurable. We note that Assumption 1 essentially agrees with the non-stochastic hybrid basic assumptions in \cite[Assumption 6.5]{2012teel}, when $G$ does not depend on $v$. A relationship route is presented in Fig. \ref{item3b} for Assumption 1-3b).


\subsection{Networked Control Configuration}\label{subsection2.3}




This paper utilizes stochastic hybrid analysis tools to facilitate more efficient analysis and ensure more reliable conclusions. The networked control system under consideration, with state $x_p\in \mathbb{R}^{n_{p}}$, is represented as the stochastic hybrid formalism (\ref{4.5shs:1}) through the emulation method \cite{c2007}: A controller with state $x_c\in \mathbb{R}^{n_{c}}$ is first designed in advance that stabilizes the system in the absence of multiple networks that operate asynchronously and independently (i.e., decentralized), then implemented over the attacked networks to maintain the stability by designing the decentralized triggered strategies which generate transmission instants $\{t^{i}_{j}\}_{j\in \mathbb{Z}_{\geq0}}$ satisfying $0\leq t^{i}_{j}< t^{i}_{j+1}$ for the $i$th-network, $i\in \bar{N}$. In what follows, two sources of randomness are considered in multiple decentralized networks: the networked communication is constrained by the stochastic network delays and also by Poisson pulsing denial-of-service (Pp-DoS) attacks, where ``pulsing'' here means each attack pulse lasts for a negligible short period of time (each attack's duration is negligible). The network delay denotes the continuous time from a transmission instant (sampled data enters the network) to the corresponding update instant (sampled data leaves the network), and is supposed to be subject to certain continuous probability distribution; while the attacks' cardinal number is a discrete random variable supposed to be subject to a Poisson distribution. We assume there is no communication delay or sampling delay outside the networks.




\begin{figure}[!htb]
\setlength{\unitlength}{0.7mm}
$$\begin{picture}(160,72)

\put(81,66){$\mathcal {P}$}
\put(70,64){\line(1,0){25}}
\put(70,72){\line(1,0){25}}
\put(70,64){\line(0,1){8}}
\put(95,64){\line(0,1){8}}
\put(100,70){$x_{pi}(t)$}
\put(95,68){\line(1,0){18}}

\put(60,68){\vector(1,0){10}}
\put(7,68){\line(0,-1){47}}

\put(47,66){$\mathcal {C}$}
\put(35,64){\line(1,0){25}}
\put(35,72){\line(1,0){25}}
\put(35,64){\line(0,1){8}}
\put(60,64){\line(0,1){8}}
\put(7,68){\vector(1,0){28}}
\put(7,70.5){$\hat{x}_{pi\ell}(t^{i}_{j}+\tau^{i}_{j})$}

\put(58,52.5){$\text{Acknowledgement}$}
\multiput(50,50)(1,0){55}{\line(1,0){0.5}}
\multiput(50,58)(1,0){55}{\line(1,0){0.5}}
\multiput(50,50)(0,1){8}{\line(0,1){0.5}}
\multiput(105,50)(0,1){8}{\line(0,1){0.5}}
\multiput(55,63)(0,-1){6}{\line(0,1){0.5}}
\put(55,60){\vector(0,-1){2}}
\multiput(100,49)(0,-1){6}{\line(0,1){0.5}}
\put(100,46){\vector(0,-1){2}}

\put(7,21){\line(1,0){8}}
\put(9,47){$t^{i}_{j}+\tau^{i}_{j}$}
\put(15,2){\line(0,1){42}}
\multiput(15,39)(1,0){25}{\line(1,0){0.5}}
\multiput(15,26)(1,0){25}{\line(1,0){0.5}}
\put(26,30){$\mathcal {N}_{1}$}
\put(28,20){$\vdots$}
\put(26,9){$\mathcal {N}_{N}$}
\put(27.5,42){$\tau^{i}_{j}$}
\put(26.5,42){\vector(-1,0){11.5}}
\put(32.5,42){\vector(1,0){7.5}}
\put(39,47){$t^{i}_{j}$}
\put(40,2){\line(0,1){2}}
\put(40,4){\line(2,1){7}}
\put(40,11){\line(0,1){15}}
\put(40,26){\line(2,1){7}}
\put(40,33){\line(0,1){11}}
\multiput(15,18)(1,0){25}{\line(1,0){0.5}}
\multiput(15,4)(1,0){25}{\line(1,0){0.5}}
\put(1,-3){$\text{Delayed}$ $\text{networks}$ $\text{under}$ $\text{attacks}$}

\put(44,32){$s_{1}$}
\put(46,26){\line(1,0){32}}
\put(56,28.5){$x_{p1}(t^{1}_{j})$}
\put(54,6.5){$x_{pN}(t^{N}_{j})$}
\put(46,4){\line(1,0){32}}
\put(44,10){$s_{N}$}

\multiput(40,33)(1,-1){5}{\line(1,-1){0.5}}
\put(43,30.2){\vector(1,-1){2}}
\multiput(40,11)(1,-1){5}{\line(1,-1){0.5}}
\put(43,8.2){\vector(1,-1){2}}


\put(113,21){\vector(-1,0){8}}
\put(105,2){\line(0,1){42}}
\multiput(85,26)(1,0){20}{\line(1,0){0.5}}
\put(85,44){\line(1,0){20}}
\put(93,40){$x_{p11}$}
\put(95,33){$\vdots$}
\put(93,28){$x_{p1\ell}$}
\put(95,20){$\vdots$}
\put(92,14.5){$x_{pN1}$}
\put(95,7.5){$\vdots$}
\put(92,3.5){$x_{pN\ell}$}
\put(85,2){\line(0,1){2}}
\put(85,4){\line(-2,1){7}}
\put(85,11){\line(0,1){15}}
\put(85,26){\line(-2,1){7}}
\put(85,33){\line(0,1){11}}
\multiput(85,18)(1,0){20}{\line(1,0){0.5}}

\put(85,2){\line(1,0){20}}
\put(80,-3){$\text{Triggered}$ $\text{strategies}$}

\put(78,32){$q_{1}$}
\multiput(85,33)(-1,-1){5}{\line(-1,-1){0.5}}
\put(82.2,30.2){\vector(-1,-1){2}}
\put(76,10){$q_{N}$}
\multiput(85,11)(-1,-1){5}{\line(-1,-1){0.5}}
\put(82.2,8.2){\vector(-1,-1){2}}

\put(113,21){\line(0,1){47}}

\end{picture}$$
\caption{Networked control configuration with plant $\mathcal {P}$, controller $\mathcal {C}$ and multiple decentralized networks $\mathcal {N}_{i}$ subject to stochastic network delays and Pp-DoS attacks, $i\in \bar{N}$. See the beginning of Subsection \ref{subsection2.3} for details on other parameters.}\label{ack}
\end{figure}


\begin{figure*}
\centering\subfigure[]{
\begin{minipage}[t]{0.47\textwidth}
\label{xp12:subfig:a} 
\includegraphics[width=3.2in,height=2.3in]{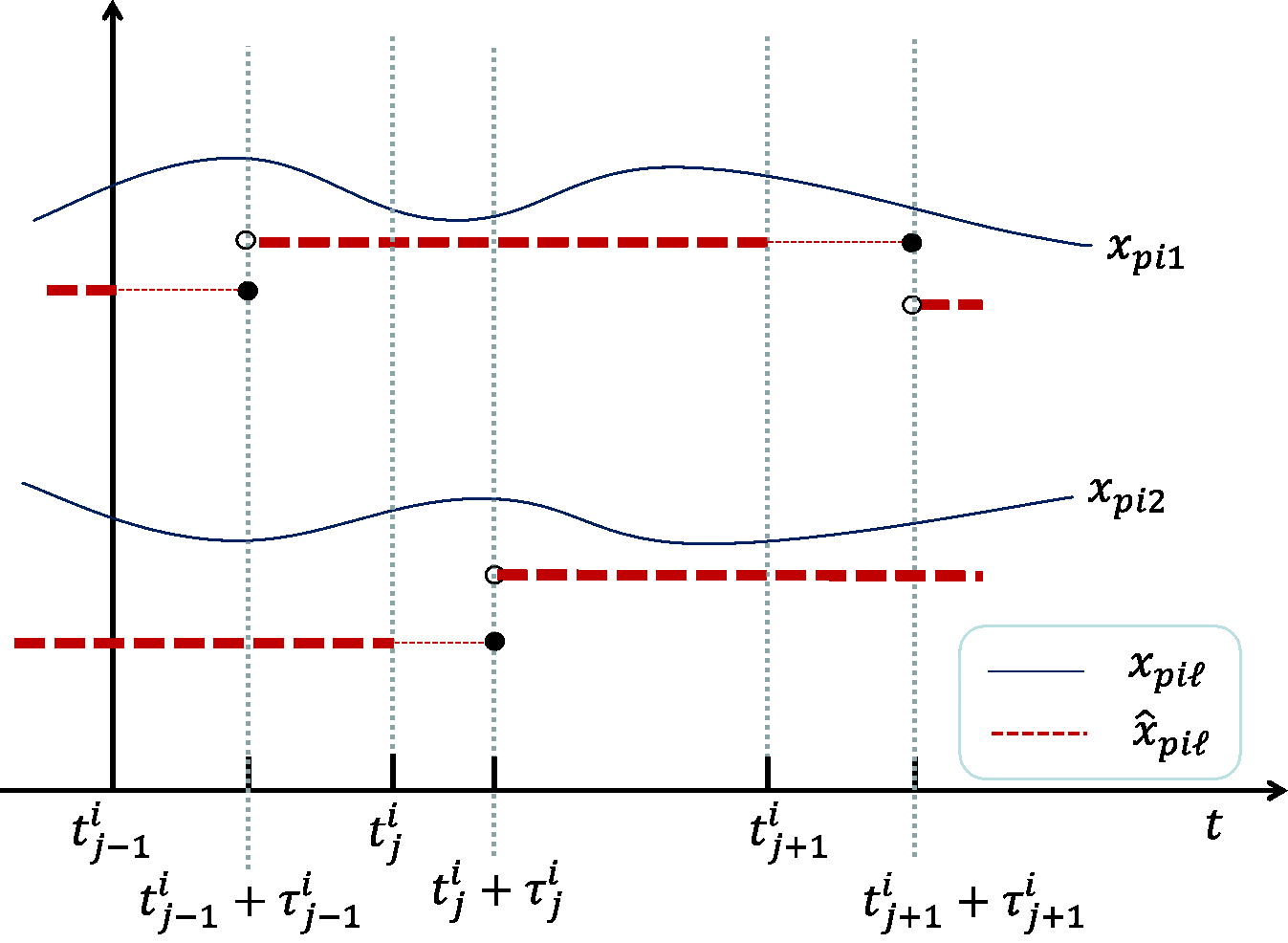}
\end{minipage}}
\centering \subfigure[]{
\begin{minipage}[t]{0.47\textwidth}
\label{xp12:subfig:b} 
\includegraphics[width=3.2in,height=2.3in]{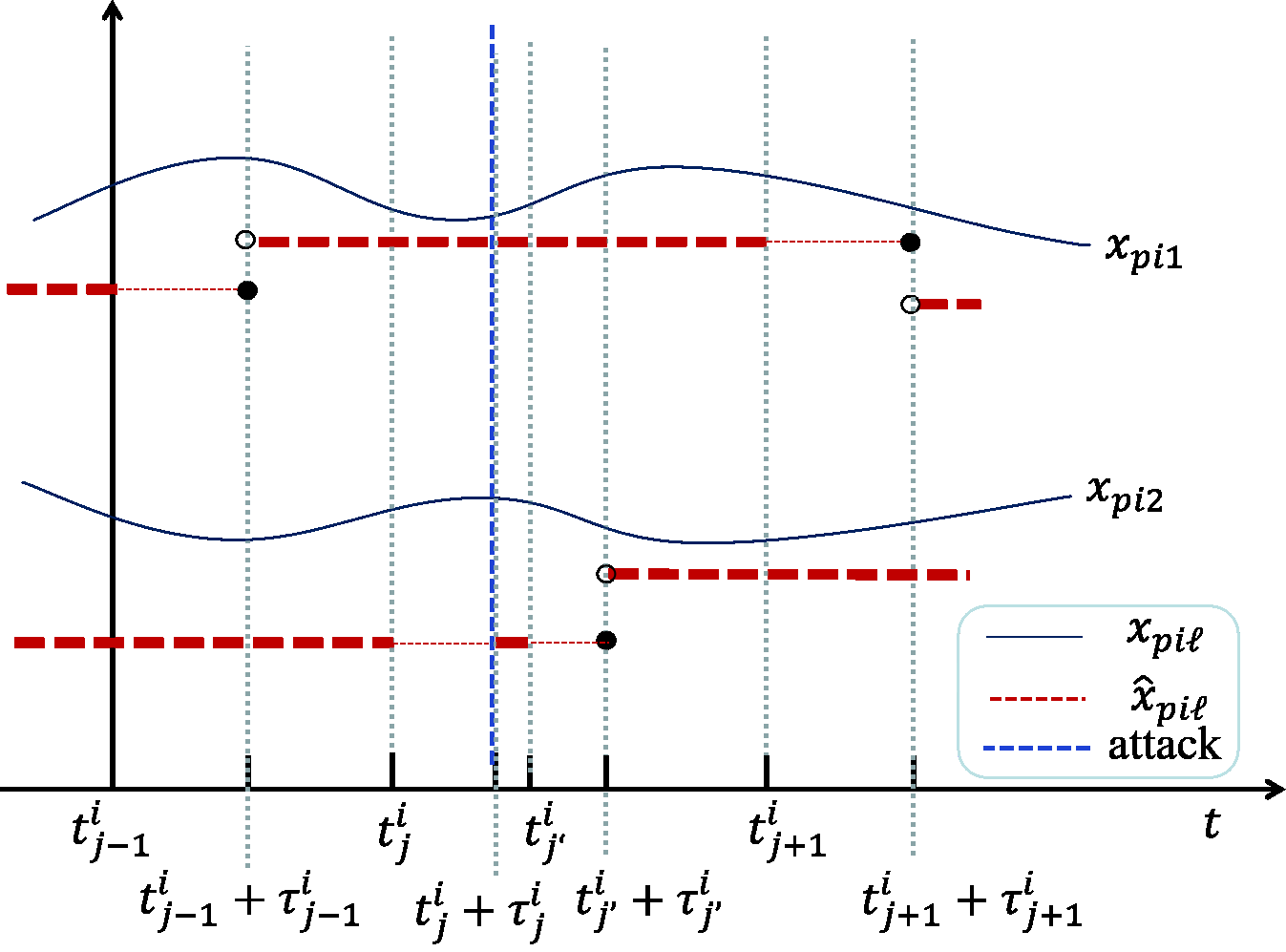}%
\end{minipage}}
\caption{A typical evolution of $x_{pi\ell}$ and $\hat{x}_{pi\ell}$ for $\ell=2$, where $t^{i}_{j-1}$, $t^{i}_{j}$ and $t^{i}_{j+1}$ are three consecutive transmission instants for $x_{pi}=(x_{pi1},x_{pi2})$ in the absence of attacks. Notations $\bullet$ and $\circ$ represent that the piecewise constant $\hat{x}_{pi\ell}$ is left-continuous. (a) No attacks in $[t^{i}_{j},t^{i}_{j}+\tau^{i}_{j}]$. (b) A Pp-DoS attack occurs at update instant $t^{i}_{j}+\tau^{i}_{j}$. See explanations above the model (\ref{asc1}).}\label{xp12}
\end{figure*}

In $N$ networks, define $x_{p} := (x_{p1}, x_{p2},\ldots,x_{pN})$, $\hat{x}_{p} = (\hat{x}_{p1}, \hat{x}_{p2},\ldots,\hat{x}_{pN})$, $(x_{pi1},\ldots,x_{pi\ell})=:x_{pi}\in \mathbb{R}^{n_{p,i}}$, $\ell\in \mathbb{Z}_{>0}$. Then the network-induced error, denoted as $e:=(e_{1},\ldots,e_{i},\ldots,e_{N})$ for $i\in \bar{N}$, can be defined as $\hat{x}_{p}-x_{p} =:e\in \mathbb{R}^{n_{p}}$ and $\hat{x}_{pi}-x_{pi} =:e_{i}\in \mathbb{R}^{n_{p,i}}$. Then Fig. \ref{ack} shows the decentralized networked control configuration: The communication of sensor data between $\mathcal {P}$ and $\mathcal {C}$ is performed via multiple networks $\mathcal {N}_1,\mathcal {N}_2,\ldots, \mathcal {N}_N$ that operate asynchronously and independently. The sampled state $x_{pi}(t^{i}_{j})$ is transmitted over the network $\mathcal {N}_i$ to the controller, and the update of $\hat{x}_{pi}$ at the controller side occurs at the time delay of $\tau^{i}_{j}$ time units after the successful transmission attempts, e.g., at $t^{i}_{j}+\tau^{i}_{j}$ where $\{\tau^{i}_{j}\}_{j\in \mathbb{Z}_{\geq0}}$ are a sequence of i.i.d. continuous random variables for each $i\in \bar{N}$.


In Fig. \ref{ack}, an acknowledgement scheme enables the decentralized triggered strategies to have knowledge about the reception of packages at the controller side, where toggle $q_{i}$ is controlled by the decentralized triggered strategies, i.e., the $i$th-strategy generates the transmission instants $t^{i}_{j}$ in terms of $e_{i}$, for each $i\in \bar{N}$ and each $j\in \mathbb{Z}_{\geq0}$, while toggle $s_{i}$ is controlled by the medium access protocol $h_{i}(j,e_{i}(t^{i}_{j}))$. Independent of the plant and controller dynamics \cite{c2007}, \cite{2004a}, the protocol $h_{i}$ typically employed in case $\mathcal {N}_i$ is shared by multiple (sensor) nodes, when only one node is granted access to $\mathcal {N}_i$ at each transmission instant, $i\in \bar{N}$. A node here refers to a group of sensor signals, which are always transmitted in a single packet \cite{walsh2001}. That is, on the basis of $\ell,j$ and the networked error $e_{i}$, the function $h_{i}$ is an update function that is related to the protocol that determines which node (the single packet with the highest priority \cite{2004a}, i.e., a subset of the entries of $x_{pi}$, denoted as $x_{pi\ell}$) is granted access to the communication channel $\mathcal {N}_i$ at each transmission instant $t^{i}_{j}$, for each $j\in \mathbb{Z}_{\geq0}$ and $\ell\in \mathbb{Z}_{>0}$. As a result, $h_{i}$ decides certain $\ell\in \mathbb{Z}_{>0}$ for $x_{pi\ell}$ that can be granted access to $\mathcal {N}_i$ to update $\hat{x}_{pi\ell}$ at the controller side, so that the corresponding components (instead of the entire vector) of $e_{i}$ are set to a new value at the update instants \cite{walsh2001}. On the other hand, the controller is implemented using the zero-order-hold devices, to hold $\hat{x}_{pi\ell}$ as a piecewise constant segmented at its update instants, where $\hat{x}_{pi\ell}$ is left-continuous at its update instants (rather than its transmission instants, due to the network delays). See Fig. \ref{xp12:subfig:a} for an example: \\
At transmission instants $t^{i}_{j-1}$ and $t^{i}_{j+1}$, $x_{pi1}$ is granted access to the communication channel $\mathcal {N}_i$, then $t^{i}_{j-1}+\tau^{i}_{j-1}$ and $t^{i}_{j+1}+\tau^{i}_{j+1}$ are two consecutive update/jump/reset instants of $\hat{x}_{pi1}$ at the controller side, respectively, so $\hat{x}_{pi1}$ remains constant during $(t^{i}_{j-1}+\tau^{i}_{j-1},t^{i}_{j+1}+\tau^{i}_{j+1}]$; while $x_{pi2}$ is granted access to $\mathcal {N}_i$ at transmission instant $t^{i}_{j}$ (so $t^{i}_{j}$ is an update instant of $\hat{x}_{pi2}$), resulting in the update/jump/reset of $\hat{x}_{pi2}$ at $t^{i}_{j}+\tau^{i}_{j}$ on the controller side. When a Pp-DoS attack exists in Fig. \ref{xp12:subfig:b}, the promised update instant $t^{i}_{j}+\tau^{i}_{j}$ does not take effect (be discarded), then another new transmission instant $t^{i}_{j'}$ is arranged for the update instant $t^{i}_{j'}+\tau^{i}_{j'}$, to update $\hat{x}_{pi2}$ on the controller side. It is worth noting that in what follows, despite the use $x_{pi}$/$\hat{x}_{pi}$ as the analysis and computation object, it will ultimately be specific and applied to $x_{pi\ell}$/$\hat{x}_{pi\ell}$ relying on $h_{i}$ for implementation \cite[Exams. 1-2]{2004a}, especially when we consider the jump of $\hat{x}_{pi}$. In other words, the jump of $\hat{x}_{pi}$ at the controller side is ultimately the jump of $\hat{x}_{pi\ell}$ determined by $h_{i}$.

%





In terms of the emulation-based networked control configuration introduced above for multiple networks, now we present the differential equations of the continuous-time system state $x_p\in \mathbb{R}^{n_{p}}$ for the plant $\mathcal {P}$, and that of the controller state $x_c\in \mathbb{R}^{n_{c}}$ for the dynamic state-feedback controller $\mathcal {C}$, i.e.,
\begin{subequations}\label{asc1}
\begin{align}
&\mathcal {P}:~\dot{x}_p= f_p(x_p,u),\label{asc1A}\\
&\mathcal {C}:\dot{x}_c=f_c(x_c,\hat{x}_{p}),~u=g_c(x_c,\hat{x}_{p}),\label{asc1B}
\end{align}
\end{subequations}
where notation $u\in \mathbb{R}^{n_{u}}$ denotes the control input. Functions $f_p$ and $f_c$ are supposed to be continuous; and (supposed to be) continuously differentiable function $g_{c}$ can be relaxed to be discontinuous, since Assumption 1 can also be ensured by getting the convex closure of discontinuous functions, see \cite[pg. 74-75]{2012teel}. Denote $\mathcal {E}$ as the closed equilibrium point set of the closed-loop $(x_p,x_c)$-system (\ref{asc1}).

\begin{figure*}[!htb]
\centering\subfigure[]{
\begin{minipage}[t]{0.47\textwidth}
\label{hatxp:subfig:a} 
\includegraphics[width=3.2in,height=1.9in]{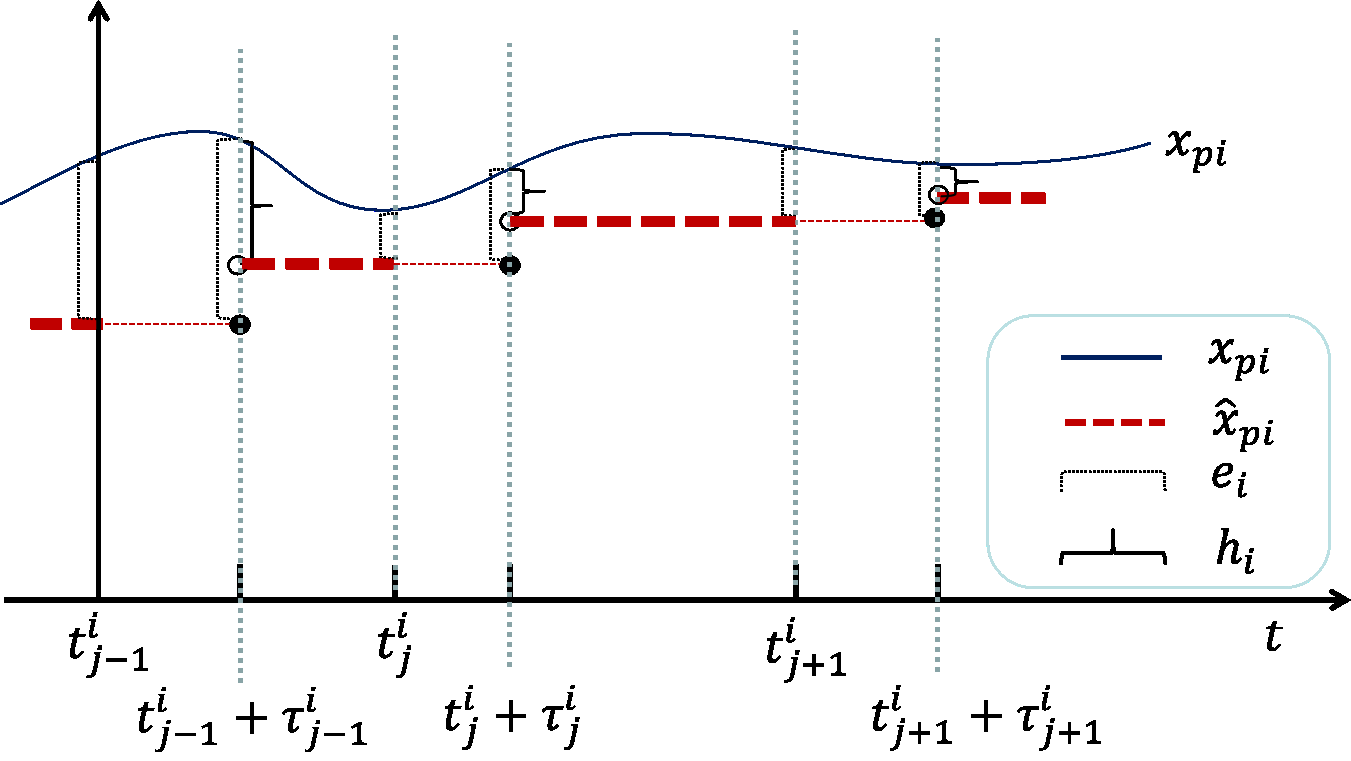}
\end{minipage}}
\centering \subfigure[]{
\begin{minipage}[t]{0.47\textwidth}
\label{hatxp:subfig:b} 
\includegraphics[width=3.2in,height=1.9in]{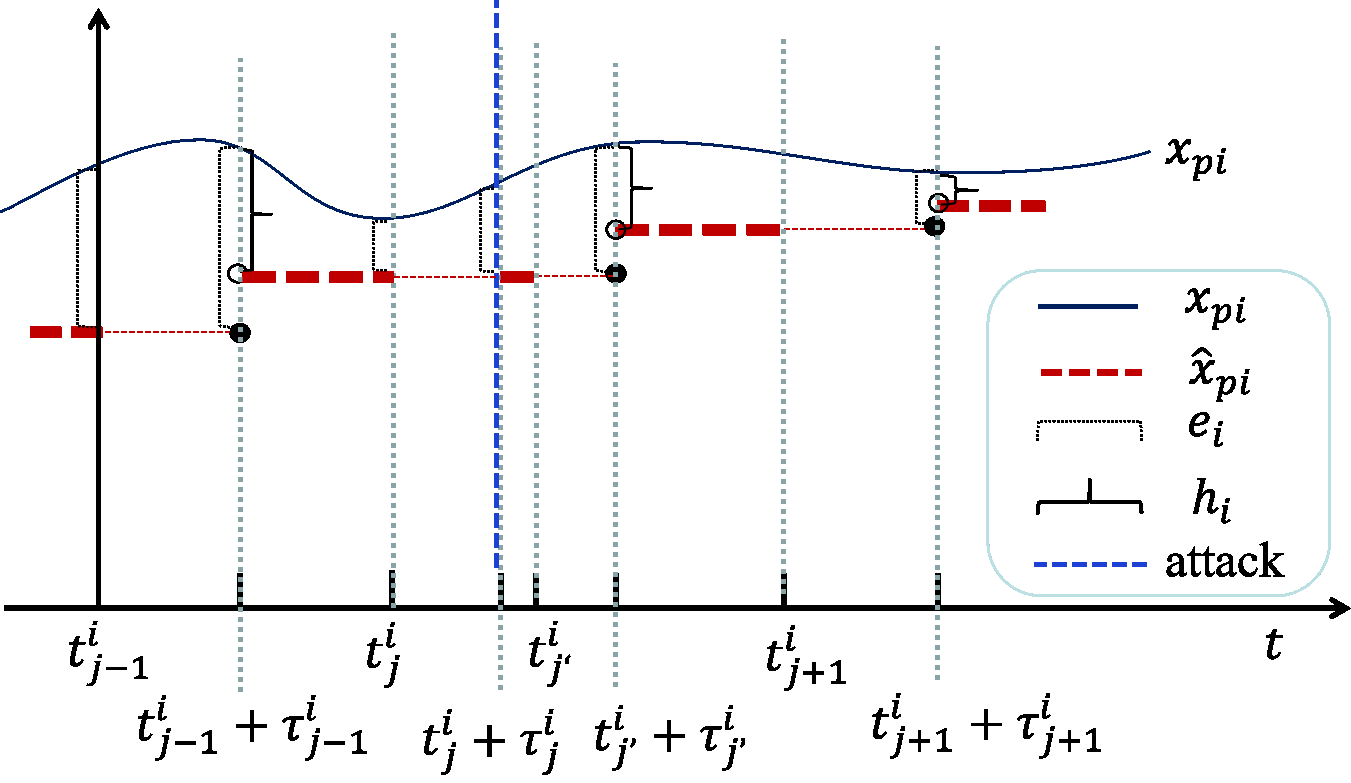}
\end{minipage}}
\caption{Evolution illustrations for $x_{pi}$ and $\hat{x}_{pi}$, where $t^{i}_{j-1}$, $t^{i}_{j}$ and $t^{i}_{j+1}$ are three consecutive transmission instants for $x_{pi}$ in the absence of attacks. Notations $\bullet$ and $\circ$ represent that the piecewise constant $\hat{x}_{pi}$ is left-continuous. (a) No attacks in $[t^{i}_{j},t^{i}_{j}+\tau^{i}_{j}]$. (b) A Pp-DoS attack occurs at update instant $t^{i}_{j}+\tau^{i}_{j}$. See explanations below the system model (\ref{asc1}).}\label{hatxp}
\end{figure*}

At the controller side, the networked value $\hat{x}_{p}\in \mathbb{R}^{n_{p}}$ is implemented by the zero-order-hold devices to hold its each element $\hat{x}_{pi}$ as a piecewise constant, when $(\hat{x}_{pi1},\cdots, \hat{x}_{pi\ell})$ in Fig. \ref{xp12} are regarded as a whole $\hat{x}_{pi}$ in Fig. \ref{hatxp}. See Fig. \ref{hatxp} for simplicity of exposition: Between two consecutively successful update instants, $\hat{x}_{pi}$ remains a constant, e.g., $\dot{\hat{x}}_{pi}(t)=0$ holds for $t\in (t^{i}_{j-1}+\tau^{i}_{j-1},t^{i}_{j}+\tau^{i}_{j}]$ in Fig. \ref{hatxp:subfig:a}, then the controller side receives and updates the networked value $\hat{x}_{pi}$ ($\hat{x}_{pi}$ has a jump/reset) respectively at successful update instants $t^{i}_{j-1}+\tau^{i}_{j-1}$, $t^{i}_{j}+\tau^{i}_{j}$ and $t^{i}_{j+1}+\tau^{i}_{j+1}$, for $i\in \bar{N}$ and $j\in \mathbb{Z}_{\geq0}$; however, when a Pp-DoS attack occurs at the promised update instant $t^{i}_{j}+\tau^{i}_{j}$ in Fig. \ref{hatxp:subfig:b}, the promised update instant does not take effect, so the value $\hat{x}_{pi}$ at the controller side will not update at $t^{i}_{j}+\tau^{i}_{j}$ ($\hat{x}_{pi}$ does not have a jump/reset), but will continue (often assumed) to maintain/utilize the current constant until another new update instant $t^{i}_{j'+1}+\tau^{i}_{j'+1}$, i.e., $\dot{\hat{x}}_{pi}(t)=0$ for $t\in (t^{i}_{j-1}+\tau^{i}_{j-1},t^{i}_{j'}+\tau^{i}_{j'}]$, see Fig. \ref{hatxp:subfig:b}. As analyzed for Fig. \ref{ack}, on the basis of $\ell,j$ and $e_{i}$, the protocol $h_{i}$ determines which node (i.e., $x_{pi\ell}$) is granted access to $\mathcal {N}_i$ at each transmission instant, so that the corresponding components (instead of the entire vector) of $e_{i}$ are set to a new value at the update instants, see Fig. \ref{xp12} for an example.

Throughout the paper, we have the convention for a worst-case scenario: if the Pp-DoS attack occurs within $(t^{i}_{j},t^{i}_{j}+\tau^{i}_{j})$, then it happens exactly when a packet passes by in the network. Following Fig. \ref{ack}, the $i$th-triggered strategy continuously monitors $x_{pi}(t)$, then the sampled data $x_{pi}(t^{i}_{j})$ is tried to be transmitted via network $\mathcal {N}_i$ at $t^{i}_{j}$ (e.g., when the $i$th-strategy is triggered and generates transmission attempt $t^{i}_{j}$), then\\
(Ia) if no attack occurs during $[t^{i}_{j},t^{i}_{j}+\tau^{i}_{j}]$: the sampled value $x_{pi}(t^{i}_{j})$ (or a subset of the entries of $x_{pi}$ chosen via $h_{i}$) is transmitted via $\mathcal {N}_i$ for generating its networked value $\hat{x}_{pi}$, and the networked value will be available at the controller side at $t^{i}_{j}+\tau^{i}_{j}$, then controller updates its knowledge of $\hat{x}_{pi}$, i.e., $\hat{x}_{pi}$ has a jump with $\hat{x}^{+}_{pi}(t^{i}_{j}+\tau^{i}_{j})=x_{pi}(t^{i}_{j})+h_{i}(j,e_{i}(t^{i}_{j}))$ \cite{2004a}, \cite{q2010hemeels}, see Fig. \ref{hatxp:subfig:a};\\
(Ib) if a Pp-DoS attack occurs during $[t^{i}_{j},t^{i}_{j}+\tau^{i}_{j}]$, then the network communication is plagued which causes the packet loss of the sampled data $x_{pi}(t^{i}_{j})$, thereby no new value $\hat{x}_{pi}$ is available at the controller side at $t^{i}_{j}+\tau^{i}_{j}$. As a result, the current value $\hat{x}_{pi}$ kept at the controller side will not be updated ($\hat{x}_{pi}$ does not have a jump) until another successful update instant $t^{i}_{j'}+\tau^{i}_{j'}$ after the corresponding successful transmission attempt $t^{i}_{j'}$, see Fig. \ref{hatxp:subfig:b}. In the attacked transmission/update process, an acknowledgement scheme enables the triggered strategy to have knowledge about the reception of packages at the controller side, so that the triggered strategy can generate another transmission attempt $t^{i}_{j'}$ as soon as possible.

To justify items (Ia)-(Ib) where the cross/confusion use of sample packets potentially caused by network delays is avoided, here we consider the small-delay case where the reception/update of the new sampled data packet (at the controller side) occurs before the next transmission is due, e.g., $t^{i}_{j}+\tau^{i}_{j}<t^{i}_{j+1}$. Sometimes, a long delay that grows beyond the inter-transmission intervals may result in packet reordering, essentially amounts to a packet dropout (especially when the receiver discards outdated arrivals): This scenario may cause confusion, when simultaneously considering the Pp-DoS attacks that also result in packet dropouts. To do so, for network $\mathcal {N}_i$, $i\in \bar{N}$, a sequence of continuous random variables $\{\tau^{i}_{j}\}_{j\in \mathbb{Z}_{\geq0}}$ subject to certain continuous probability distribution are upper-bounded by a maximum allowable delay $\tau^{i}_{mad}$ (i.e., a bound on the tolerable delays), as characterized in the following assumption.

\textbf{Assumption 2}. In each network $\mathcal {N}_i$, $i\in \bar{N}$, the time delay $\tau^{i}_{j}$ is subject to a continuous probability distribution and satisfies $0\leq \tau^{i}_{j} \leq \tau^{i}_{mad} < \inf_{j\in \mathbb{Z}_{\geq0}}(t^{i}_{j+1}-t^{i}_{j})$ for two consecutive transmission instants $t^{i}_{j}$ and $t^{i}_{j+1}$.



Consider the network-induced error $\hat{x}_{p}-x_{p}=e\in \mathbb{R}^{n_{p}}$ defined in Subsection \ref{subsection2.3}, where $e=(e_{1},\ldots,e_{i},\ldots,e_{N})$ and $\hat{x}_{pi}-x_{pi}=e_{i}\in \mathbb{R}^{n_{p,i}}$ (see Fig. \ref{hatxp}). Then it holds from the system equation (\ref{asc1A}) that
\begin{equation}\label{s4e}
\begin{split}
\dot{e}_{i} =-\dot{x}_{pi}, t\in (t^{i}_{j}+\tau^{i}_{j},t^{i}_{j+1}+\tau^{i}_{j+1}],
\end{split}
\end{equation}
and that
\begin{equation}\label{e4e}
\begin{split}
\dot{e}(t) =-f_p(x_p(t),g_c(x_c(t),x_{p}(t)+e(t))),
\end{split}
\end{equation}
for $t\in \bigcap_{i\in \bar{N}}(t^{i}_{j}+\tau^{i}_{j},t^{i}_{j+1}+\tau^{i}_{j+1}]$, where $t^{i}_{j}+\tau^{i}_{j}$ and $t^{i}_{j+1}+\tau^{i}_{j+1}$ are two consecutively successful update instants for network $\mathcal {N}_i$, since $\hat{x}_{pi}$ remains a constant in $(t^{i}_{j}+\tau^{i}_{j}, t^{i}_{j+1}+\tau^{i}_{j+1}]$, see Fig. \ref{hatxp:subfig:a}. Based on the items (Ia)-(Ib) above Assumption 2, the update of $\hat{x}_{pi}(t^{i}_{j}+\tau^{i}_{j})$ at the controller side can be described as
\begin{equation}\label{ac3}
\begin{split}
&\hat{x}^{+}_{pi}(t^{i}_{j}+\tau^{i}_{j})\\
=&\left\{\begin{split}
&x_{pi}(t^{i}_{j})+h_{i}(j,e_{i}(t^{i}_{j})),\hbox{no~attack~in}~[t^{i}_{j},t^{i}_{j}+\tau^{i}_{j}],\\
&\hat{x}_{pi}(t^{i}_{j}+\tau^{i}_{j}),\hbox{attack~in}~[t^{i}_{j},t^{i}_{j}+\tau^{i}_{j}],
\end{split}\right.
\end {split}
\end{equation}
for $i\in \bar{N}$ and $j\in \mathbb{Z}_{\geq0}$, where $h_{i}$ implies that the update equations in (\ref{ac3}) imply that the jump of $\hat{x}_{pi}$ is ultimately the jump of $\hat{x}_{pi\ell}$ for certain $\ell\in \mathbb{Z}_{>0}$, see \cite[Exams. 1-2]{2004a}.






The update equations in (\ref{ac3}) without the cross/confusion use of sample packets, which are obtained from items (Ia)-(Ib) and simultaneously checked from Fig. \ref{hatxp:subfig:a}-\ref{hatxp:subfig:b}, can be justified by the fact: Assumption 2 implies $t^{i}_{j}+\tau^{i}_{j}< t^{i}_{j+1}$, so the update of $\hat{x}^{+}_{pi}(t^{i}_{j}+\tau^{i}_{j})$ follows the sample value $x_{pi}(t^{i}_{j})$, rather than the most latest sampled values $x_{pi}(t^{i}_{j-1})$ (should have been used to update $\hat{x}_{pi}$ at $t^{i}_{j-1}+\tau^{i}_{j-1}$ on controller before $t^{i}_{j}$ due to $t^{i}_{j-1}+\tau^{i}_{j-1}< t^{i}_{j}$) or $x_{pi}(t^{i}_{j+1})$ (should be sampled after $t^{i}_{j}+\tau^{i}_{j}$ due to $t^{i}_{j}+\tau^{i}_{j}< t^{i}_{j+1}$). In other words, the sampled value, transmitted in the network and may be dropped during $[t^{i}_{j},t^{i}_{j}+\tau^{i}_{j}]$, can/should only be $x_{pi}(t^{i}_{j})$ that is sampled at $t^{i}_{j}$. More specifically, there is only one transmission and the corresponding unique update within the interval $(t^{i}_{j},t^{i}_{j}+\tau^{i}_{j}]$, which will be finished before the next cycle interval $(t^{i}_{j+1},t^{i}_{j+1}+\tau^{i}_{j+1}]$, i.e., the update at $t^{i}_{j}+\tau^{i}_{j}$ will be executed and completed, (strictly) before the next transmission at $t^{i}_{j+1}$, with no cross/confusion use of sample packets, see Fig. \ref{hatxp:subfig:a}.



The zero-order-hold character of the network $\mathcal {N}_i$, $i\in \bar{N}$ implies that $\hat{x}_{pi}(t^{i}_{j})=\hat{x}_{pi}(t^{i}_{j}+\tau^{i}_{j})$ holds (see Fig. \ref{hatxp:subfig:a}), together with (\ref{ac3}), implying that
\begin{equation}\label{ac4}
\begin{split}
e^{+}_{i}(t^{i}_{j}+\tau^{i}_{j})=&\hat{x}^{+}_{pi}(t^{i}_{j}+\tau^{i}_{j})-x_{pi}(t^{i}_{j}+\tau^{i}_{j})\\
=&\hat{x}^{+}_{pi}(t^{i}_{j}+\tau^{i}_{j})-x_{pi}(t^{i}_{j}+\tau^{i}_{j})\\
=& x_{pi}(t^{i}_{j})+h_{i}(j,e_{i}(t^{i}_{j}))-x_{pi}(t^{i}_{j}+\tau^{i}_{j})\\
=&h_{i}(j,e_{i}(t^{i}_{j}))+\underbrace{x_{pi}(t^{i}_{j})-\hat{x}_{pi}(t^{i}_{j})}_{-e_{i}(t^{i}_{j})}\\
&+\underbrace{\hat{x}_{pi}(t^{i}_{j}+\tau^{i}_{j})-x_{pi}(t^{i}_{j}+\tau^{i}_{j})}_{e_{i}(t^{i}_{j}+\tau^{i}_{j})}\\
=&h_{i}(j,e_{i}(t^{i}_{j}))-e_{i}(t^{i}_{j})+e_{i}(t^{i}_{j}+\tau^{i}_{j}),
\end {split}
\end{equation}
holds for the network $\mathcal {N}_i$ without attacks, then the update $e_{i}$ at $t^{i}_{j}+\tau^{i}_{j}$ can be described as
\begin{equation}\label{ac5}
\begin{split}
e^{+}_{i}(t^{i}_{j}+\tau^{i}_{j})=\left\{\begin{split}
&h_{i}(j,e_{i}(t^{i}_{j}))-e_{i}(t^{i}_{j})+e_{i}(t^{i}_{j}+\tau^{i}_{j}),\\
&~~~~~~~~~~~~\hbox{no~attack~in}~[t^{i}_{j},t^{i}_{j}+\tau^{i}_{j}],\\
&e_{i}(t^{i}_{j}+\tau^{i}_{j}),\hbox{attack~in}~[t^{i}_{j},t^{i}_{j}+\tau^{i}_{j}],
\end{split}\right.
\end {split}
\end{equation}
for $i\in \bar{N}$, $j\in \mathbb{Z}_{\geq0}$. That is, on the basis of $i$ and $e_{i}$, the function $h_{i}$ decides which node (certain $\ell\in \mathbb{Z}_{>0}$) is granted access to $\mathcal {N}_i$ at transmission instant $t^{i}_{j}$ (see Fig. \ref{xp12:subfig:a}), then the corresponding component (instead of the entire vector) of $e_{i}$ has a jump at $t^{i}_{j}+\tau^{i}_{j}$.





In the literature with triggered strategies, the standard case considered is that each network updates asynchronously and independently with one-packet transmission (i.e., all sensor signals are sent with one packet without competition, so $h_{i}=0$): In the special case of one-package transmission, Fig. \ref{xp12} will degenerate and become the same as Fig. \ref{hatxp}, since $x_{pi}=(x_{pi1},\ldots,x_{pi\ell})$ holds for $\ell=1$; then the node $x_{pi}$ which carries the sampled data for all sensors will get access to its network $\mathcal {N}_i$ and be transmitted, so it holds on the update instant that $\hat{x}^{+}_{pi}(t^{i}_{j}+\tau^{i}_{j})=x_{pi}(t^{i}_{j})$ in (\ref{ac3}), see Fig. \ref{hatxp:subfig:a} for $h_{i}=0$, which means the entire vector $e_{i}$ is reset to zero at each update instant, for all $i\in \bar{N}$ and $j\in \mathbb{Z}_{\geq0}$. Next to this standard sampled-data control setup, however, a framework can be obtained by building upon \cite{c2007}, \cite{2004a} and \cite{q2010hemeels} that allows us to study also other access protocols ($h_{i}\neq0$) such as the dynamic try-once-discard (TOD) protocols or the static round-robin (RR) within one framework without additional burden. Therefore, the modelability and rationality of this framework provide a theoretical justification of the feasibility and reliability for our work presented at this level of generality, where the dynamic TOD protocol is more appropriate especially when the transmission instants have been determined in advance by the decentralized triggered strategies.



\subsection{Pp-DoS Attack}\label{subsection2.4}

In what follows, we characterize another source of randomness, i.e., Pp-DoS attack where each attack pulse lasts for a negligible short period of time, as mentioned in the beginning of Subsection \ref{subsection2.3}: The attacks' cardinal number is a discrete random variable and supposed to be subject to Poisson distribution. Denote $\{t^{i}_{k}\}_{k\in \mathbb{Z}_{\geq0}}$ as attacks' instants for network $\mathcal {N}_i$, $i\in \bar{N}$, and $\Delta^{i}_{k}:=t^{i}_{k+1}-t^{i}_{k}$ as the inter-attack time, i.e., the time between two consecutive attack instants. The cardinal number of occurred attacks at network $\mathcal {N}_i$, $i\in \bar{N}$, is denoted as $N^{i}(t)$ used to count the attacks' total number during $[0,t]$. As a result, the discrete random variable $N^{i}$ subject to Poisson distribution implies that the inter-attack time $\Delta^{i}_{k}$ is subject to exponential distribution as shown in the following.





\textbf{Assumption 3.}\label{ass-1}
Given a time interval $[0,t]$ with length $t$, the cardinal number $N^i(t)$ is subject to Poisson distribution with parameter $\lambda^i_{exp} t$ satisfying $\lambda^i_{exp}\geq1$, i.e., $N^i(t)\sim P(\lambda^i_{exp} t)$, for each $i\in \bar{N}$.

Now we characterize the inter-attack time $\Delta^i_{k}$ for all $k\in \mathbb{Z}_{\geq0}$, in terms of the discrete random variable $N^i$ subject to Poisson distribution for each $i\in \bar{N}$.

\textbf{Proposition 1.}\label{Prop-1}
Let Assumption 3 hold. For each $i\in \bar{N}$, the inter-attack time $\Delta^i_{k}$ is subject to exponential distribution with parameter $\lambda^i_{exp}$, i.e., $\Delta^i_{k}\sim Exp(\lambda^i_{exp})$, for all $k\in \mathbb{Z}_{\geq0}$.

\textbf{Proof:} Assumption 3 implies $P(N^i(t)=k)=\frac{(\lambda^i_{exp} t)^{k}}{k!}*e^{-\lambda^i_{exp} t}$ for $k\in \mathbb{Z}_{\geq0}$, and $\{\Delta^i_{k}\geq t\}$ implies that no attacks exist during $[0,t]$, i.e., $\{\Delta^i_{k}\geq t\}=\{N^i(t)=0\}$. Denote $F_{\Delta^i_{k}}(t)$ as the distribution function of nonnegative random variable $\Delta^i_{k}$. Then it holds $F_{\Delta^i_{k}}(t)=P(\Delta^i_{k}\leq t)=0$ for $t<0$, while $F_{\Delta^i_{k}}(t)=P(\Delta^i_{k}\leq t)=1-P(\Delta_{k}>t)
=1-P(N^i(t)=0)
=1-\frac{(\lambda^i_{exp} t)^{k}}{k!}e^{-\lambda^i_{exp} t}|_{k=0}
=1-e^{-\lambda^i_{exp} t}$ for $t\geq0$, thus $\Delta^i_{k}\sim Exp(\lambda^i_{exp})$ holds. \qed

We note that the inter-attack times $\Delta^i_{k}$'s are independent for all $k\in \mathbb{Z}_{\geq0}$: According to the properties of the Poisson distribution, events (i.e., attack occurs) in a Poisson process are independent, meaning that the occurrence of events in one time period is independent of the occurrence of events in other time periods. Therefore, the time between two consecutive attack instants is also independent and follows an exponential distribution, i.e. $\{\Delta^i_{k}\}_{k\in \mathbb{Z}_{\geq0}}$ are a sequence of i.i.d. discrete random variables for each $i\in \bar{N}$. In what follows and thereafter, the independence property of the Poisson distribution enables us to use $t^i_{k}=t^{i}_{j}+\tau^{i}_{j}$ as an equivalent replacement of $t^i_{k}$'s $\in[t^{i}_{j},t^{i}_{j}+\tau^{i}_{j}]$ for the convenience of modeling and analysis: first, due to the no cross/confusion use of sample packets within different transmission-update intervals, one Pp-DoS attack or multiple Pp-DoS attacks within $[t^{i}_{j},t^{i}_{j}+\tau^{i}_{j}]$ will lead to one loss of the unique data packet, thus only making one (same) effect on the failed update of $\hat{x}_{pi}$ at $t^{i}_{j}+\tau^{i}_{j}$ on the controller side; second, for one Pp-DoS attack, occurring in $[t^{i}_{j},t^{i}_{j}+\tau^{i}_{j}]$ can be seen as occurring at the update instant $t^{i}_{j}+\tau^{i}_{j}$, as they make the same attack effect on preventing $\hat{x}_{pi}$ on the controller side from being updated at $t^{i}_{j}+\tau^{i}_{j}$.

\subsection{Triggered Strategy}\label{subsection2.5}
Following the design philosophy in \cite{q28} and \cite{c2007}, a state-dependent event-triggered strategy for generating transmission instants $\{t^i_{j}\}_{j\in \mathbb{Z}_{\geq0}}$ in network $\mathcal {N}_i$, $i\in \bar{N}$, is designed as
\begin{equation}\label{tg1}
\begin{split}
t^i_{j+1}=\inf\{t:~t> t^i_{j}+\tau^{m_i,i}_{miet}|\chi_i(t)<0,j\in \mathbb{Z}_{\geq0}\},
\end{split}
\end{equation}
with boolean variable $m_i\in \{0,1\}$, meaning that and the next event can only be triggered after at least $\tau^{m_i,i}_{miet}>0$ has elapsed since the latest instant $t^i_{j}$, i.e., $t^i_{j+1}-t^i_{j}\geq \tau^{m_i,i}_{miet}$, since the time regularization is adopted by taking the enforced minimum inter-event(transmission) time $\tau^{m_i,i}_{miet}$. We remark that the boolean variable $m_i(t)\in \{0,1\}$ is theoretically time-related and used to keep track of whether the most recent update attempt at time $t$ was successful or not, i.e., (i) when $m_i(t)=0$: the update attempt on the controller side succeeds at $t:=t^{i}_{j}+\tau^{i}_{j}$ since no attack occurs during $t\in[t^{i}_{j},t^{i}_{j}+\tau^{i}_{j}]$, i.e., $t^i_{k}\notin [t^{i}_{j},t^{i}_{j}+\tau^{i}_{j}]$ for any $j,k\in \mathbb{Z}_{\geq0}$; (ii) $m_i(t)=1$: the invalid update on the controller side at $t:=t^{i}_{j}+\tau^{i}_{j}$, due to $t^i_{k}\in [t^{i}_{j},t^{i}_{j}+\tau^{i}_{j}]$ for certain $j,k\in \mathbb{Z}_{\geq0}$, i.e., the $k^{th}$ attack exactly occurs at transmission attempt $t^i_{k}=t^i_{j}$ or the update instant $t^i_{k}=t^{i}_{j}+\tau^{i}_{j}$, or even when a packet passes by coincidentally within $(t^i_{j},t^i_{j}+\tau^{i}_{j})$ in terms of the convention for a worst-case scenario. The triggered function $\chi_i(t)$ is designed to evolve according to
\begin{equation}\label{a42}
\begin{split}
&\dot{\chi}_i(t)=\Psi_i(o_i(t)),t\in (t^i_{j}, t^i_{j+1}],\\
&\frac{d}{d\tau^{i}_{e}}\phi_{l_{i},i} (m_i, \tau^i_{e})=f_{\phi_{l_{i},i}}(\tau^i_{e}, m_i, \phi_{l_{i},i}),t\in(t^i_{j}, t^i_{j+1}],\\
&
\chi^{+}_i(t^i_{j})=\left\{\begin{split}&\tilde{\chi}_i(\Delta^i_{k},e_i),t^i_{k}\notin [t^{i}_{j},t^{i}_{j}+\tau^{i}_{j}], \forall~ j,k\in \mathbb{Z}_{\geq0},\\
&\chi_i(t^i_{j}),t^i_{k}=t^{i}_{j}+\tau^{i}_{j},\exists ~k,j\in \mathbb{Z}_{\geq0},
\end{split}\right.
\end{split}
\end{equation}
where $o_i:=(m_i,e_i, \tau^i_{e}, \phi_{l_{i},i}(m_i, \tau^i_{e}), \chi_i)$ represent the local information available at the $i$th event-triggered strategy, the networked error $e_{i}$ satisfies (\ref{s4e}) and (\ref{ac4}), the inter-attack time $\Delta^{i}_{k}$ satisfies Proposition \ref{Prop-1}, $\tau^i_{e}\in\mathbb{R}_{\geq0}$ denotes the time elapsed since the last transmission attempt, functions $\phi_{l_{i},i}\in[\lambda_{i},\lambda^{-1}_{i}]$ designed later is to derive minimum inter-transmission time $\tau^{m_i,i}_{miet}$ with $\lambda_{i}\in(0,1)$, and notations $\Psi_i, f_{\phi_{l_{i},i}}, l_{i}, \tilde{\chi}_i$ will be specified later.




Referring to Fig. \ref{hatxp}, we briefly summarize the $i$th event-triggered strategy (\ref{tg1}) which generates the transmission instants $\{t^i_{j}\}_{j\in \mathbb{Z}_{\geq0}}$ for network $\mathcal {N}_i$, $i\in \bar{N}$:\\
$~$if $t-t^i_{j-1}\geq \tau^{m_i,i}_{miet}$ and $\chi_i(t)\leq0$ then\\
--generate transmission attempt $t^i_{j}$ in terms of (\ref{tg1});\\
$~~~~~~~~$if $t^i_{k}\notin [t^{i}_{j},t^{i}_{j}+\tau^{i}_{j}]$ (or $m_{i}=0$) for all $k\in \mathbb{Z}_{\geq0}$ then\\
--network $\mathcal {N}_i$ transmits $x_{pi}(t^i_{j})$ at $t^i_{j}$, the $i$th-event-triggered strategy updates $\chi_i(t_{j})$: $\chi^{+}_i(t^i_{j})=\tilde{\chi}_i(\Delta^i_{k},e_i)$; then the controller side updates $\hat{x}_{pi}(t^i_{j})$ at $t^i_{j}+\tau^{i}_{j}$ in terms of (\ref{ac3});\\
$~~~~~~~~$else if $t^i_{k}=t^{i}_{j}+\tau^{i}_{j}$ (or $m_i=1$) for certain $k\in \mathbb{Z}_{\geq0}$ then\\
--discard update instant $t^i_{j}+\tau^{i}_{j}$, the triggered strategy starts of another rule to generate new transmission instant $t^i_{j'}\geq t^i_{j}+\tau^{1,i}_{miet}$ (at least $\tau^{1,i}_{miet}$ after the latest transmission instant $t^i_{j}$), then transmit at $t^i_{j'}$ and update at $t^i_{j'}+\tau^{i}_{j'}$;\\
$~$else if $t-t^i_{j-1}< \tau^{m_i,i}_{miet}$ or $\chi_i(t)>0$ then\\
--not generate transmission attempt instant;\\
$~$end if$~~~~$end if$~~~~$end if$~~~~$end if

Now we analyze the necessity and feasibility of the parameter $\tau^{m_i,i}_{miet}$ that time-regularizes the decentralized event-triggered strategy (\ref{tg1}). \\
\emph{Necessity:} We note that designing two different waiting times $\tau^{0,i}_{miet}$ and $\tau^{1,i}_{miet}$ is to time-regularize the decentralized event-triggered strategy (\ref{tg1}), which can exclude Zeno behavior (infinitely many impulses over a finite period of time): Indeed, $m_i\in\{0,1\}$ denotes two modes of each network, which will not appear at the same time and have no overlap on the time scale, so Zeno solution will not exhibit for each network if no Zeno behavior exists in each mode; also, since the multiple decentralized networks operate asynchronously and independently,
Zeno solution will not exhibit for the entire system if no Zeno behavior exists in each network.\\
\emph{Feasibility:} We also note that the state-dependent triggered strategy (\ref{tg1}) implies we have no prior knowledge of transmission instants $\{t^i_{j}\}_{j\in \mathbb{Z}_{\geq0}}$ that are automatically generated only according to the actual system needs, so Assumption 2 is hard to realize theoretically. However, the parameter $\tau^{m_i,i}_{miet}$ reverses the situation and makes Assumption 2 feasible. More specifically, $\tau^{i}_{mad} >\tau^{m_i,i}_{miet}$ or $\tau^{i}_{mad}=\tau^{m_i,i}_{miet}$ is impossible in term of ``$\inf$'' in Assumption 2 and in (\ref{tg1}): Indeed, notation ``$\inf$'' in (\ref{tg1}) implies that $\tau^{m_i,i}_{miet}$ maybe exactly the lower bound for certain $j$, i.e., $\tau^{m_i,i}_{miet}=t^{i}_{j+1}-t^{i}_{j}$, then $\tau^{i}_{mad} >\tau^{m_i,i}_{miet}$ or $\tau^{i}_{mad}=\tau^{m_i,i}_{miet}$ will contradict Assumption 2. That is, Assumption 2 and the event-triggered strategy (\ref{tg1}) imply that $\tau^{i}_{mad}$ satisfies $\tau^{i}_{mad} <\tau^{m_i,i}_{miet}$. In summary, any attack satisfying Assumptions 2-3 will not prevent the triggered strategy from generating new transmission instant $t^i_{j'}\geq t^i_{j}+\tau^{1,i}_{miet}$, which justifies the feasibility of the parameter $\tau^{m_i,i}_{miet}$.

\section{Mathematical Formulation}\label{section3}

%

Before further analyzing the stability property of system (\ref{asc1}) using stochastic hybrid analysis tools, we need to model (\ref{asc1}) as the stochastic hybrid formalism (\ref{4.5shs:1}). To do so, we first define the following notations: $x:=(\tilde{x},e,\tau_{e},k, s, l,m, \chi)\in \mathbb{X}$, $\tilde{x}:=(x_{p},x_{c})\in \mathbb{R}^{n_{\tilde{x}}}$, $\mathbb{X}:=
\mathbb{R}^{n_{\tilde{x}}}\times\mathbb{R}^{n_p}\times\mathbb{R}^{N}_{\geq0}\times\mathbb{Z}^{N}_{\geq0}\times\mathbb{R}^{n_{p}}
\times\{0,1\}^{N}\times\{0,1\}^{N}\times \mathbb{R}^{N}_{\geq0}$, $\tau_{e}:=(\tau^{1}_{e},\ldots,\tau^{N}_{e})\in \mathbb{R}^{N}_{\geq0}$ with $\tau^{i}_{e}\in\mathbb{R}_{\geq0}$ capturing the time
elapsed since the latest transmissions,
$k:=(k_{1},\ldots,k_{N})\in \mathbb{Z}^{N}_{\geq0}$ with $k_i\in \mathbb{Z}_{\geq0}$ keeping track of the total
amount of transmissions over time,
$s:=(s_{1},\ldots,s_{N})\in\mathbb{R}^{n_p}$ with $s_i\in \mathbb{R}^{n_{p,i}}$ serving as a memory variable to store the value $h_{i}(j,e_{i}(t^{i}_{j}))-e_{i}(t^{i}_{j})$ at the moment of a transmission at time $t^{i}_{j}$ and is used to model the update event at time $t^{i}_{j}+\tau^{i}_{j}$ (see (\ref{ac4})),
$l:=(l_{1},\ldots,l_{N})\in \{0,1\}^{N}$ with $l_i= 0$ ($l_i= 1$) keeping track of whether the next event is transmission (or update),
$m:=(m_{1},\ldots,m_{N})\in \{0,1\}^{N}$ with $m_i= 0$ ($m_i= 1$) indicating whether the attack occurs or not (below (\ref{tg1})) and thus keeping track of whether the next update is successful (or not), $\chi:=(\chi_{1},\ldots,\chi_{N})\in \mathbb{R}^{N}_{\geq0}$ with $\chi_i$ denoting the triggered function (see (\ref{a42})) for the $i$th network $\mathcal {N}_i$, $i\in \bar{N}$. Combining (\ref{asc1}), (\ref{e4e}), (\ref{ac5}) and (\ref{a42}) implies the flow map in (\ref{4.5shs:1}), i.e.,
\begin{equation}\label{e2}
\begin{split}
&F(x)=(f(\tilde{x},e), g(\tilde{x},e), \textbf{1}_{N},\textbf{0}_{n_{p}}, \textbf{0}_{N},\textbf{0}_{N}, \textbf{0}_{N},\Psi(o))),\\
&f=(f_p(x_p,g_c(x_c,x_{p}+e)),f_c(x_c,x_{p}+e)), \\
&g=-f_p(x_p,g_c(x_c,x_{p}+e)),\Psi(o):=(\Psi_i(o_i))_{i\in \bar{N}}, \\
&o:=(m,e, \tau_{e}, \phi(m, \tau_{e}), \chi), \\
&o_i=(m_i,e_i, \tau^i_{e}, \phi_{l_{i},i}(m_i, \tau^i_{e}), \chi_i), \\
&f_{\phi}:=(f_{\phi_{l_{i},i}}(\tau^i_{e}, m_i, \phi_{l_{i},i}))_{i\in \bar{N}},
\end{split}
\end{equation}
with the jump set $C:=\bigcap_{i\in \bar{N}}C_{i}$, $C_{i}:=\{x\in\mathbb{X}|((\tau^{i}_{e}\leq\tau^{m_{i},i}_{miet}~\vee~\chi_{i}\geq0)\wedge(l_{i}=0))\vee
((0\leq \tau^{i}_{e} \leq \tau^{i}_{mad})\wedge(l_{i}=1))\}$;
and also implies the jump map in (\ref{4.5shs:1}), i.e., $G(x, v^{+})=\bigcup_{i\in \bar{N}}G_{i}(x, v^{+})$ and
\begin{equation}\label{2e2}
\begin{split}
&G_{i}(x, v^{+})=\{G_{l_{i},m_{i},i}(x, v^{+})\}\\
=&\left\{\begin{split}
& \{G_{0,0,i}(x, v^{+})\},x\in  D_{i} \wedge l_{i}=0 \wedge  m_{i}=0,\\
&\{G_{1,0, i}(x, v^{+})\},x\in D_{i} \wedge l_{i}=1 \wedge  m_{i}=0,\\
&\{G_{1,1,i}(x, v^{+})\},x\in D_{i} \wedge l_{i}=1 \wedge  m_{i}=1,
\end{split}\right.
\end{split}
\end{equation}
with the jump set $D:=\bigcup_{i\in \bar{N}}D_{i}$, $D_{i}:=\{x\in\mathbb{X}|(\tau^{i}_{e}\geq\tau^{m_{i},i}_{miet}\wedge\chi_{i}\leq0\wedge l_{i}=0)\vee
(l_{i}=1)\}$, under the convention that $G_{i}(x, v^{+}) = \emptyset$ when $x\notin D_{i}$. Before showing the jump maps in (\ref{2e2}), we introduce a matrix $\Gamma_{i}\in \mathbb{R}^{N\times N}$ whose entries are zero except the $ii$-th (diagonal) entry being one, a matrix $I_{N\times N}$ whose entries are zero except the diagonal entries being one, and a diagonal matrix $\bar{\Gamma}_{i}\in \mathbb{R}^{n_{p}\times n_{p}}$ consisting of diagonal elements being one for the index corresponding to the network $\mathcal {N}_i$ and zero for the other networks $\mathcal {N}_1,\ldots, \mathcal {N}_{i-1}, \mathcal {N}_{i+1},\ldots, \mathcal {N}_N$. For example, it holds that $(\bar{\Gamma}_{i})_{jj}$ is equal to 1 for $\sum^{i-1}_{\jmath=1}n_{p,\jmath} <j\leq \sum^{i}_{\jmath=1}n_{p,\jmath}$ while 0 otherwise. Now we show the jump maps in (\ref{2e2}), i.e.,
\begin{equation*}\label{e23}
\begin{split}
&G_{0,0,i}(x, v^{+})=(\tilde{x},e,(I_{N}-\Gamma_{i})\tau_{e},k+\Gamma_{i}\textbf{1}_{N},\\
&\bar{\Gamma}_{i}(h(k,e)-e)+(I_{n_{p}}-\bar{\Gamma}_{i})s,l+\Gamma_{i}\textbf{1}_{N},(I_{N\times N}-\Gamma_{i})m,\\
&(I_{N\times N}-\Gamma_{i})\chi+\Gamma_{i}\tilde{\chi}),\\
&G_{1,0,i}(x, v^{+})=(\tilde{x},\bar{\Gamma}_{i}s+e,\tau_{e},k,(I_{n_{p}}-\bar{\Gamma}_{i})s,l-\Gamma_{i}\textbf{1}_{N},\\
&(I_{N\times N}-\Gamma_{i})m, \chi),\\
&G_{1,1,i}(x, v^{+})=(\tilde{x},e, \tau_{e},k,\bar{\Gamma}_{i}(h(k,e)-e)+(I_{n_{p}}-\bar{\Gamma}_{i})s,\\
&l-\Gamma_{i}\textbf{1}_{N},(I_{N\times N}-\Gamma_{i})m+\Gamma_{i}I_{N},\chi),
\end{split}
\end{equation*}
for $h(k,e):=(h_{i}(k_{i},e_{i}))_{i\in \bar{N}}$ where $h_{i}(k_{i},e_{i})$ corresponds to the term $h_{i}(j,e_{i})$ in (\ref{ac5}), implying $G_{0,0,i}$ corresponds to a successful transmission, $G_{1,0,i}$ corresponds to a successful update, while $G_{1,1,i}$ corresponds to a failed update. That is, the goal of this paper is to make the closed set $\mathcal {A}:=\{x\in \mathbb{X}|\tilde{x}=\mathcal {E}, e=s=\textbf{0}_{n_{p}},\chi=\textbf{0}_{N}\}$ uniformly globally asymptotically stable in probability for the SHS (\ref{4.5shs:1}) with (\ref{e2}) and (\ref{2e2}).




This paper assumes that almost each stochastic hybrid trajectory starts in the set $\mathbb{X}_{0}=\{x\in\mathbb{X}| \tau_{e}\geq\tau^{0,i}_{miet},s=0,\chi=0\}$. The initial set can be freely chosen and is deterministic once chosen, which does not influences the global sense: No matter how far the solutions start from $\mathcal {A}$ within a distance $\delta>0$, the probability $\rho$ of having solutions escaping $\mathcal {A}+\varepsilon\mathbb{B}^{\circ}$ can be arbitrarily small for certain $\varepsilon>0$ (see definition below). We note that choosing $\mathbb{X}_{0}$ does not impose any constraints on initial states $x_{p}$ and $x_{c}$, or initial networked value $\hat{x}_{p}$ at the controller side, but rather reflects the initialization of variables in the triggered strategy (\ref{tg1})-(\ref{a42}). In the following, we report the concept for the uniform global asymptotic stability in probability (UGASp) of a closed set for SHS (\ref{4.5shs:1}), which is with respect to the probability where the graphs of solutions simultaneously have three subproperties, i.e., uniform global stability in probability (UGSp) composed of uniform Lyapunov stability in probability (ULySp) and uniform Lagrange stability in probability (ULaSp), and uniform global attractivity in probability (UGAp). To do so, we define $\Gamma_{\geq \tau}:=\{(s, t)\in \mathbb{R}^{2}: s+t \geq \tau\}$, $\mathcal {S}_{r}(K)$ as the set of solutions to (\ref{4.5shs:1}) with initial conditions belonging to $K$, and $\mbox{graph}(\textbf{x}(\omega)):=\{(t,j,z)\in\mathbb{R}_{\geq0}\times \mathbb{Z}_{\geq0}\times\mathbb{R}^{n}:(t,j)\in \mbox{dom}~\textbf{x}_{\omega}, z=\textbf{x}_{\omega}(t,j), \textbf{x}(\omega):=\textbf{x}_{\omega}\}$, where $\omega$ can be dropped on a few occasions without confusions.

\textbf{Definition 1} \cite{3.1shs}. The closed set $\mathcal {A}\subseteq \mathbb{X}$ is said to be uniformly globally asymptotically stable in probability if the following holds.\\
\indent (a) [UGSp] The closed set $\mathcal {A}\subseteq \mathbb{X}$ is said to be uniformly globally stable in probability if the following holds.\\
   \indent\indent (a1) [ULySp] If there exists $\mu>0$ such that, for each solution of $\dot{x}\in F(x)$ with $x\in C\cap(\mathcal {A}+\mu \mathbb{B})$, there are no finite escape times almost surely and for any $\varepsilon, \rho>0$, there exists $\delta>0$ such that
\begin{equation}\label{4.5u}
\begin{split}
&\textbf{x}\in\mathcal {S}_{r}(\mathcal {A}+\delta \mathbb{B})\Longrightarrow\\
&\mathbb{P}(\mbox{graph}(\textbf{x})\subseteq \mathbb{R}^{2}\times(\mathcal {A}+\varepsilon \mathbb{B}))\geq1-\rho.
\end{split}
\end{equation}
\indent \indent (a2) [ULaSp] If for each solution of (\ref{4.5shs:1A}) there are no finite escape times almost surely and for any $\delta, \rho>0$ there exists $\varepsilon>0$ such that (\ref{4.5u}) holds.\\
\indent (b) [UGAp] The closed set $\mathcal {A}\subseteq \mathbb{X}$ is said to be uniformly globally attractive in probability if for each solution of (\ref{4.5shs:1A}), there are no finite escape times for (\ref{4.5shs:1A}) and for each $\delta>0$, $\varepsilon>0$ and $\rho>0$, there exists $\tau>0$ such that
\begin{equation*}\label{4.5shs:4}
\begin{split}
&x\in\mathcal {S}_{r}(\mathcal {A}+\delta \mathbb{B})\Longrightarrow\mathbb{P}(\mbox{graph}(\textbf{x})\cap(\Gamma_{\geq \tau}\times \mathbb{R}^{2})\\
&~~~~~~~~~~~~~~~~~~~~~~~~~\subset \mathbb{R}^{2}\times(\mathcal {A}+\varepsilon \mathbb{B}^{\circ}))\geq1-\rho.
\end{split}
\end{equation*}


We note that the measurability of the set $\{\omega \in \Omega|\mbox{graph}(\textbf{x})\subset \mathbb{R}^{2}\times(\mathcal {A}+\varepsilon \mathbb{B})\}$ comes from the closed set $(\mathcal {A}+\varepsilon \mathbb{B})$ \cite[Thm. 14.3(i)]{Rock1998} and the measurability of $\omega\mapsto\mbox{graph}(\textbf{x}(\omega))$ with closed values in Assumption 1 \cite[Prop. 14.11 (b)\&(d)]{Rock1998}; while the measurability of the set $\{\omega \in \Omega|\mbox{graph}(\textbf{x})\cap(\Gamma_{\geq \tau}\times \mathbb{R}^{2})\subset \mathbb{R}^{2}\times(\mathcal {A}+\varepsilon \mathbb{B}^{\circ})\}$ comes from the measurability of $\omega\mapsto \mbox{graph}(\textbf{x})\cap(\Gamma_{\geq \tau}\times \mathbb{R}^{2})$ \cite[Prop. 14.11 (a)]{Rock1998} and the open set $(\mathcal {A}+\varepsilon \mathbb{B}^{\circ})$ \cite[Thm. 14.3(h)]{Rock1998}. The phrase ``relative to $\varpi$'' is typically dropped in Definition 1 when defining $\varpi(x):=|x|_{\mathcal {A}}$ (a continuous mapping, \cite[Lemma 2]{3.6shs}) for $x\in C\cup D$, i.e., the Euclidean distance to the closed set $\mathcal {A}$, see \cite{3.1shs} for more information.


\subsection{Constraints}\label{subsection3.1}


We first report the required preliminaries in view of \cite{c2007} to make it feasible for designing event-triggered conditions. Define $\bar{e}_i:= (e_1,\ldots,e_{i-1}, e_{i+1},\ldots,e_N)\in\mathbb{R}^{n_{p}-n_{p,i}}$.


\textbf{Assumption 4.} For $i\in \bar{N}$ and $l_{i}\in \{0,1\}$, there exist a function $W_{i}: \mathbb{Z}_{\geq0}\times \{0,1\}\times \mathbb{R}^{n_{p,i}}\times \mathbb{R}^{n_{p,i}}\rightarrow \mathbb{R}_{\geq0}$ with $W_{i}(k_i,l_{i},e_{i},s_{i})$ locally Lipschitz in its latter two arguments, a continuous function $H_{l_{i},i}: \mathbb{R}^{n_{\tilde{x}}}\times \mathbb{R}^{n_{p}-n_{p,i}}\rightarrow \mathbb{R}_{\geq0}$, functions $\underline{c}_{i},\bar{c}_{i}\in \mathcal {K}_{\infty}$, constants $\lambda_{i}\in(0,1)$ and $L_{l_{i},i}>0$, such that\\
\indent 1) $\underline{c}_{i}(|(e_{i},s_{i})|)\leq W_{i}\leq \bar{c}_{i}(|(e_{i},s_{i})|)$,
  $\forall e_{i}\in \mathbb{R}^{n_{p,i}}$;\\
\indent 2) $ W_{i}(k_i+1,1,e_{i},h_{i}(k_{i},e_{i})-e_{i}) \leq \lambda_{i}W_{i}(k_i,0,e_{i},s_{i})$,
  $\forall e_{i}\in \mathbb{R}^{n_{p,i}}$;\\
\indent 3) $ W_{i}(k_i,0,s_{i}+e_{i},0) \leq W_{i}(k_i,1,e_{i},s_{i})$ or $ W_{i}(k_i,0,e_{i},h_{i}(k_{i},e_{i})-e_{i}) \leq W_{i}(k_i,1,e_{i},s_{i})$,
  $\forall e_{i}\in \mathbb{R}^{n_{p,i}}$;\\ 
\indent 4) $\left\langle \frac{\partial W_{i}}{\partial e_{i}}, g_{i}(\tilde{x},e_{i})\right\rangle\leq L_{l_{i},i} W_{i}+H_{l_{i},i}(\tilde{x},\bar{e}_{i})$, a.a. $e_{i}\in \mathbb{R}^{n_{p,i}}$, $\forall\tilde{x}\in \mathbb{R}^{n_{\tilde{x}}}$;\\
and a locally Lipschitz function $V: \mathbb{R}^{n_{\tilde{x}}}\rightarrow\mathbb{R}_{\geq0}$, a function $\tilde{\varrho}_{i}: \mathbb{R}^{n_{p,i}}\rightarrow \mathbb{R}_{\geq0}$, a function $\sigma_{l_{i},i}: \mathbb{R}_{\geq0}\rightarrow \mathbb{R}_{\geq0}$ satisfying $\sigma_{l_{i},i}(W_{i})\geq W^{2}_{i}\rho_{l_{i},i} $ for $\rho_{l_{i},i}>0$, functions $\underline{\alpha},
\overline{\alpha} \in \mathcal {K}_{\infty}$, constants $\tilde{\rho},\gamma_{l_{i},i}>0$ with $\gamma^{2}_{l_{i},i}\geq  \rho_{l_{i},i}$, such that\\
\indent 5) $\underline{\alpha}(|\tilde{x}|)\leq V(\tilde{x})\leq \overline{\alpha}(|\tilde{x}|)$, $\forall\tilde{x}\in \mathbb{R}^{n_{\tilde{x}}}$;\\
\indent 6) $\left\langle \nabla V(\tilde{x}, f(\tilde{x},e)\right\rangle
\leq - \tilde{\rho}V(\tilde{x})
+\sum^{N}_{i=1}(-\tilde{\varrho}_{i}(x_{p,i})-H_{l_{i},i}^{2}-\sigma_{l_{i},i}(W_{i})
+\gamma^{2}_{l_{i},i}W^{2}_{i})$, a.a. $\tilde{x}\in \mathbb{R}^{n_{\tilde{x}}}$, $\forall e_{i}\in \mathbb{R}^{n_{p,i}}$.



Note that items 1)-3) do not make any reference to the NCS (\ref{asc1})-(\ref{e4e}), and they capture the intrinsic properties of the access protocol itself, meaning that the discrete-time system (\ref{ac5}) or the protocol is uniformly globally exponentially stable when we refer to the reset system (\ref{ac5}) as a discrete-time system induced by the protocol $h_{i}$, see \cite[Definition 7]{2004a}. For the reset system (\ref{ac5}), item 2) characterizes the reset at a transmission instant ($l_{i}$ resets from 0 to 1 and $k_i$ increases 1), after which the protocol works for the transmission within the $i$th network during $(t^{i}_{j},t^{i}_{j}+\tau^{i}_{j}]$; while item 3) is relaxed ($\lambda_{i}=1$), since it is for the reset at a successful/faied update instant ($l_{i}$ resets from 1 to 0 and $k_i$ remains unchanged), after which the $i$th network is idle during $(t^{i}_{j}+\tau^{i}_{j},t^{i}_{j+1}]$ or $(t^{i}_{j}+\tau^{i}_{j},t^{i}_{j'}]$ (see Fig. \ref{xp12:subfig:b}). That is, the reset system (\ref{ac5}) will undergo two different modes (on-working and idle), and this is why two cases $l_{i}=\{0,1\}$ need to be applied here.

Additionally, items 1) and 5) characterize the radial unboundedness of functions $W_{i}$ and $V$, i.e., they tend to infinity when any of their arguments tends to infinity, in order to ensure that solutions cannot escape to infinity, see \cite[Lemma 2]{3.6shs}. Item 4) characterizes the exponential growth of the network-induced error $e_{i}$ between the real value and the last networked value (i.e., between two consecutive and successful update instants); while item 6) shows the $\mathcal {L}_{2}$-gain property from $(W_{1},\ldots,W_{N})$ to $(H_{l_{1},1},\ldots,H_{l_{N},N})$ of $\tilde{x}$-system. Given this, the controller usually needs to be designed to ensure an input-to-state stable (ISS)-like property for the closed-loop $\tilde{x}$-system with respect to the input error $e$, which can be realized by the robustness (tolerable well) to (nonvanishing) exogenous input disturbances when ignoring the network. In view of \cite{c2007} and \cite{2004a}, function $W_{i}$ is usually designed to be a non-negative definite, quadratic function about $e$, while function $V$ is usually designed to be a non-negative definite, quadratic function about $\tilde{x}$. Refer to \cite[Propositions 4-5]{2004a} for designing $W_{i}$ and verifying items 2) and 3), or \cite[Exams. 3-4]{2004a} for designing $W_{i}$ and verifying item 4) with respect to a linear NCS system; for strong nonlinear systems, one can usually design $W_{i}$ and $V$ using systems' kinematic and dynamic characteristics, refer to \cite[Exam. 1]{D2009} for verifying items 4) and 6) with respect to a nonlinear planar control system, or Sec. \ref{exam} in this paper for the design and verifications of items 4) and 6) with respect to attitude systems of rigid bodies.


To determine the minimum inter-transmission time $\tau^{m_{i},i}_{miet}$ in (\ref{tg1}), we specify function $f_{\phi_{l_{i},i}}(\tau^{i}_{e}, m_{i}, \phi_{l_{i},i}): \mathbb{R}_{\geq0}\times\{0,1\}\times\mathbb{R}_{\geq0}\rightarrow\mathbb{R}$ for (\ref{a42}), i.e., $\frac{d}{d\tau^{i}_{e}}\phi_{l_{i},i} (m_i, \tau^i_{e})=f_{\phi_{l_{i},i}}:=$
\begin{equation}\label{e11}
\begin{split}
\left\{\begin{split}
&(m_{i}-1)(2L_{l_{i},i}\phi_{l_{i},i}+\gamma_{l_{i},i}(\phi^{2}_{l_{i},i}+1)),\tau^{i}_{e}\in [0,\tau^{m_{i},i}_{miet}],\\
&0,\tau^{i}_{e}>\tau^{m_{i},i}_{miet},
\end{split}\right.
\end{split}
\end{equation}
where $L_{l_{i},i},\rho_{l_{i},i}$ satisfy Assumption 4. To ensure the local Lipschitz of the certification candidate established later using $\phi_{l_{i},i}$, it holds $\lim_{\tau^{i}_{e}\rightarrow \tau^{m_{i},i}_{miet}}\phi_{l_{i},i}(\tau^{i}_{e})=\phi_{l_{i},i}(\tau^{m_{i},i}_{miet})$, i.e.,
piecewise continuous function $\phi_{l_{i},i}$ is continuous at $\tau^{i}_{e}$.
In view of \cite{c2007}, $\tau^{m_{i},i}_{miet}$ can be chosen no more than the maximally allowable transmission interval bound $\tau^{l_{i},i}_{mati}$ (inversely proportional to the channel bandwidth) given as
\begin{equation}\label{e12}
\tau^{l_{i},i}_{mati}:=
\left\{\begin{split}
&\frac{1}{L_{l_{i},i}r_{l_{i},i}}\arctan\tilde{\Gamma}_{l_{i},i}, \gamma_{l_{i},i} >L_{l_{i},i},\\
&\frac{1}{L_{l_{i},i}}\frac{1-\lambda_{i}}{1+\lambda_{i}}, \gamma_{l_{i},i}=L_{l_{i},i},\\
&\frac{1}{L_{l_{i},i}r_{l_{i},i}}\mbox{arctanh}\tilde{\Gamma}_{l_{i},i},  \gamma_{l_{i},i}<L_{l_{i},i},
\end{split}\right.
\end{equation}
where $\lambda_{i}\in (0,1)$ satisfies (\ref{a42}) and Assumption 4,
$r_{l_{i},i}:=\sqrt{|\frac{\gamma^{2}_{l_{i},i}}{L^2_{l_{i},i}}-1|}$ and $\tilde{\Gamma}_{l_{i},i}:=\frac{r_{l_{i},i}(1-\lambda_{i})}{\frac{2\lambda_{i}}{1
+\lambda_{i}}(\frac{\gamma_{l_{i},i}}{L_{l_{i},i}}-1)+1+\lambda_{i}}$.
Next we present the following lemma for (\ref{e12}).

\textbf{Lemma 1.} \cite{c2007}
Given $\tau^{l_{i},i}_{mati}$ in (\ref{e12}) for $l_{i}\in\{0,1\}$ and $i\in \bar{N}$. The solution to
$\dot{\tilde{\phi}}_{l_{i},i}=-2L_{l_{i},i}\tilde{\phi}_{l_{i},i}-\gamma_{l_{i},i}(\tilde{\phi}^{2}_{l_{i},i}+1)$ satisfies $\tilde{\phi}_{l_{i},i}(t)\in[\lambda_{i},{\lambda}^{-1}_{i}]$ for all $t\in [0,\tau^{l_{i},i}_{mati}]$, with $\tilde{\phi}_{l_{i},i}(0)=\lambda^{-1}_{i}$ and $\tilde{\phi}_{l_{i},i}(\tau^{l_{i},i}_{mati})=\lambda_{i}$.



Let $\phi^{l_{i},i}_{miet}:=\tilde{\phi}_{l_{i},i}(\tau^{0,i}_{miet})$ with the convention that $\phi_{l_i,i}(1, \tau^i_{e})=\phi^{l_{i},i}_{miet}$ holds for $m_{i}=1$ due to $\dot{\phi}_{l_{i},i} (1, \tau^i_{e})=0$ in (\ref{e11}), then it holds that (i) $\{m_{i}=1~\vee~\tau^{i}_{e}>\tau^{0,i}_{miet}\}$ implies $\phi_{l_{i},i}=\phi^{l_{i},i}_{miet}$; (ii) $\lambda_{i}\leq\phi^{l_{i},i}_{miet}\leq\phi_{l_{i},i}\leq\lambda^{-1}_{i}$; (iii) $\chi_{i}\geq0$. To simplify the analysis and calculation, we focus on $\tau^{i}_{mati}:=\min_{l_{i}\in\{0,1\}}\{\tau^{l_{i},i}_{mati}\}$ by dropping of parameter $l_{i}$, then constant $\tau^{m_{i},i}_{miet}$ can be chosen no more than $\tau^{i}_{mati}$. Next we present constraints in terms of Assumptions 2-4 to derive the UGASp of the closed set $\mathcal {A}$.





\textbf{Condition 1.} For $i\in \bar{N}$ and $m_{i}\in\{0,1\}$, it holds that $0<\tau^{m_{i},i}_{miet}\leq\tau^{i}_{mati}$, $0<\tau^{1,i}_{miet}\leq\tau^{0,i}_{miet}$, $\tau^{i}_{mad}+\frac{1}{\vartheta_{i}}\tau^{m_{i},i}_{miet}\leq\tau^{m_{i},i}_{miet}$, and $E(\tau^{1,i}_{miet}-\tau^{i}_{j})\leq E(\Delta^{i}_{k})$, for $\vartheta_{i}>1$ and $k\in \mathbb{Z}_{\geq0}$.



In fact, the constraint using $\vartheta_{i}>1$ in Condition 1 is a specific explanation and justification for $``<''$ in Assumption 2, since $\tau^{m_i,i}_{miet}$ is actually a lower-bound on the minimum inter-event time of network $\mathcal {N}_i$, i.e., $\tau^{m_i,i}_{miet}\leq \inf_{j\in \mathbb{Z}_{\geq0}}(t^{i}_{j+1}-t^{i}_{j})$.


\textbf{Condition 2.} For $i\in \bar{N}$ and $k\in \mathbb{Z}_{\geq0}$, it holds that
1) $\gamma_{0,i} \phi^{0,i}_{miet}\geq\lambda^{2}_{i}\gamma_{1,i}\phi_{1,i}(0,0)
  +10\lambda^{2}_{i}\gamma_{1,i}\phi^{1,i}_{miet}\int_{\mathbb{R}}\Delta^i_{k}\mu(d\Delta^i_{k})$
  for $\tau^{i}_{e}\geq \tau^{0,i}_{miet}$;
2) $\gamma_{1,i}\phi_{1,i}(m_{i},\tau^{i}_{e})\geq \gamma_{0,i}\phi_{0,i}(m_{i},\tau^{i}_{e})$ for $\tau^{i}_{e}\in [0,\tau^{i}_{mad}]$, where $\phi_{l_i,i}$ follows (\ref{e11}) for some fixed initial conditions that satisfy $\gamma_{1,i}\phi_{1,i}(0,0)\geq \gamma_{0,i}\phi_{0,i}(0,0)>\lambda^{2}_{i}\gamma_{1,i}\phi_{1,i}(0,0)>0$, with $\lambda_{i}\in(0,1)$ as in Lemma 1.


\textbf{Condition 3.}
For $i\in \bar{N}$, it holds $\frac{\varpi^{0}}{(\varpi^{0}+ \varpi^{1})}> \lambda^{i}_{exp}\tau^{1,i}_{miet}$, where $\varpi^{0}:=\min\{\tilde{\rho},\min_{i\in \bar{N}, l_{i}\in\{0,1\}}\{\frac{\lambda_{i}\rho_{l_{i},i}}{\gamma_{l_{i},i}}\}\}$, $\varpi^{1}:= \max_{i\in \bar{N}, l_{i}\in\{0,1\}}\frac{\bar{\gamma}_{l_i,i}-\rho_{l_{i},i}}{\gamma_{l_i,i}\phi^{l_{i},i}_{miet}}$ and $\bar{\gamma}_{l_i,i}:=\gamma_{l_i,i}(2\phi^{l_{i},i}_{miet} L_{l_i,i}$ $+\gamma_{l_i,i}(1+(\phi^{l_{i},i}_{miet})^{2}))$.



\textbf{Condition 4.} Consider a function $\delta_{\chi_i}(\chi_i)\in \mathcal {K}_{\infty}$ that satisfies $\delta_{\chi_i}(\chi_i)\geq\varpi^{0}\chi_i$ for $\chi_i\geq0$, with $\varpi^{0}$ as in Condition 3. For triggered function $\chi_{i}$ in (\ref{a42}), it holds that 1)
$\tilde{\chi}_{i}:=10\Delta^i_{k}\lambda^{2}_{i}\gamma_{1,i}\phi^{1,i}_{miet}W^{2}_{i}(k_i,0,e_{i},s_{i})$; 2) for $m_{i}=0$,
\begin{equation}\label{t2}
\begin{split}
\Psi_{i}:=\left\{\begin{split}
&\tilde{\varrho}_{i}(x_{p,i})-\delta_{\chi_i}(\chi_{i}),  0\leq\tau^{i}_{e}\leq\tau^{0,i}_{miet},\\
&\tilde{\varrho}_{i}(x_{p,i})-\bar{\gamma}_{l_i,i}W^{2}_{i}-\delta_{\chi_i}(\chi_{i}), \tau^{i}_{e}>\tau^{0,i}_{miet},
\end{split}\right.
\end{split}
\end{equation}
where $\bar{\gamma}_{l_{i},i}$ is defined in Condition 3, while for $m_{i}=1$,
\begin{equation}\label{t22}
\begin{split}
\Psi_{i}:=\left\{\begin{split}
&\tilde{\varrho}_{i}(x_{p,i})-\delta_{\chi_i}(\chi_{i}),  0\leq\tau^{i}_{e}\leq\tau^{i}_{j},\\
&-\frac{\chi_{i}(\tau^{i}_{j})-0}{\frac{1}{\vartheta_{i}}\tau^{1,i}_{miet}},
\tau^{i}_{j}<\tau^{i}_{e}\leq\tau^{i}_{j}+\frac{1}{\vartheta_{i}}\tau^{1,i}_{miet},\\
&-1,\tau^{i}_{e}>\tau^{i}_{j}+\frac{1}{\vartheta_{i}}\tau^{1,i}_{miet},
\end{split}\right.
\end{split}
\end{equation}
where $\tilde{\varrho}_{i}(x_{p,i})$ satisfies Assumption 4.

\begin{figure*}[!htb]
\centering
\includegraphics[width=6in,height=4in]{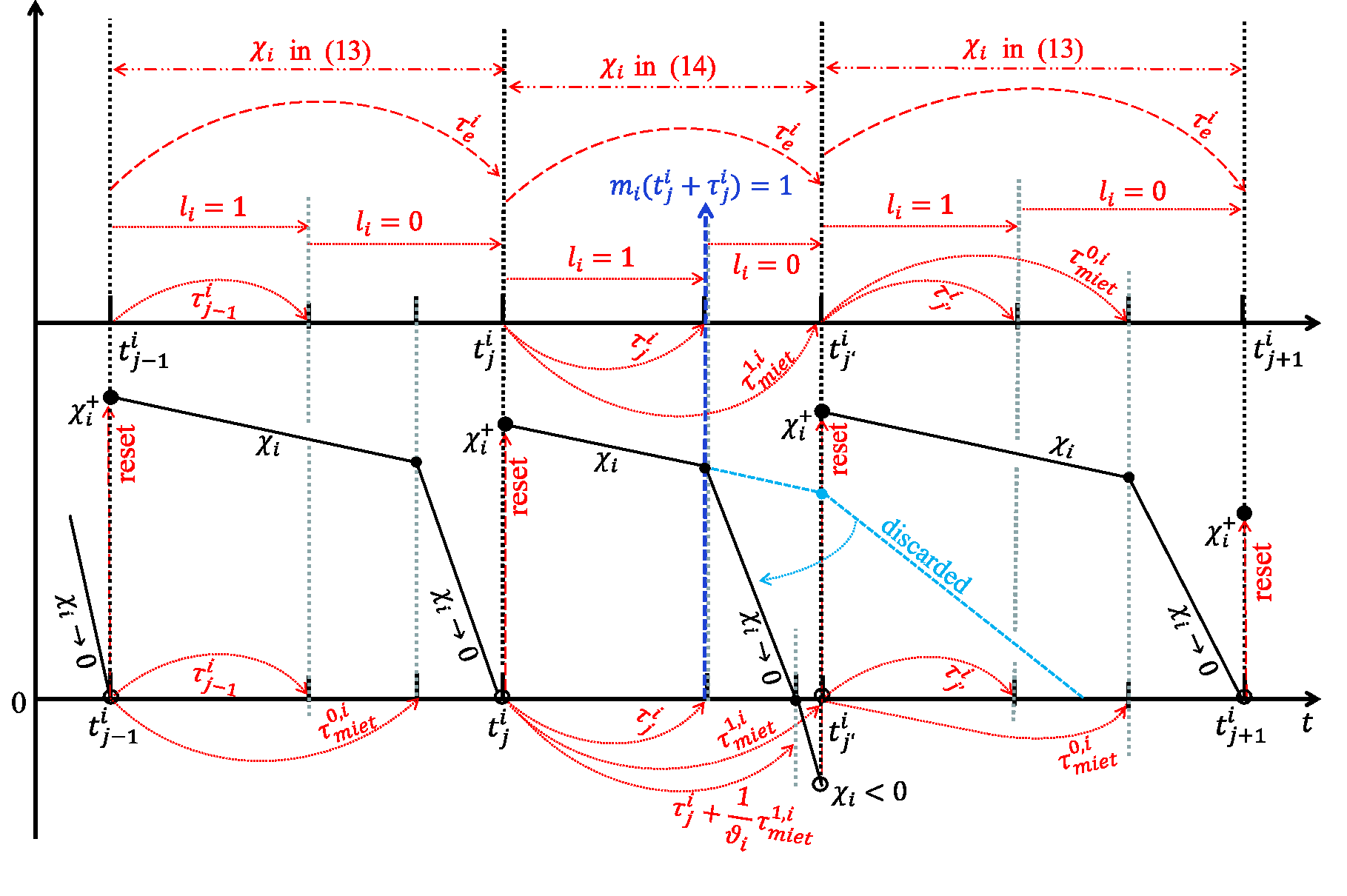}
\caption{Impact of an attack $m(t^{i}_{j}+\tau^{i}_{j})=1$ on the triggered strategy (\ref{tg1}) under Conditions 1 and 4.}\label{whole}
\end{figure*}

As analyzed in Section \ref{subsection2.5}, the triggered strategy (\ref{tg1}) schedules a new transmission attempt at least $\tau^{1,i}_{miet}$ after one attack, and this can be ensured by Condition 4. Indeed, $\chi(\tau^{i}_{e})>0$ (strictly more than zero, or else the event-triggered strategy is exactly the periodic triggered strategy with a period $\tau^{m_{i},i}_{miet}$) should hold during $\tau^{i}_{e}\in[0, \tau^{i}_{j}]$, where $\chi(\tau^{i}_{e})$ evolves according to $\Psi_{i}=\tilde{\varrho}_{i}(x_{p,i})-\delta_{\chi_i}(\chi_{i})$ during $\tau^{i}_{e}\in[0, \tau^{i}_{j}]$, so surely $\chi(\tau^{i}_{e}=\tau^{i}_{j})>0$ holds (i.e., $\chi$ is strictly positive at update attempt $t^i_{j}+\tau^{i}_{j}$); if $t^{i}_{k}=t^i_{j}+\tau^{i}_{j}$ (i.e., $m_{i}(t^i_{j}+\tau^{i}_{j})=1$) holds for certain $k\in \mathbb{Z}_{\geq0}$, based on the acknowledgement scheme (see Fig. \ref{ack}), the ETS starts of another rule using $\Psi_{i}=-\frac{\chi(\tau^{i}_{j})-0}{\frac{1}{\vartheta_{i}}\tau^{1,i}_{miet}}$ during $(\tau^{i}_{j},\tau^{i}_{j}+\frac{1}{\vartheta_{i}}\tau^{1,i}_{miet}]$, in order to generate a new transmission instant $t^i_{j'}= (t^i_{j}+\tau^{i}_{j})+\frac{1}{\vartheta_{i}}\tau^{1,i}_{miet}$ (i.e., the impact of attacks on the strategy), which can indeed be ensured by $\Psi_{i}=-\frac{\chi(\tau^{i}_{j})-0}{\frac{1}{\vartheta_{i}}\tau^{1,i}_{miet}}$: $\chi$ decreases sharply from $\chi(\tau^{i}_{j})$ to zero within $\frac{1}{\vartheta_{i}}\tau^{1,i}_{miet}$ units of times, i.e., $\chi(\tau^{i}_{j}+\frac{1}{\vartheta_{i}}\tau^{1,i}_{miet})=0$ (at most zero) should hold when generating transmission attempt $t_{j'}$; then function $\chi$ at $\chi=0$ strictly decreases so that $\chi$ is strictly negative for $\tau^{i}_{e}>\tau^{i}_{j}+\frac{1}{\vartheta_{i}}\tau^{1,i}_{miet}$ due to $\Psi=-1$. We note that despite $\chi<0$ holds for $\tau^{i}_{e}>\tau^{i}_{j}+\frac{1}{\vartheta_{i}}\tau^{1,i}_{miet}$, (\ref{tg1}) will not generate a new transmission instant $t_{j'}$ at $\tau^{i}_{mad}+\frac{1}{\vartheta_{i}}\tau^{1,i}_{miet}$ until $\tau^{1,i}_{miet}$ after $t^i_{j}$ (see Fig. \ref{hatxp}), unless $\tau^{i}_{j}+\frac{1}{\vartheta_{i}}\tau^{m_{i},i}_{miet}=\tau^{m_{i},i}_{miet}$ holds, i.e., $t_{j'}=t^i_{j}+\tau^{1,i}_{miet}$. As a result, when an attack $t^{i}_{k}$ occurs at $t^i_{j}+\tau^{i}_{j}$, (\ref{tg1}) forcibly/automatically generates a new transmission instant $t_{j'}$ at $\tau^{1}_{\mbox{miet}}$ after the attack (i.e., the impact of attacks on the strategy). See Fig. \ref{whole} for the above analysis: To simplify the characterization, $\Psi_i$ in Condition 4 is simplified respectively during $0\leq\tau^{i}_{e}\leq\tau^{0,i}_{miet}$, $\tau^{i}_{e}>\tau^{0,i}_{miet}$ and $0\leq\tau^{i}_{e}\leq\tau^{i}_{j}$.


Note that the new attempt at $t_{j'}$ will succeed almost surely due to $E(\tau^{1,i}_{miet}-\tau^{i}_{j})\leq E(\Delta^{i}_{k})$ (no attacks occur almost surely during the $(\tau^{1,i}_{miet}-\tau^{i}_{j})$ units of time after the most recently occurred attack, see Condition 1. According to the above analysis, the constraint $\tau^{1,i}_{miet}\leq\tau^{0,i}_{miet}$ in Condition 1 means that a new transmission attempt can be scheduled more often after the most recently occurred attack (in the case when starting of the new rule), than that in the normal case using $\tau^{0,i}_{miet}$ without attack. Indeed, the above new schedule rule when an attack occurs at $t^i_{j}+\tau^{i}_{j}$ is to compensate for the attack-caused disadvantages using $\tau^{i}_{j}+\frac{1}{\vartheta_{i}}\tau^{m_{i},i}_{miet}\leq\tau^{m_{i},i}_{miet}$: if we do not adopt the new rule, the triggered strategy (\ref{tg1}) implies that a new transmission attempt $t_{j'}$ should/can only be generated at least after $\tau^{m_{i},i}_{miet}$ since $t_{j}$; while the new rule renders that a new transmission attempt $t_{j'}$ will be forcibly/automatically generated at $\tau^{i}_{j}+\frac{1}{\vartheta_{i}}\tau^{m_{i},i}_{miet}\leq\tau^{m_{i},i}_{miet}$, i.e., at most at $\tau^{m_{i},i}_{miet}$ since $t_{j}$, where parameter $\vartheta_{i}>1$ determines `how early' that $t_{j'}$ is forcibly/automatically generated.

As mentioned before Assumption 4, asymptotically stabilizing the closed-loop $(x_{p},x_{c})$-system by the designed control law (via Assumption 4 when $e\equiv0$; in the absence of network and attacks) is the premise of designing Condition 4; as well, $(x_{p},x_{c})$-system can be $\mathcal {L}_{2}$-gain stable (robust; tolerable well) with respect to network-induced errors (via Assumption 4 when $e\neq0$). To tolerate attacks before undergoing instability of the closed-loop systems, or more specifically, to prevent next attack from occurring exactly at the update instant arriving after the forcibly/automatically generated transmission instant due to one attack (i.e., being attacked twice or multiple consecutive update instants, resulting in the controller never receiving new data from the closed-loop system), it is required to limit the frequency of attacks, i.e., $E(\tau^{1,i}_{miet}-\tau^{i}_{j})\leq E(\Delta^{i}_{k})$ (in Condition 1). Otherwise, the stability can be lost no matter what control law and triggered strategy (in Condition 4) are adopted.

Hence, the premise of designing triggered strategy (Condition 4) is Condition 1 with the two-step design approach: designing a control law using standard control methodologies to achieve stabilities (via Assumption 4 respectively when $e\equiv0$ or $e\neq0$); then designing the triggered strategy with the appropriate channel bandwidth (in Condition 4) and attacks' frequencies (by Condition 1) to maintain the closed-loop asymptotic stability when an attacked network is inserted into the feedback loop. We note that Condition 4 designs the triggered strategy for $m_{i}=0$ by treating the network induced-error terms as vanishing perturbations in (\ref{u5.2}) characterized by parameter $\varpi^{0}$ to maintain the asymptotic stability as opposed to the role of the controller (to achieve stabilities); while Condition 3 compensates for $m_{i}=1$ in Condition 4 where the network induced-error terms cannot be treated as vanishing perturbations in (\ref{u5.3}) characterized by parameter $\varpi^{1}$.

\subsection{Stability Guarantee}\label{subsection3.2}

In all that follows, define $\mathcal {V}:=\cup_{\omega\in \Omega, i\in \mathbb{Z}_{\geq1}}\textbf{v}_{i}(\omega)$. A function $\mathcal {U}: \hbox{dom} \mathcal {U}\rightarrow \mathbb{R}$ is a certification candidate for $\mathcal {H}$ (\ref{4.5shs:1}), denoted as $\mathcal {U}\in \mathcal {D}(\mathcal {H})$, if\\
C1. $C\cup D\cup G(D\times \mathcal {V})\subset \hbox{dom} \mathcal {U}$;\\
C2. $0\leq \mathcal {U}(x)$ for $x\in C\cup D\cup G(D\times \mathcal {V})$;\\
C3. the quantity $\int_{\mathbb{R}^{m}}\sup_{g\in G(x,v)}\mathcal {U}(g)\mu(dv)$ is well-defined for each
$x\in D$, using the convention that $\sup_{g\in G(x,v)}\mathcal {U}(g) = 0$ when
$G(x,v) =\emptyset$, justified by the preceding item C2.



Based on Assumption 1, any upper semicontinuous function for $\mathcal {H}$ that satisfies C1-C2 is a certification candidate for $\mathcal {H}$, see \cite[Lem. 4.1]{3.1shs}. Sharper than upper semicontinuity, the certification candidate $\mathcal {U}$ can be partially Lipschitz for $\mathcal {H}$ if $\mathcal {U}$ is locally Lipschitz on an open set containing $C\setminus L_{\mathcal {U}}(0)$ as well as continuous on $C$, where $L_{\mathcal {U}}(0):=\{x\in\hbox{dom} \mathcal {U}: \mathcal {U}(x)=0\}$. We note that the certification candidate considered here is locally Lipschitz with respect to the flow set $C$ (due to the existence of several key instants, e.g., $\tau^{i}_{e}=\tau^{1,i}_{miet}$), rather than continuously differentiable; as shown in Assumption 4, even if ``locally Lipschitz'' is assumed rather than ``continuously differentiable'' (see \cite[Lemmas 5-6]{wangwei2015auto}, \cite[Lemmas 1-2]{2014tracking}), we can still use the standard directional derivative for a.a. $\tilde{x}\in \mathbb{R}^{n_{\tilde{x}}}$ in Assumption 4-6), or in the later computation using Hybrid Dynkin's formula \cite[Lem. 9]{4.5shs} since the locally Lipschitz candidate is continuously differentiable when we consider it in segments (with several key instants as segmentation points, e.g., $\tau^{i}_{e}=\tau^{1,i}_{miet}$). Next we present stability properties of the closed set $\mathcal {A}$ for the SHS (\ref{4.5shs:1}) with (\ref{e2}) and (\ref{2e2}).



\textbf{Proposition 2.}\label{prop2}
Let Assumptions 1-4 and Conditions 1-4 hold for the SHS (\ref{4.5shs:1}) with (\ref{e2}) and (\ref{2e2}). Then there exists a certification candidate $\mathcal{U}(x)$ and $\underline{\beta}_{\mathcal{U}},\overline{\beta}_{\mathcal{U}}\in \mathcal {K}_{\infty}$, such that $\underline{\beta}_{\mathcal{U}}(|x|_{\mathcal {A}})\leq\mathcal{U}(x)$ holds for $x\in C\cup D\cup G(D\times \mathcal {V})$, and $\mathcal{U}(x)\leq\overline{\beta}_{\mathcal{U}}(|x|_{\mathcal {A}})$ holds for $x\in C\cup D$.


\textbf{Proof:} Now we show that there exists a certification candidate which is also partially Lipschitz, relative to the closed set $\mathcal {A}$. For $x\in C\cup D$, define $\mathcal{U}(x):=V(\tilde{x})+\sum^{N}_{i=1}(\gamma_{l_{i},i}\phi_{l_{i},i} W^{2}_{i}+\chi_{i})$, then $\mathcal{U}$ is a certification candidate, i.e., $\mathcal{U}(x)\in \mathcal {D}(\mathcal {H})$. Indeed, it holds from (\ref{2e2}) that $G(D\times \mathcal {V})\subset \mathbb{X}$, which implies C1 holds, i.e., $C\cup D\cup G(D\times \mathcal {V})\subset \hbox{dom} \mathcal {U}$; as well, C2 holds from the fact that $\mathcal{U}(x)=0$ iff $x\in\mathcal {A}$, i.e., $L_{\mathcal{U}}(0)=\mathcal {A}$, and $\mathcal{U}(\mathbb{X}\setminus\mathcal {A})>0$. Assumption 4 and Condition 4 imply that function $\mathcal{U}(x)$ is locally Lipschitz on an open set containing $C\setminus L_{\mathcal{U}}(0)$ and continuous on its domain; based on Assumption 1, \cite[Lem. 4.1]{3.1shs} implies that any upper semicontinuous (weaker than locally Lipschitz) function for $\mathcal {H}$ that satisfies C1-C2 is a certification candidate for $\mathcal {H}$. Indeed, in C3 holds due to the upper semicontinuous function $\mathcal {U}(\cdot)$: $-\mathcal {U}(\cdot)$ is a normal integrand since it is upper semicontinuous \cite[Exam. 14.30]{Rock1998}, and in terms of \cite[Exam. 14.32\&Thm. 14.37]{Rock1998}, the measurability of $v\mapsto G(x,v)$ and the outer semicontinuity of $x\mapsto G(x,v)$ in Assumption 1 (Fig. \ref{item3b}) imply that the quantity $\int_{\mathbb{R}^{m}}\sup_{g\in G(x,v)}\mathcal {U}(g)\mu(dv)$ is well-defined for each $x\in D$. Therefore, the partially Lipschitz function $\mathcal{U}(x)$ is a certification candidate relative to the closed set $\mathcal {A}$ for $\mathcal {H}$. Then we show the radial unboundedness of $\mathcal{U}(x)$. Define
$\underline{\beta}_{\mathcal{U}}(|x|):= \min\{\underline{\alpha}(\frac{|\tilde{x}|}{2}), \sum^{N}_{i=1}\gamma_{l_{i},i}\lambda_{i}\underline{c}^{2}_{i}(\frac{|(e_{i},s_{i})|}{2})\}+ \sum^{N}_{i=1}\ln(1+\chi_{i})$ and $\overline{\beta}_{\mathcal{U}} (|x|):=\max\{2\overline{\alpha}(|\tilde{x}|),\sum^{N}_{i=1}2\frac{\gamma_{l_{i},i}}{\lambda_{i}}\bar{c}^{2}_{i}(|(e_{i},s_{i})|)\}
+\sum^{N}_{i=1}(\exp(\chi_{i})-1)$. Then it holds in vies of Assumption 4 for all $x\in \mathbb{X}$ that $\underline{\beta}_{\mathcal{U}}(|x(t,j)|_{\mathcal {A}})\leq\mathcal{U}(x)$ for $x\in C\cup D\cup G(D\times \mathcal {V})$, while $\mathcal{U}(x)\leq\overline{\beta}_{\mathcal{U}}(|x(t,j)|_{\mathcal {A}})$ for $x\in C\cup D$. In view of Assumption 4 and $\ln(\cdot), \exp(\cdot)\in \mathcal {K}_{\infty}$, functions $\underline{\beta}_{\mathcal{U}}$ and $\overline{\beta}_{\mathcal{U}}$ are \emph{of class} $\mathcal {K}_{\infty}$. As a result, $\mathcal{U}(x)$ is a partially Lipschitz and radially unbounded certification candidate relative to the closed set $\mathcal {A}$ for the SHS (\ref{4.5shs:1}) with (\ref{e2}) and (\ref{2e2}).\qed



\textbf{Theorem 1.}\label{th3}
For the SHS (\ref{4.5shs:1}) with (\ref{e2}) and (\ref{2e2}) under Assumptions 1-4 and Conditions 1-4, the closed set $\mathcal {A}$ is uniformly globally stable in probability.



\textbf{Proof}: Consider the partially Lipschitz candidate $\mathcal{U}(x)=V(\tilde{x})+\sum^{N}_{i=1}(\gamma_{l_{i},i}\phi_{l_{i},i} W^{2}_{i}+\chi_{i})$ defined in Proposition 2.

For the jump case of $x\in D\cap \mathbb{X}$: For $x\in  D_{i} ~\wedge~l_{i}=0~\wedge~ m_{i}=0$, it holds from (\ref{2e2}) that $x^{+}\in\{G_{0,0,i}(x, v^{+})\}$ with $G_{0,0,i}(x, v^{+})=(\tilde{x},e,(I_{N}-\Gamma_{i})\tau_{e},k+\Gamma_{i}\textbf{1}_{N},
\bar{\Gamma}_{i}(h(k,e)-e)+(I_{n_{p}}-\bar{\Gamma}_{i})s,l+\Gamma_{i}\textbf{1}_{N},(I_{N\times N}-\Gamma_{i})m,
(I_{N\times N}-\Gamma_{i})\chi+\Gamma_{i}\tilde{\chi})$, i.e., $(\tilde{x},e_{i},\tau^{i}_{e},k_{i}, s_{i}, l_{i},m_{i}, \chi_{i})^{+}=(\tilde{x},e_{i},0,k_{i}+1, h_{i}-e_{i}, 1,0, \tilde{\chi}_{i})$. Given this, for $x\in  D_{i} ~\wedge~l_{i}=0~\wedge~ m_{i}=0$ (i.e., $\tau^{i}_{e}\geq \tau^{0,i}_{miet}$), it holds from Hybrid Dynkin's formula \cite[Lem. 9]{4.5shs}, Assumption 4-2) and Condition 2-1) that
\begin{equation}\label{jum1}
\begin{split}
&\Delta_{\mathbb{X}} \mathcal{U}(x)
=\int_{\mathbb{R}}\sup_{g\in G(x,v)\cap \mathbb{X}}\mathcal{U}(g)\mu(dv)-\mathcal{U}(x)\\
=&\sum^{N}_{i=1}(\gamma_{1,i}\phi_{1,i}(0,0) W^{2}_{i}(k_i+1,1,e_{i},h_{i}-e_{i})+\tilde{\chi}_{i})\\
&-\sum^{N}_{i=1}(\gamma_{0,i}\phi_{0,i}(0,\tau^{i}_{e}) W^{2}_{i}(k_i,0,e_{i},s_{i})+0)\\
\leq&\int_{\mathbb{R}}\sup_{g\in G(x,v)\cap \mathbb{X}}\sum^{N}_{i=1}(\gamma_{1,i}\phi_{1,i}(0,0) \lambda^{2}_{i}W^{2}_{i}(k_i,0,e_{i},s_{i})\\
&+\tilde{\chi}_{i}) \mu(dv)
-\sum^{N}_{i=1}(\gamma_{0,i}\phi_{0,i}(0,\tau^{i}_{e}) W^{2}_{i}(k_i,0,e_{i},s_{i})+0)\\
\leq&0.
\end{split}
\end{equation}
Similarly, it holds from (\ref{2e2}) that $\Delta_{\mathbb{X}} \mathcal{U}(x)\leq0$ respectively for $x\in  D_{i} ~\wedge~l_{i}=1~\wedge~ m_{i}=0$ and for $x\in  D_{i} ~\wedge~l_{i}=1~\wedge~ m_{i}=1$, under Assumption 4-3) and Condition 2-2). Therefore, it holds
\begin{equation}\label{jum3}
\begin{split}
\Delta_{\mathbb{X}} \mathcal{U}(x)\leq0,~ x\in D\cap \mathbb{X}, 
\end{split}
\end{equation}
i.e., the certification candidate $\mathcal {U}(x)$ is non-increased in expected value during jumps along solutions.

For the flow case of $x\in C\cap \mathbb{X}$: When $m_{i}=0$: If $(\tau^{i}_{e}\in [0,\tau^{i}_{mad}])\wedge (l_{i}=1)$ or $(\tau^{i}_{e}\in (\tau^{i}_{mad},\tau^{0,i}_{miet}])\wedge (l_{i}=0)$, (\ref{e11}) becomes $\frac{d}{d\tau^{i}_{e}}\phi_{l_{i},i}=-2L_{l_{i},i}\phi_{l_{i},i}-\gamma_{l_{i},i}(\phi^{2}_{l_{i},i}+1)$. With $2\gamma\phi W_{i}H_{l_{i},i}\leq\gamma^{2}_{l_{i},i}\phi^{2}_{l_{i},i}W^{2}_{i}+H^{2}_{l_{i},i}$, it holds from Hybrid Dynkin's formula \cite[Lem. 9]{4.5shs}, Assumptions 4-4) \& 4-6) that
\begin{equation}\label{flow1}
\begin{split}
&\mathcal {L}_{\mathbb{X}} \mathcal{U}(x)
=\sup_{f\in F(x)}\langle \nabla \mathcal{U}(x), f\rangle\\
\leq&- \tilde{\rho}V
+\sum^{N}_{i=1}(-\tilde{\varrho}_{i}(x_{p,i})-H_{l_{i},i}^{2}-\sigma_{l_{i},i}(W_{i})
+\gamma^{2}_{l_{i},i}W^{2}_{i})\\
&+\sum^{N}_{i=1}(\gamma_{l_{i},i}(-2L_{l_{i},i}\phi_{l_{i},i}-\gamma_{l_{i},i}(\phi^{2}_{l_{i},i}+1))W^{2}_{i}
\\&+2\gamma_{l_{i},i}\phi_{l_{i},i} W_{i} (L_{l_{i},i}W_{i}+H_{l_{i},i})
+\sum^{N}_{i=1}\Psi_{i}\\
\leq& -\tilde{\rho}V
+\sum^{N}_{i=1}(-\tilde{\varrho}_{i}(x_{p,i})-\sigma_{l_{i},i}(W_{i})+\Psi_{i})\\
\leq& -\tilde{\rho}V
+\sum^{N}_{i=1}(-\sigma_{l_{i},i}(W_{i})-\delta_{\chi_{i}}(\chi_{i}))\\
\leq&-\varpi^{0}\mathcal{U}(x),
\end{split}
\end{equation}
since $\phi_{l_{i},i}\leq \lambda^{-1}_{i}$ holds from Lemma 1, $\sigma_{l_{i},i}(W_{i})\geq \rho_{l_{i},i}W^{2}_{i}$ holds from Assumption 4-6), $\Psi_{i}=\tilde{\varrho}_{i}(x_{p,i})-\delta_{\chi_{i}}(\chi_{i})$ is defined in Condition 4 for $0\leq\tau^{i}_{e}\leq\tau^{0,i}_{miet}$ satisfying $\varpi^{0}\chi_{i}\leq\delta_{\chi_i}(\chi_{i})$, where $\varpi^{0}=\min\{\tilde{\rho},\min_{i\in \bar{N}, l_{i}\in\{0,1\}}\{\frac{\lambda_{i}\rho_{l_{i},i}}{\gamma_{l_{i},i}}\}\}$ is defined in Condition 3. If $(\tau^{i}_{e}>\tau^{0,i}_{miet})\wedge(l_{i}=0)$, (\ref{e11}) becomes $\frac{d}{d\tau^{i}_{e}}\phi_{l_{i},i}=0$ and $\phi_{l_{i},i}=\phi^{l_{i},i}_{miet}$, then similarly, it holds $\mathcal {L}_{\mathbb{X}} \mathcal{U}(x)\leq-\varpi^{0}\mathcal{U}(x)$, where $\varpi^{0}$ and $\bar{\gamma}_{l_{i},i}$ are defined in Condition 3, $\Psi_{i}=\tilde{\varrho}_{i}(x_{p,i})-\bar{\gamma}_{l_i,i}W^{2}_{i}-\delta_{\chi_i}(\chi_{i})$ is defined in Condition 4 for $\tau^{i}_{e}>\tau^{0,i}_{miet}$. Therefore, it holds
\begin{equation}\label{u5.2}
\begin{split}
\mathcal {L}_{\mathbb{X}} \mathcal{U}(x)\leq-\varpi^{0}\mathcal{U}(x),x\in C\cap\mathbb{X},
\end{split}
\end{equation}
i.e., the candidate $\mathcal {U}(x)$ is strictly decreased almost surely during flows along the system solutions outside the attractor $\mathcal {A}$, called the stable mode ($\forall~i\in \bar{N}$ such that $m_{i}=0$). When $m_{i}=1$: (\ref{e11}) becomes $\frac{d}{d\tau^{i}_{e}}\phi_{l_{i},i}=0$ with $\phi_{l_i,i}=\phi^{l_{i},i}_{miet}$ (see below Lemma 1); Condition 4 implies $\Psi_{i}=\tilde{\varrho}_{i}(x_{p,i})-\delta_{\chi_{i}}(\chi_{i})$ for $0\leq\tau^{i}_{e}\leq\tau^{i}_{j}$ ($l_{i}=1$), $\Psi_{i}=-\frac{\chi(\tau^{i}_{mad})-0}{\frac{1}{\vartheta_{i}}\tau^{1,i}_{miet}}$ for $\tau^{i}_{j}<\tau^{i}_{e}\leq\tau^{i}_{j}+\frac{1}{\vartheta_{i}}\tau^{1,i}_{miet}$ ($l_{i}=0$), and $\Psi=-1$ for $\tau^{i}_{e}>\tau^{i}_{j}+\frac{1}{\vartheta_{i}}\tau^{1,i}_{miet}$ ($l_{i}=0$), so it holds
\begin{equation}\label{u5.3}
\begin{split}
\mathcal {L}_{\mathbb{X}} \mathcal{U}(x)\leq\varpi^{1}\mathcal{U}(x), x\in C\cap\mathbb{X},
\end{split}
\end{equation}
where $\varpi^{1}_{i}= \max_{i\in \bar{N}, l_{i}\in\{0,1\}}\frac{\bar{\gamma}_{l_i,i}-\rho_{l_{i},i}}{\gamma_{l_i,i}\phi^{l_{i},i}_{miet}}>0$ is defined in Condition 3. That is, the candidate $\mathcal{U}(x)$ may not decrease during flows along the system solutions even outside the attractor $\mathcal {A}$, called unstable mode ($\exists~i\in \bar{N}$ such that $m_{i}=1$).


As shown above, for the flow case, the state trajectories of $\tilde{x}$ and $e$ as shown in the radial unboundedness of $\mathcal{U}(x)$, (\ref{u5.2}) and (\ref{u5.3}), and the fact that the trajectories of $\tau_{e},k, s, l,m, \chi,\phi$ do not exhibit finite escape-times, together exclude the finite escape-times for (\ref{4.5shs:1A}) almost surly. Next we characterize the stable and unstable modes by bridging the gap between the two modes' parameters ($\varpi^{0}$ and $\varpi^{1}$) and attack characteristics (attacks' number $N^{i}(t)$ and attack-over time interval $\Delta^{i}_{k}$). To do so,we need to collect the time at which $m_{i}=0$ holds and at which $m_{i}=1$ holds, then prolong almost each pulsing attack's duration using constant $\tau^{1,i}_{miet}$, since the value $\hat{x}_{pi}(t^{i}_{j})$ at the controller side is successfully updated at $\frac{1}{\vartheta_{i}}\tau^{1,i}_{miet}$ after a Pp-DoS attack occurring at $t^{i}_{k}=t^{i}_{j}+\tau^{i}_{j}$ (see Condition 4). For this reason, we decompose the continuous time axis into attack-active and attack-over parts, and relate the total time intervals of the attack-active parts to $E[N^{i}(t)]$ and $\tau^{1,i}_{miet}$. For a solution $x_{i}$ within $[T_1, T_2]$ satisfying $0\leq T_1\leq T_2$, we define time collections for attack-over parts as
\begin{equation}\label{t1}
\begin{split}
&\bar{\Theta}_{x_{i}}(T_{1}, T_{2}):=\{\tilde{t}\in(T_{1}, T_{2})|
t^i_{k}\notin [t^{i}_{j},t^{i}_{j}+\tau^{i}_{j}), \\
&~~~~~~~~~\forall~ k,j\in \mathbb{Z}_{\geq0},
 (\tilde{t},j)\in \mbox{dom}~x \Rightarrow m_{i}(\tilde{t})=0\},
\end{split}
\end{equation}
so solution $x_{i}$ is in the stable mode satisfying (\ref{u5.2}) if $t\in\bar{\Theta}_{x_{i}}$; then we define time collections for attack-active parts during which solution $x_{i}$ is in the unstable mode satisfying (\ref{u5.3}), i.e., $\bar{\Xi}_{x_{i}}(T_{1}, T_{2}):=[T_{1}, T_{2}]\setminus\bar{\Theta}_{x_{i}}(T_{1}, T_{2})$. We can write $\bar{\Xi}_{x_{i}}(T_{1}, T_{2})$ and $\bar{\Theta}_{x_{i}}(T_{1}, T_{2})$ as
\begin{equation}\label{t3}
\begin{split}
\left\{\begin{split}
\bar{\Xi}_{x_{i}}(T_{1}, T_{2})=&\bigcup_{k\in \mathbb{Z}_{\geq0}}\tilde{Z}^{i}_{k}\cap[T_{1}, T_{2}],\\
\bar{\Theta}_{x_{i}}(T_{1}, T_{2})=&\bigcup_{k\in \mathbb{Z}_{\geq0}}\tilde{W}^{i}_{k-1}\cap[T_{1}, T_{2}],
\end{split}\right.
\end{split}
\end{equation}
where $\tilde{Z}^{i}_{k}$ and $\tilde{W}^{i}_{k}$ are respectively defined as
\begin{equation}\label{t4}
\begin{split}
&\tilde{Z}^{i}_{k}:=\left\{\begin{split}
&[\zeta^{i}_{k},\zeta^{i}_{k}+\upsilon^{i}_{k}), \upsilon^{i}_{k}>0,\\
&\{\zeta^{i}_{k}\}, \upsilon^{i}_{k}=0,
\end{split}\right.\\
&\tilde{W}^{i}_{k}:=\left\{\begin{split}
&[\zeta^{i}_{k}+\upsilon^{i}_{k},\zeta^{i}_{k+1}), \upsilon^{i}_{k}>0,\\
&(\zeta^{i}_{k},\zeta^{i}_{k+1}), \upsilon^{i}_{k}=0,
\end{split}\right.
\end{split}
\end{equation}
and $\upsilon^{i}_{k}$ denotes the elapsed time from $\zeta^{i}_{k}$ to the next
successful transmission. For $\upsilon^{i}_{k}=0$, it holds $\tilde{Z}^{i}_{k}=\{\zeta^{i}_{k}\}$ and $\tilde{W}^{i}_{k}=(\zeta^{i}_{k},\zeta^{i}_{k+1})$, implying that the attack is a pulse.
Let $\zeta^{i}_{0}:=t^{i}_{0}$ for $k=0$, where $\tilde{W}^{i}_{-1}=[0,\zeta^{i}_{0})$ if $t_{0}>0$ (the first attack does not occur at time zero), otherwise $\tilde{W}^{i}_{-1}=\emptyset$. Then the attack-active parts prolonged by $\tau^{1,i}_{miet}$ imply
\begin{equation}\label{t5}
|\bar{\Xi}_{x_{i}}(T_{1}, T_{2})|\leq  E[N^{i}(T_1, T_2)]\tau^{1,i}_{miet},
\end{equation}
where the memoryless property (exponential distribution) of the time interval $\Delta^{i}_{k}$ in Proposition 1 and the inherent property of Poisson distribution (i.e., $N^{i}(\cdot)$ is a cardinal number) in Assumption 3 imply that $[0,t]$ can be replaced by $[T_{1},T_{2}]$, then it holds $E[N^{i}(T_1, T_2)]= \lambda^{i}_{exp}(T_{2}-T_{1})$. Based on (\ref{jum3}), (\ref{u5.2}) and (\ref{u5.3}), it holds
\begin{equation}\label{u0-17}
\begin{split}
&E[\mathcal{U}(x_{i}(t,j))]\\
\leq&\exp(- \varpi^{0} (t-\zeta^{i}_{k}-\upsilon^{i}_{k}))E[\mathcal{U}(x_{i}
(\zeta^{i}_{k}+\upsilon^{i}_{k},j))]
\end{split}
\end{equation}
for $(t,j)\in (\mbox{dom}~x_{i})\cap (\tilde{W}^{i}_{k}\times \mathbb{Z}_{\geq0})$ and $k\in \mathbb{Z}_{\geq0}\cup\{-1\}$, while it holds that
\begin{equation}\label{u0-18}
\begin{split}
E[\mathcal{U}(x_{i}(t,j))]
\leq \exp (\varpi^{1}(t-\zeta^{i}_{k}) ) E[\mathcal{U}(x_{i}(\zeta^{i}_{k},j))]
\end{split}
\end{equation}
for $(t,j)\in (\mbox{dom}~x_{i})\cap (\tilde{Z}^{i}_{k}\times \mathbb{Z}_{\geq0})$ and $k\in \mathbb{Z}_{\geq0}$. The right-hand sides of (\ref{u0-17}) and (\ref{u0-18}) reflect the upper bounds of function $\mathcal{U}$ in expected value respectively for the stable mode and the unstable mode, which together implies a new upper bound of $\mathcal{U}$ in expected value for all $(t,j)\in \mbox{dom}~x_{i}$, i.e.,
\begin{equation}\label{t7}
\begin{split}
E[\mathcal{U}(x_{i}(t,j))]\leq \Upsilon_{i}(0,t)\mathcal{U}(x_{i}(0, 0)),
\end{split}
\end{equation}
with $\Upsilon_{i}(s, t):=e^{-\varpi^{0}|\bar{\Theta}_{x_{i}}(s,t)|}e^{\varpi^{1}|\bar{\Xi}_{x_{i}}(s,t)|}$. Indeed, it holds from (\ref{u0-17}) and (\ref{u0-18}) that
\begin{equation*}\label{t7-1}
\begin{split}
&E[\mathcal{U}(x_{i}(t,j))]\\
\leq&\exp(- \varpi^{0} (t-\zeta^{i}_{k}-\upsilon^{i}_{k}))E[\mathcal{U}(x_{i}
(\zeta^{i}_{k}+\upsilon^{i}_{k},j))]\\
& \Downarrow_{\tilde{Z}^{i}_{k}}\\
\leq&\exp(- \varpi^{0} (t-\zeta^{i}_{k}-\upsilon^{i}_{k}))
\\&\exp (\varpi^{1}(t-\zeta^{i}_{k}) ) E[\mathcal{U}(x_{i}
(\zeta^{i}_{k},j-1))]\\
& \Downarrow_{\tilde{W}^{i}_{k-1}}\\
\leq&\exp(- \varpi^{0} (t-\zeta^{i}_{k}-\upsilon^{i}_{k}))
\exp (\varpi^{1}(t-\zeta^{i}_{k}) ) \\
&\exp(- \varpi^{0} (t-\zeta^{i}_{k-1}-\upsilon^{i}_{k-1}))E[\mathcal{U}(x_{i}
(\zeta^{i}_{k-1}+\upsilon^{i}_{k-1},j-2))]\\
& \Downarrow_{\tilde{Z}^{i}_{k-1}}\\
\leq&\exp(- \varpi^{0} (t-\zeta^{i}_{k}-\upsilon^{i}_{k}))
\exp (\varpi^{1}(t-\zeta^{i}_{k}) )\\
&\exp(- \varpi^{0} (t-\zeta^{i}_{k-1}-\upsilon^{i}_{k-1}))
\\&\exp (\varpi^{1}(t-\zeta^{i}_{k-1}) ) E[\mathcal{U}(x_{i}
(\zeta^{i}_{k-1},j-3))]\\
\leq&\cdots \cdots
\Downarrow_{\tilde{W}^{i}_{k'}\dashrightarrow\tilde{Z}^{i}_{k'}
\dashrightarrow\tilde{W}^{i}_{k'-1}\dashrightarrow\cdots}\\
\leq&\exp(- \varpi^{0}|\bar{\Theta}_{x}(s,t)|)\exp(\varpi^{1}|\bar{\Xi}_{x}(s,t)|)E[\mathcal{U}(x_{i}(0, 0))]\\
=& \Upsilon_{i}(0,t)\mathcal{U}(x_{i}(0, 0)),
\end{split}
\end{equation*}
where $x_{i}(0,0)\in \mathbb{X}_{0}$ is deterministic. According to Assumption 3 and (\ref{t1})-(\ref{t5}), it holds
$\Upsilon_{i}(T_{1}, T_{2})
\leq e^{-\varpi_{1}(T_{2}-T_{1})}\exp((\varpi_{0}+\varpi_{1})\ast E[N^{i}(T_1, T_2)]\ast\tau^{1,i}_{miet})
=e^{-\hat{\beta}_{i}(T_{2}-T_{1})}$,
where $\hat{\beta}_{i}:=\varpi_{1}-(\varpi_{0}+\varpi_{1})\lambda^{i}_{exp}\tau^{1,i}_{miet}$. Hence, Condition 3 implies $\hat{\beta}_{i}>0$, so it holds from (\ref{t7}) that $E[\mathcal{U}(x_{i}(t,j))]\leq e^{-\hat{\beta}_{i}t}\mathcal{U}(x_{i}(0,0))$ and thus
\begin{equation}\label{u7}
\begin{split}
E[\mathcal{U}(x(t,j))]\leq e^{-\hat{\beta} t}\mathcal{U}(x(0,0)),
\end{split}
\end{equation}
with $\hat{\beta}:=\inf_{i\in \bar{N}}\{\hat{\beta}_{i}\}$. In view of \cite[(10)]{4.5shs} and Proposition 2, it holds from (\ref{u7}) that $\mathbb{P}(\mbox{graph}(\textbf{x})\subseteq \mathbb{R}^{2}\times(\mathcal {A}+\varepsilon \mathbb{B}))\geq1-\overline{\beta}_{\mathcal{U}}(\delta)/\underline{\beta}_{\mathcal{U}}(\varepsilon)$. Define $\overline{\beta}_{\mathcal{U}}(\delta)/\underline{\beta}_{\mathcal{U}}(\varepsilon):=\rho$, then it holds for $\forall$ $\varepsilon, \rho>0$, $\exists~ \delta>0$, as well as for $\forall$ $\delta, \rho>0$, $\exists ~\varepsilon>0$ that $\textbf{x}\in\mathcal {S}_{r}(\mathcal {A}+\delta \mathbb{B})\Longrightarrow$
$\mathbb{P}(\mbox{graph}(\textbf{x})\subseteq \mathbb{R}^{2}\times(\mathcal {A}+\varepsilon \mathbb{B}))\geq1-\rho$. Consequently, Definition 1-(a) implies the UGSp of closed set $\mathcal {A}$ for the SHS (\ref{4.5shs:1}) with (\ref{e2}) and (\ref{2e2}).

%


\textbf{Theorem 2.}\label{th3}
For the SHS (\ref{4.5shs:1}) with (\ref{e2}) and (\ref{2e2}) under Assumptions 1-4 and Conditions 1-4, the closed set $\mathcal {A}$ is uniformly globally attractive in probability.


\textbf{Proof}: Following the radial unboundedness of $\mathcal{U}(x)$ and the upper bound of the expected value in (\ref{u7}), we obtain
$E[\bar{\beta}_{\mathcal{U}}(|x(t,j)|_{\mathcal {A}})]
\leq E[\mathcal{U}(x(t,j))]\leq e^{-\hat{\beta} t}\mathcal{U}(x(0,0))
\leq e^{-\hat{\beta}t}\underline{\beta}_{\mathcal{U}}(|x(0,0)|_{\mathcal {A}})$ with $\hat{\beta}=\inf_{i\in \bar{N}}\{\hat{\beta}_{i}\}$,
which has an implication for the probability that the distance of the solution to $\mathcal {A}$ is large after a prescribed amount of time, see \cite[(50)]{5.2shs}. In other words, for each $\delta>0$, $\varepsilon>0$ and $\rho>0$, there exists $\tau>0$ such that
$\textbf{x}\in\mathcal {S}_{r}(\mathcal {A}+\delta \mathbb{B})\Longrightarrow$
$P[(\textbf{T}_{\tau},\textbf{J}_{\tau})\in \hbox{dom}(\textbf{x}), |\textbf{x}(\textbf{T}_{\tau},\textbf{J}_{\tau})|_{\mathcal {A}}\geq \varepsilon]
\leq e^{-\hat{\beta} t}\frac{\underline{\beta}_{\mathcal{U}}(|x(0,0)|_{\mathcal {A}})}{\bar{\beta}_{\mathcal{U}}(\varepsilon)}\leq \rho$,
i.e.,
$\textbf{x}\in\mathcal {S}_{r}(\mathcal {A}+\delta \mathbb{B})\Longrightarrow
 \mathbb{P}(\mbox{graph}(\textbf{x})\cap(\Gamma_{\geq \tau}\times \mathbb{R}^{2})\subset \mathbb{R}^{2}\times(\mathcal {A}+\varepsilon \mathbb{B}^{\circ}))\geq1-\rho$, implying the UGAp of $\mathcal {A}$ in terms of Definition 1-(b).  \qed


\textbf{Theorem 3.}\label{th3}
For the SHS (\ref{4.5shs:1}) with (\ref{e2}) and (\ref{2e2}) under Assumptions 1-4 and Conditions 1-4, the closed set $\mathcal {A}$ is uniformly globally asymptotically stable in probability.


\textbf{Proof}: Proposition 2, Theorems 1-2 and Definition 1 conclude Theorem 3. \qed

\begin{figure*}[!htb]
\centering\subfigure[]{
\begin{minipage}[t]{0.3\textwidth}
\label{normq:subfig:a} 
\includegraphics[width=2.0in,height=1.5in]{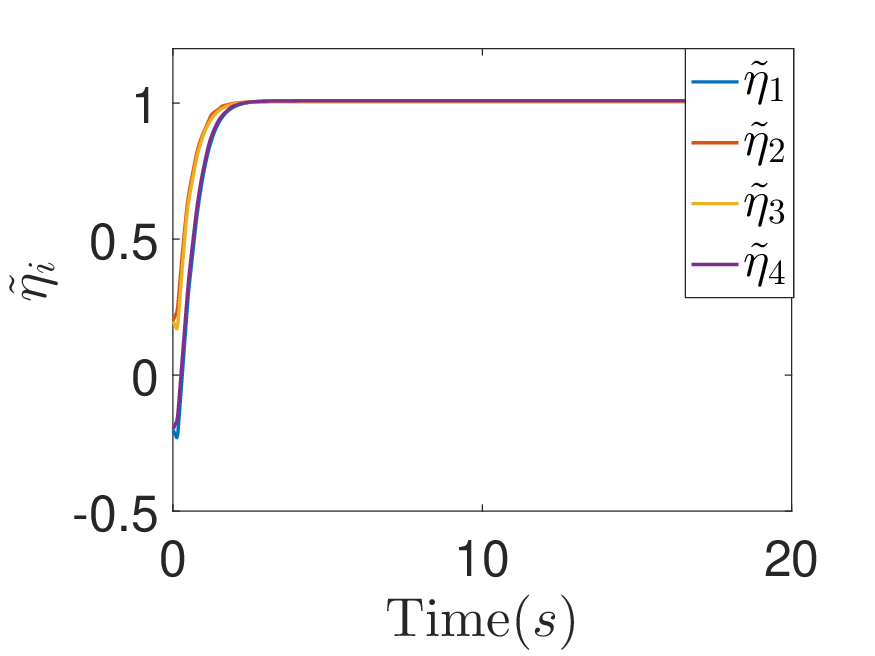}
\end{minipage}}
\centering \subfigure[]{
\begin{minipage}[t]{0.3\textwidth}
\label{normq:subfig:b} 
\includegraphics[width=2.0in,height=1.5in]{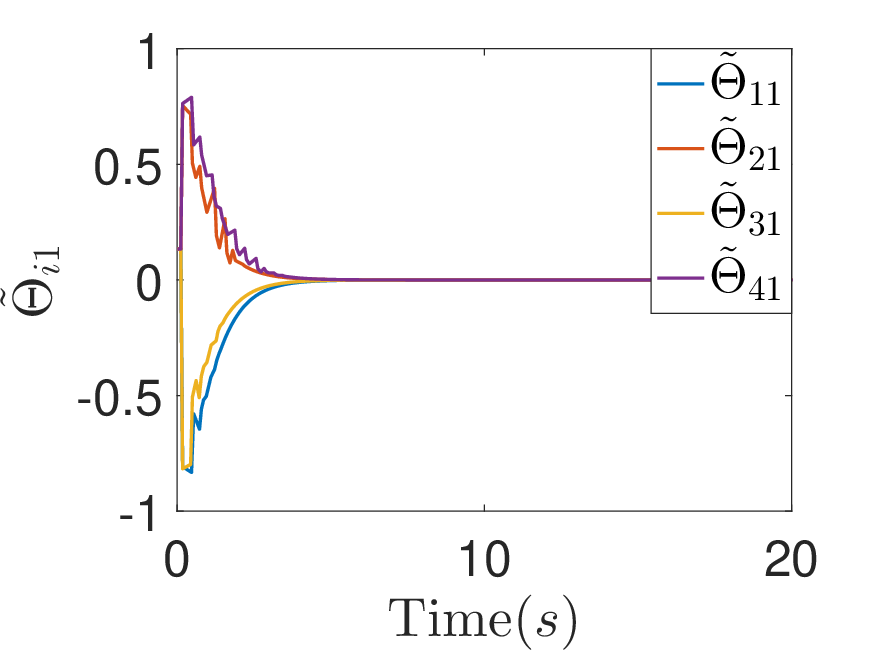}
\end{minipage}}
\centering \subfigure[]{
\begin{minipage}[t]{0.3\textwidth}
\label{normq:subfig:c} 
\includegraphics[width=2.0in,height=1.5in]{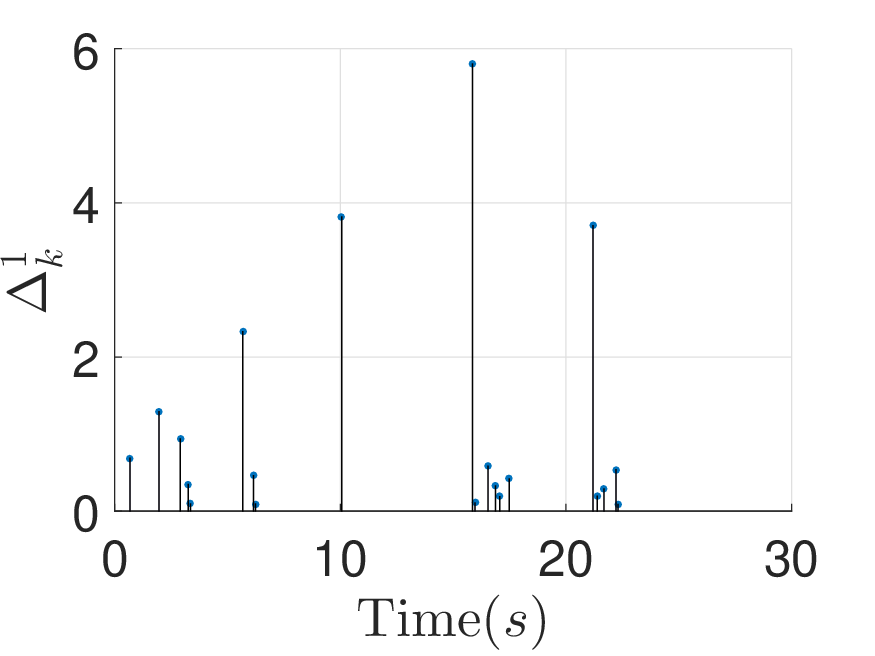}
\end{minipage}}
\centering\subfigure[]{
\begin{minipage}[t]{0.3\textwidth}
\label{normq:subfig:d} 
\includegraphics[width=2.0in,height=1.5in]{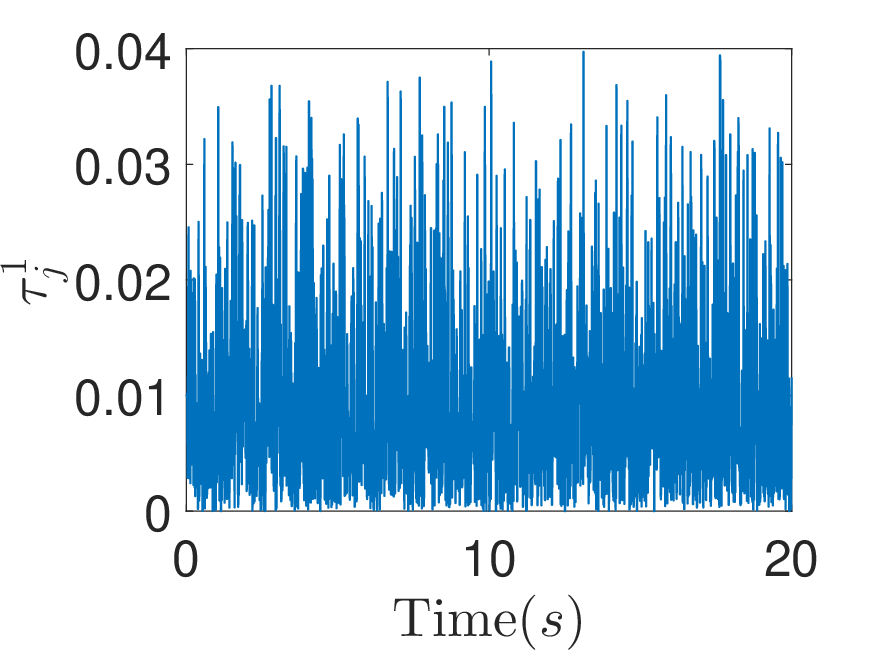}
\end{minipage}}
\centering \subfigure[]{
\begin{minipage}[t]{0.3\textwidth}
\label{normq:subfig:e} 
\includegraphics[width=2.0in,height=1.5in]{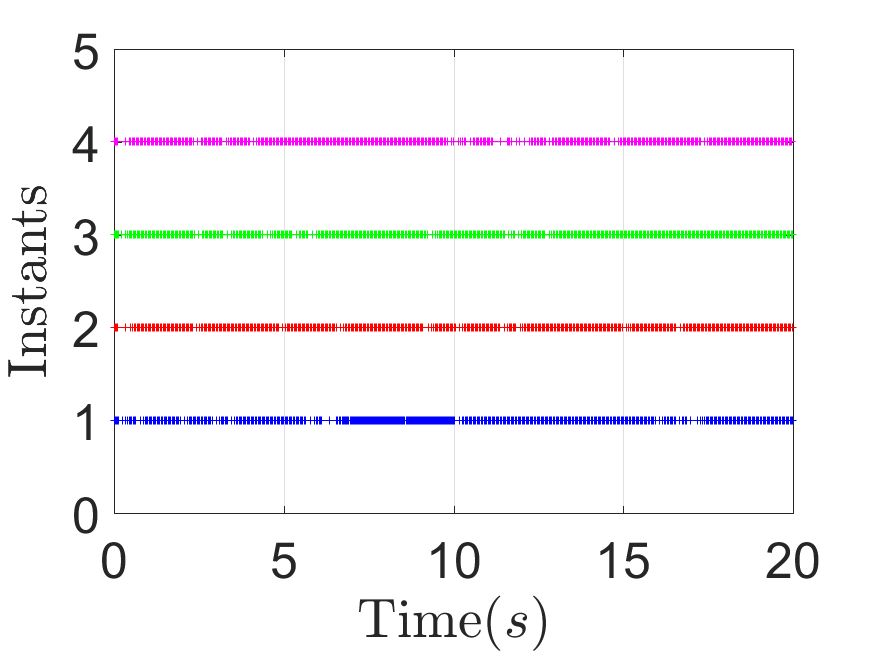}
\end{minipage}}
\centering \subfigure[]{
\begin{minipage}[t]{0.3\textwidth}
\label{normq:subfig:f} 
\includegraphics[width=2.0in,height=1.5in]{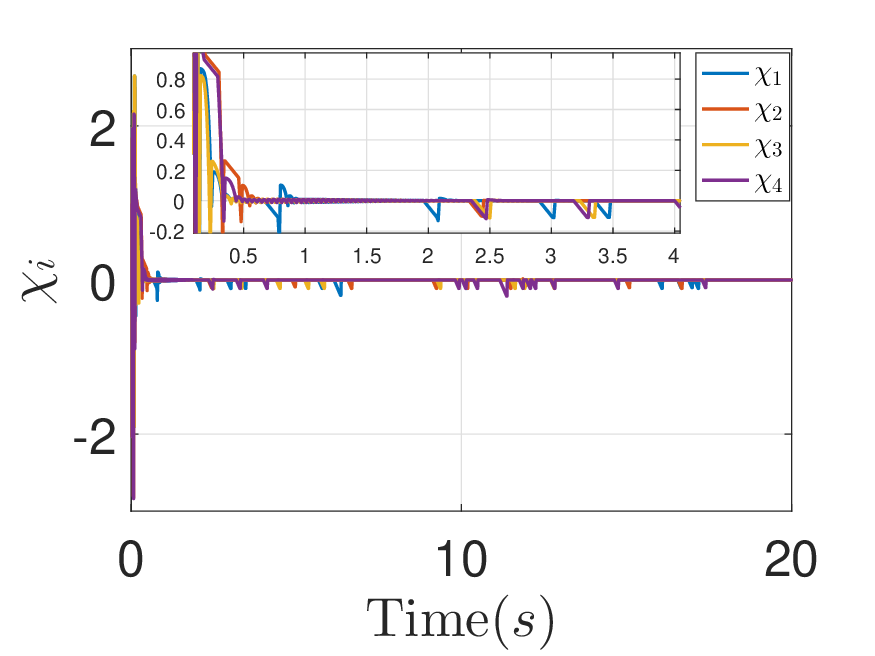}
\end{minipage}}
\caption{Under the network $\mathcal {N}_i$ for $i\in \{1,2,3,4\}$. (a) Trajectories of $\tilde{\eta}_{i}$, where $\tilde{q}_{i}=(\tilde{\eta}_{i}, \tilde{\epsilon}_{i})$.
(b) Trajectories of $\tilde{\Theta}_{i1}$, where $\tilde{\Theta}_{i}:=(\tilde{\Theta}_{i1}, \tilde{\Theta}_{i2}, \tilde{\Theta}_{i3})$.
(c) Pp-DoS attacks in $\mathcal {N}_1$ satisfying $\Delta^{1}_{k}\sim Exp(1)$.
(d) Network delays $\tau^{1}_{j}$, where $0\leq \tau^{1}_{j} \leq \tau^{1}_{mad} < 0.016$ and $\tau^{1}_{j}\sim Exp(100)$ for $j\in \mathbb{Z}_{\geq0}$.
(e) Event-triggered instants $\{t^i_{j}\}_{j\in \mathbb{Z}_{\geq0}}$ in terms of (\ref{tg1}), where $\tau^{0,i}_{miet}=0.029$ and $\tau^{1,i}_{miet}=0.016$.
(f) Trajectories of event-triggered function $\chi_i$ satisfying Condition 4.}\label{normq}
\end{figure*}

\section{Simulations}\label{exam}
Imagine a group of small rotorcraft-like aerial vehicles, each with its own flexible wing, connected in parallel to form a larger composite aircraft. The flexible wings allow for a seamless connection between the individual vehicles, creating a high aspect ratio aircraft with improved efficiency and maneuverability. This innovative cluster formation enables the aircraft to perform a wide range of tasks, from surveillance and reconnaissance to cargo transportation and search and rescue missions. Flexible combined rotorcraft-like aerial vehicles takes low-cost, highly reliable, easy to group small and medium-sized rotorcraft-like aerial vehicles as the basic unit, and several single machines are connected through flexible structures to form a collaborative assembly with large aspect ratio. Flexible combination is a characteristic of combined rotorcraft-like aerial vehicles besides low flight speed, high aspect ratio, and flexible connection, and the flexible structure is achieved by using elastic materials or flexible mechanisms with limited degrees of freedom deformation in certain directions. This utility model provides a combined track ejection rack joint platoon track ejection rack based on the flexible combined rotorcraft-like aerial vehicles, which includes several single rotorcraft-like aerial vehicles ejection racks that are connected in parallel, with the characteristics of adjustable launch angle, adjustable launch speed, good synchronization, and strong universality.

Consider the rotorcraft-like aerial vehicle model $\dot{R}_{i}=R_{i}S(\Theta_{i})$ and $J_{i} \dot{\Theta}_{i} =S(J_{i}\Theta_{i})\Theta_{i}+u_{\tau,i}$, $i\in\bar{N}:=\{1,2,3,4\}$, with attitude matrix  $R_{i}\in SO(3):=\{R_{i}\in \mathbb{R}^{3\times 3}: R^{T}_{i}R_{i}=I, |R_{i}|=1\}$,
angular velocity $\Theta_{i}\in \mathbb{R}^{3}$, skew-symmetric matrix $S(\cdot)\in \mathbb{R}^{3\times 3}$
and torque controller $u_{\tau,i}\in \mathbb{R}^{3}$. Denote unit quaternions as $(\eta_{i},\epsilon^{T}_{i})^{T}:=q_{i}$ and $(\tilde{\eta}_{i},\tilde{\epsilon}^{T}_{i})^{T}:=\tilde{q}_{i}$, satisfying $q_{i}, \tilde{q}_{i}\in \mathcal{S}^{3}:=\{y\in \mathbb{R}^{1+3}: y^{T}y =1\}$. Define the attitude error as $\tilde{q}_{i}:=q^{-1}_{d}\otimes q_{i}$ with the desired attitude $q_{d}\in \mathcal{S}^{3}$, and the angular velocity error as $\tilde{\Theta}_{i}:=\Theta_{i}-\bar{\Theta}_{d,i}$ where the desired angular velocity $\Theta_{d}\in \mathbb{R}^{3}$ satisfies $\bar{\Theta}_{d,i}:=\tilde{R}^{T}_{i}\Theta_{d}$, $\tilde{R}_{i}:=\mathcal {R}(\tilde{q}_{i})$ and
$\mathcal {R}(\tilde{q}_{i}):=I_{3}+2\tilde{\eta}_{i}S(\tilde{\epsilon}_{i})+2S^{2}(\tilde{\epsilon}_{i})$,
see \cite{2011tac}. The attitude kinematic and dynamic system equations of the rotorcraft-like aerial vehicles can be written as
\begin{equation*}\label{e5.2}
	\begin{split}
		\dot{\tilde{q}}_{i}=&\frac{1}{2}\left(\begin{array}{c}
			-\tilde{\epsilon}^{T}_{i}\\
			\tilde{\eta}_{i} I+S(\tilde{\epsilon}_{i})
		\end{array}\right)
		\tilde{\Theta}_{i},\\
		J_{i}\dot{\tilde{\Theta}}_{i}=&\Sigma(\tilde{\Theta}_{i},\bar{\Theta}_{d,i})\tilde{\Theta}_{i}-S(\bar{\Theta}_{d,i})J_{i}\bar{\Theta}_{d,i}
		-J_{i}\tilde{R}^{T}_{i}\dot{\Theta}_{d}+u_{\tau,i},
	\end{split}
\end{equation*}
where
$\Sigma(\tilde{\Theta}_{i},\bar{\Theta}_{d,i}):=S(J_{i}\tilde{\Theta}_{i})+S(J_{i}\bar{\Theta}_{d,i})-S(\bar{\Theta}_{d,i})J
-J_{i}S(\bar{\Theta}_{d,i})$ is skew symmetric, then it holds $\dot{\tilde{\eta}}_{i}=-\frac{1}{2}\tilde{\epsilon}^{T}_{i}\tilde{\Theta}_{i}$ and $\dot{\tilde{\epsilon}}_{i}=\frac{1}{2}(\tilde{\eta}_{i} I+S(\tilde{\epsilon}_{i}))\tilde{\Theta}_{i}$. Consider the feedback $\tilde{\Theta}_{i}(\tilde{q}_{i}):=-h_{i}\tilde{\epsilon}_{i}$ for the kinematic systems, while the torque input $u_{\tau,i}:=S(\bar{\Theta}_{d,i})J_{i}\bar{\Theta}_{d,i}+J_{i}\tilde{R}^{T}_{i}\dot{\Theta}_{d}-\kappa_{i} h_{i}\tilde{\epsilon}_{i}-K_{i}\tilde{\Theta}_{i}$ for the dynamic systems with control gains $\kappa_{i}>0$ and $K_{i}\in \mathbb{R}^{3\times3}$ satisfying $K^{T}_{i}=K_{i}>0$. The control variable $h_{i}\in\{-1,1\}$ satisfies $\dot{h}_{i}=0$ for $h_{i}\tilde{\eta}_{i}\geq-\underline{\delta}_{i}$ and $h^{+}_{i}\in -\overline{\mbox{sgn}}(h_{i})$ for $h_{i}\tilde{\eta}_{i}\leq-\underline{\delta}_{i}$, where $\overline{\mbox{sgn}}(h_{i})=1$ if $h_{i}>0$, $\overline{\mbox{sgn}}(h_{i})=-1$ if $h_{i}<0$, and $\underline{\delta}_{i}\in (0,1)$. To make the following analysis feasible, here we make the convention that the strategy (\ref{tg1}) is also triggered to generate transmission attempts once $h_{i}$ resets.


Let $J_{i}:=\mbox{diag}\{0.13, 0.13, 0.04\}$, $K_{i}:=0.013I$, $\kappa_{i}:=3$ and $\underline{\delta}_{i}:=0.45$. Let $q_{d}(0):=(1,0,0,0)$ and $\Theta_{d}(0):=0$ with $\dot{\Theta}_{d}\equiv0$. Define initial values as $q_{1}(0):=(-0.2, \sqrt{1-0.04}\vec{\upsilon})$, $q_{2}(0):=(0.2, -\sqrt{1-0.04}\vec{\upsilon})$, $q_{3}(0):=(0.2, \sqrt{1-0.04}\vec{\upsilon})$, $q_{4}(0):=(-0.2, -\sqrt{1-0.04}\vec{\upsilon})$, and $\Theta_{i}(0):=0.2\vec{\upsilon}$, where $\vec{\upsilon}:=(1,2,3)/\sqrt{14}$. In view of \cite{2004a}, we apply the standard sampled-data protocol with $h_{i}=0$ for $i\in\{1,2,3,4\}$, considering the strong nonlinearity of attitude kinematic and dynamic systems. Define $\tilde{x}:=(\tilde{q}_{i},J_{i}\tilde{\Theta}_{i}, h_{i})_{i\in \bar{N}}\in \mathbb{X}'$ with $\mathbb{X}':=(\mathcal{S}^{3})^{N} \times (\mathbb{R}^{3})^{N}\times \{-1,1\}^{N}$. The goal is to stabilize $\mathcal {A}:=(\mathcal {A}_{p,i})_{i\in \bar{N}}\times \{e=\textbf{0}_{N}\}$, where $\mathcal {A}_{p,i}:=\{\tilde{x}\in \mathbb{X}'|\tilde{\eta}_{i}=\overline{\mbox{sgn}}(h_{i}), \tilde{\Theta}_{i}=0\}$.

Choose non-negative definite functions $W_{i}:=|e_{i}|$ and $V(\tilde{x}):=\sum^{N}_{i=1}(2\kappa_{i}(1-h_{i}\tilde{\eta}_{i})+\frac{1}{2}\tilde{\Theta}^{T}_{i} J_{i}\tilde{\Theta}_{i})$. In view of Proposition 2, Assumptions 4-1) and 4-5) hold by choosing concave function $\ln(\cdot)\in\mathcal {K}_{\infty}$ and convex function $\exp(\cdot)\in\mathcal {K}_{\infty}$.  To simplify the computations, we
consider Assumption 4-6) in the absence of the network, then it holds
\begin{equation*}\label{vo}
	\begin{split}
		&\dot{V}(\tilde{x})\\
		=&\sum^{N}_{i=1}(-2\kappa_{i} h_{i} \dot{\tilde{\eta}}_{i}+\tilde{\Theta}^{T}_{i}(\Sigma(\tilde{\Theta}_{i},\bar{\Theta}_{d,i})\tilde{\Theta}_{i}-\kappa_{i} h_{i} \tilde{\epsilon}_{i}-K_{i}\tilde{\Theta}_{i}))\\
		=&\sum^{N}_{i=1}(-\kappa_{i}( h_{i} )^{2}\tilde{\epsilon}^{T}_{i}\tilde{\epsilon}_{i}-\tilde{\Theta}^{T}_{i}K_{i}\tilde{\Theta}_{i}
		+\frac{1}{2}\kappa_{i}(\tilde{\Theta}^{2}_{i}+\tilde{\epsilon}^{2}_{i})
		\\
		=&\sum^{N}_{i=1}(-\frac{1}{10}(\kappa_{i}( h_{i} )^{2}\tilde{\epsilon}^{T}_{i}\tilde{\epsilon}_{i}+\tilde{\Theta}^{T}_{i}K_{i}\tilde{\Theta}_{i})\\
		&-\frac{9}{10}(\kappa_{i}\tilde{\epsilon}^{T}_{i}\tilde{\epsilon}_{i}+\lambda_{\min}(K_{i})\tilde{\Theta}^{T}_{i}\tilde{\Theta}_{i})
		+\frac{1}{2}\kappa_{i}(\tilde{\Theta}^{2}_{i}+\tilde{\epsilon}^{2}_{i})
		\\
		\leq&- \tilde{\rho}V(\tilde{x})
		+\sum^{N}_{i=1}(-\tilde{\varrho}_{i}(x_{p,i})-H_{l_{i},i}^{2}-\sigma_{l_{i},i}(W_{i})
		+\gamma^{2}_{l_{i},i}W^{2}_{i}),
	\end{split}
\end{equation*}
where $H_{l_{i},i}:=(\frac{1}{2}|h_{i}|+J^{-1}_{i}\kappa_{i})|\tilde{\epsilon}_{i}|+|J^{-1}_{i}K_{i}\tilde{\Theta}_{i}|
=0.65|\tilde{\epsilon}_{i}|+0.2|\tilde{\Theta}_{i}|$ for $\tilde{\varrho}_{i}(x_{p,i}):=0\cdot|e_{i}|$, since $-\kappa_{i} h_{i} \tilde{\Theta}^{T}_{i}\tilde{\epsilon}_{i}\leq\frac{1}{2}\kappa_{i}(\tilde{\Theta}^{2}_{i}+\tilde{\epsilon}^{2}_{i})$ holds following Yong's inequality, where
\begin{equation*}\label{vo-2}
	\begin{split}
		&-\frac{9}{10}(\kappa_{i}\tilde{\epsilon}^{T}_{i}\tilde{\epsilon}_{i}+\lambda_{\min}(K_{i})\tilde{\Theta}^{T}_{i}\tilde{\Theta}_{i})
		+\frac{1}{2}\kappa_{i}(\tilde{\Theta}^{2}_{i}+\tilde{\epsilon}^{2}_{i})\\
		=&-(0.9-0.5)\kappa_{i}|\tilde{\epsilon}_{i}|^{2}-(0.9\lambda_{\min}(K_{i})-0.5\kappa_{i})|\tilde{\Theta}_{i}|^{2}\\
		=&-0.6|\tilde{\epsilon}_{i}|^{2}-1.05|\tilde{\Theta}_{i}|^{2}\\
		\leq&-0.5525|\tilde{\epsilon}_{i}|^{2}-0.17|\tilde{\Theta}_{i}|^{2}\\
		=&-((\frac{1}{2}| h_{i}|+J^{-1}_{i}\kappa_{i})^{2}+(\frac{1}{2}| h_{i} |+J^{-1}_{i}\kappa_{i})\ast|J^{-1}_{i}K_{i}|
		)|\tilde{\epsilon}_{i}|^{2}\\
		&-(|J^{-1}_{i}K_{i}|^{2}+(\frac{1}{2}| h _{i}|+J^{-1}_{i}\kappa_{i})\ast|J^{-1}_{i}K_{i}|)|\tilde{\Theta}_{i}|^{2}\\
		\leq&-((\frac{1}{2}| h_{i} |+J^{-1}_{i}\kappa_{i})|\tilde{\epsilon}_{i}|+|J^{-1}_{i}K_{i}|\tilde{\Theta}_{i}|)^{2}=-H^{2}_{l_{l},i},
	\end{split}
\end{equation*}
and
$(\gamma^{2}_{i}-\rho_{l_{i},i})\geq\sup_{\tilde{x}\in \mathbb{X}'}\{\kappa_{i}(( h_{i} )^{2}|\tilde{\epsilon}_{i}|+| h_{i} |\tilde{\Theta}^{2}_{i}), \lambda_{\max}(K_{i})|\tilde{\Theta}_{i}|\}$,
$\tilde{\rho}V(\tilde{x})\leq\inf_{\tilde{x}\in \mathbb{X}'}\{\frac{1}{10}(\kappa_{i} h^{2}_{i}\tilde{\epsilon}^{T}_{i}\tilde{\epsilon}_{i}
+\tilde{\Theta}^{T}_{i}K_{i}\tilde{\Theta}_{i})\}$, so Assumption 4-6) holds. We note that the reset of control variable $h_{i}$ at the controller side satisfies $V(h^{+}_{i})<V(h_{i})$, see \cite{2011tac}, which justifies the convention that the strategy (\ref{tg1}) can be triggered once $h_{i}$ resets.


Choose $\lambda^{i}_{exp}=1$ for Assumption 3, i.e., $N^{i}(t)\sim P(1*t)$, then Proposition 1 implies $E(\Delta^{i}_{k})=\frac{1}{\lambda^{i}_{exp}}=1$. The stochastic network delays $\tau^{i}_{j}$ in network $\mathcal {N}_i$, $i\in \bar{N}$, is subject to exponential distribution with parameter $100$, i.e., $\tau^{i}_{j}\sim Exp(100)$, so it holds that $E(\tau^{i}_{j})=0.01$ for all $j\in \mathbb{Z}_{\geq0}$.
Choose $\lambda_{i}=0.1$, so by choosing $\gamma_{1,i}=L_{1,i}=10$ and $\gamma_{0,i}=L_{0,i}=5$, it implies that $\tau^{1,i}_{mati}=0.0818$ and $\tau^{0,i}_{mati}=0.1636$ respectively for $l_{i}=1$ and $l_{i}=0$, then we can choose $\tau^{i}_{mati}=\min_{l_{i}\in\{0,1\}}\{\tau^{l_{i},i}_{mati}\}=0.08$,
and thus we can choose $\tau^{0,i}_{miet}=0.029$ and $\tau^{1,i}_{miet}=0.016$,
then it holds $\phi^{1,i}_{miet}=\tilde{\phi}_{1,i}(\tau^{0,i}_{miet})=0.6869$
and $\phi^{0,i}_{miet}=\tilde{\phi}_{0,i}(\tau^{0,i}_{miet})=1.928$,
from which Conditions 1-2 are satisfied. Indeed, the equality in Condition 2-2) holds
at $\tau^{i}_{e}=0.029$, and Condition 2-2) holds during $\tau^{i}_{e}\in [0,0.029]$,
but will be violated for $\tau^{i}_{e}>0.029$, so it implies $\tau^{i}_{mad} <\tau^{0,i}_{miet}= 0.029$ or $\tau^{i}_{mad} <\tau^{1,i}_{miet}= 0.016$, from which we can choose $\tau^{i}_{mad}=0.012$
(due to $\tau^{i}_{mad}+\frac{1}{\vartheta_{i}}\tau^{m_{i},i}_{miet}\leq\tau^{m_{i},i}_{miet}$ by choosing $\vartheta_{i}:=4$).
Choose $\tilde{\rho}=1$, $\rho_{0,i}=100$ and $\rho_{0,i}=25$ satisfying $\gamma^{2}_{l_{i},i}-\rho_{l_{i},i}\geq0$ for $l_{i}\in\{0,1\}$. Condition 3 is satisfied, where $\varpi^{0}=0.5$, $\varpi^{1}=26.8685$, $\bar{\gamma}_{1,i}=284.56$ and $\bar{\gamma}_{0,i}=241.33$. Based on Condition 4, we get
$\tilde{\chi}_{i}
=10\Delta^i_{k}\lambda^{2}_{i}\gamma_{1,i}\phi^{1,i}_{miet}W^{2}_{i}(k_i,0,e_{i},s_{i})
=0.6869\Delta^i_{k} |e_{i}|^{2}$;\\
when $m_{i}=0$, we get $\Psi_{i}=-\delta_{\chi_i}(\chi_{i})=-0.5\chi_{i}$ for $0\leq\tau_{e}\leq0.029$ while $\Psi_{i}=\tilde{\varrho}_{i}(x_{p,i})-\bar{\gamma}_{l_i,i}W^{2}_{i}(e_{i})-\delta_{\chi_i}(\chi_{i})=-241.33|e_{i}|^{2}-0.5\chi_{i}$ for $\tau_{e}>0.029$, where $\delta_{\chi_i}(\chi_{i}):=\varpi^{0}\chi_{i}$; \\
when $m_{i}=1$, with $\vartheta_{i}=4$, we get $\Psi_{i}=-\delta_{\chi_i}(\chi_{i})=-\chi_{i}$, for $0\leq\tau^{i}_{e}\leq\tau^{i}_{j}$, $\Psi_{i}=-\frac{\chi_{i}(\tau^{i}_{j})-0}{\frac{1}{\vartheta_{i}}\tau^{1,i}_{miet}}=-250\chi_{i}(\tau^{i}_{j})$,
for $\tau^{i}_{j}<\tau^{i}_{e}\leq\tau^{i}_{j}+\frac{0.016}{\vartheta_{i}} $,
$\Psi_{i}=-1$ for $\tau^{i}_{e}>\tau^{i}_{j}+\frac{0.016}{\vartheta_{i}} $. As a result, the closed set $\mathcal {A}$ is stabilized in view of the trajectories of the first element of the four attitude vectors and the four angular velocity vectors, respectively shown in Fig. \ref{normq:subfig:a} and \ref{normq:subfig:b}.
In network $\mathcal {N}_{1}$, Pp-DoS attacks are shown in Fig. \ref{normq:subfig:c}, satisfying $\Delta^{1}_{k}\sim Exp(1)$ for $k\in \mathbb{Z}_{\geq0}$, while Fig. \ref{normq:subfig:d} presents the stochastic delays $\tau^{1}_{j}$ satisfying $0\leq \tau^{1}_{j} \leq \tau^{1}_{mad} < 0.012$ and $\tau^{1}_{j}\sim Exp(100)$ for $j\in \mathbb{Z}_{\geq0}$. For $i\in \{1,2,3,4\}$, Fig. \ref{normq:subfig:e} shows the event-triggered instants $\{t^i_{j}\}_{j\in \mathbb{Z}_{\geq0}}$ in terms of (\ref{tg1}), satisfying $\tau^{0,i}_{miet}=0.029$ and $\tau^{1,i}_{miet}=0.016$, while Fig. \ref{normq:subfig:f} shows the function $\chi_i$ that satisfies Condition 4.

\section{Conclusion}\label{conclusion}
This study introduces novel decentralized event-triggered state-feedback control strategies that demonstrate resilience to Pp-DoS attacks and robustness in the face of stochastic network delays. The Pp-DoS attacks, whose number is driven by Poisson process, aim to disrupt multiple asynchronous and independent communication channels, leading to instances where transmission of measurement data becomes impossible. Within the established stochastic hybrid framework for stabilizing nonlinear networked control systems under communication constraints (i.e., being attacked and with only one node accessing one network per transmission), the synthesized event-triggered strategies are designed in a decentralized, state-based manner, while accounting for specific medium access protocols. Leveraging stochastic hybrid tools to integrate attack-active parts (unstable modes) with attack-over parts (stable), we establish a radially unbounded certification candidate relative to the attractor. When tailored to the system's specific requirements, these strategies can effectively withstand Pp-DoS attacks and stochastic network delays, offering various tradeoff conditions between minimum inter-event times and maximum allowable delays without compromising stability and Zeno-freeness.

The main advance of this work is the development of the resilient decentralized event-triggered strategies for the state-feedback control under Pp-DoS attacks, especially when the network delays are stochastic and simultaneously the attacks' number is driven by Poisson process. The advantage of the advance is that these resilient strategies can automatically, asynchronously and independently respond to attacks that occur in their own networks, neither influencing each other nor harming the normal operation of the whole system itself. One disadvantage here is that we set the duration of each Pp-DoS attack to negligible mode, which may make the attack pulses miss many transmission flows and have less impact, so that the control strategy is always in a ``lucky'' security environment. Despite that we consider the worst-case scenario, i.e., if the Pp-DoS attack occurs, then it happens exactly when a packet passes by in the network, however, it s not easy to expect these attacks which occur randomly and quickly disappears (negligible durations) to happen to occur at transmission attempts $t^i_{j}$ or the update instant $t^i_{j}+\tau^{i}_{j}$, or even when a packet passes by coincidentally within $(t^i_{j},t^i_{j}+\tau^{i}_{j})$ in terms of the convention for a worst-case scenario. So in practical situations, the impact of attacks on transmission instants is almost traceless, see Fig. \ref{normq:subfig:e} in Simulation. Given this and inspired by the six different millisecond-level attack-active times in \cite{key2009}, the decentralized triggered strategies in this paper are expected to be resilient to the scenario where each attack pulse lasts for any longer acceptable period of time. Due to the immaturity of the techniques used to handle this scenario, we will explore the constraints for the acceptable period in future work.


\begin{thebibliography}{10}
\providecommand{\url}[1]{#1}
\csname url@samestyle\endcsname
\providecommand{\newblock}{\relax}
\providecommand{\bibinfo}[2]{#2}
\providecommand{\BIBentrySTDinterwordspacing}{\spaceskip=0pt\relax}
\providecommand{\BIBentryALTinterwordstretchfactor}{4}
\providecommand{\BIBentryALTinterwordspacing}{\spaceskip=\fontdimen2\font plus
\BIBentryALTinterwordstretchfactor\fontdimen3\font minus
  \fontdimen4\font\relax}
\providecommand{\BIBforeignlanguage}[2]{{%
\expandafter\ifx\csname l@#1\endcsname\relax
\typeout{** WARNING: IEEEtran.bst: No hyphenation pattern has been}%
\typeout{** loaded for the language `#1'. Using the pattern for}%
\typeout{** the default language instead.}%
\else
\language=\csname l@#1\endcsname
\fi
#2}}
\providecommand{\BIBdecl}{\relax}
\BIBdecl

\bibitem{3.1shs}
A. R. Teel, ``Lyapunov conditions certifying stability and recurrence for a
class of stochastic hybrid systems'', \emph{Annual Reviews in Control}, vol. 37, no. 1, pp. 1--24, 2013.



\bibitem{5.2shs}
A. R. Teel and J. P. Hespanha, ``Stochastic hybrid systems: A
modeling and stability theory tutorial,'' in \emph{Proceedings of the 54th
IEEE Conference on Decision and Control}, Osaka, Japan, December
15-18, 2015, pp. 3116--3136.


\bibitem{4.1shs}
A. R. Teel, A. Subbaraman, A. Sferlazza, ``Stability analysis for
stochastic hybrid systems: A survey,'' \emph{Automatica}, vol. 50, no. 10, pp. 2435--2456, 2014.



\bibitem{romain2022}
R. Postoyan, R. G. Sanfelice, W. Heemels, ``Explaining the ``mystery'' of periodicity in inter-transmission times in two-dimensional
event-triggered controlled system,'' \emph{IEEE Trans. Autom. Control}, vol. 68, no. 2, pp. 912--927, 2023.



\bibitem{pan2022}
Y. Pan, Y. Wu, H. K. Lam, ``Security-based fuzzy control for nonlinear networked control systems with
DoS attacks via a resilient event-triggered scheme,'' \emph{IEEE Transactions on Fuzzy Systems}, vol. 30, no. 10, pp. 4359--4368, 2022.



\bibitem{q2010hemeels}
W.~M.~H. Heemels, A.~R. Teel, N.~Van~de Wouw, and D.~Ne{\v{s}}i{\'c}, ``Networked control
  systems with communication constraints: Tradeoffs between transmission
  intervals, delays and performance,'' \emph{IEEE Transactions on Automatic
  Control}, vol.~55, no.~8, pp. 1781--1796, 2010.


\bibitem{q15}
V.~Dolk, D.~P. Borgers, and W.~Heemels, ``Output-based and decentralized
  dynamic event-triggered control with guaranteed $\mathcal {L}_{p}$-gain
  performance and zeno-freeness,'' \emph{IEEE Trans. Autom. Control}, vol.~62, no.~1, pp. 34--49, 2017.




\bibitem{walsh2001}
G.~C. Walsh and H.~Ye, ``Scheduling of networked control systems,'' \emph{IEEE
  Control Systems}, vol.~21, no.~1, pp. 57--65, 2001.




\bibitem{key2009}
X. Luo, E. W. W. Chan, R. K. C. Chang, ``Detecting pulsing denialof-service attacks with nondeterministic attack intervals,''
\emph{EURASIP J. Adv. Signal Process.}, vol. 2009, no. 1, pp. 1--13, 2009.




\bibitem{2002flood}
R. K. C. Chang, ``Defending against flooding-based distributed denial-of-service attacks: A tutorial,''
\emph{IEEE Commun. Mag.}, Vol. 40, no. 10, pp. 42--51, 2002.



\bibitem{Golait2016}
D. Golait, N. Voipfd Hubballi, ``Voice over ip flooding detection,'' In \emph{Proceedings of the 2016 Twenty Second
National Conference on Communication (NCC)}, Guwahati, India, 4-6 March 2016; pp. 1--6.



\bibitem{c2007}
D.~Carnevale, A.~R. Teel, D.~Ne{\v{s}}i{\'c}, ``A Lyapunov proof of an improved
  maximum allowable transfer interval for networked control systems,''
  \emph{IEEE Trans. Autom. Control}, vol.~52, no.~5, pp. 892--897, 2007.




\bibitem{2004a}
D.~Ne{\v{s}}i{\'c} and A.~Teel, ``Input-output stability properties of networked
  control systems,'' \emph{IEEE Trans. Automatic Control}, vol.~49,
  no.~10, pp. 1650--1667, 2004.


\bibitem{Rock1998}
R. T. Rockafellar, R. J.-B. Wets, \emph{Variational Analysis}: Springer, 1998.




\bibitem{Fristedt1997}
B. Fristedt, L. Gray, \emph{A modern approach to probability theory}. Birkhauser, 1997.



\bibitem{4.5shs}
A. R. Teel, ``Stochastic hybrid inclusions with diffusive flows'',
in \emph{Proc. 53rd IEEE Conf. Decision and Control}, 2014, pp. 3071--3076.




\bibitem{4.6shs}
A. R. Teel, J. P. Hespanha, and A. Subbaraman, ``A converse Lyapunov
theorem and robustness for asymptotic stability in probability'',
\emph{IEEE Trans. Autom. Control}, Vol. 59, no. 9, pp. 2426--2441, 2014.



\bibitem{3.2shs}
A. Subbaraman and A. R. Teel, ``A converse Lyapunov theorem for
strong global recurrence'', \emph{Automatica}, vol. 49, no. 10, pp. 2963--2974, 2013.



\bibitem{7.1shs}
A. Subbaraman, A. R. Teel, ``Robust global recurrence for a class of stochastic hybrid systems'',
\emph{Nonlinear Analysis: Hybrid Systems}, vol. 25, pp. 283--297, 2017.



\bibitem{3.6shs}
S. Grammatico, A. Subbaraman, and A. R. Teel, ``Discrete-time stochastic
control systems: A continuous Lyapunov function implies robustness to
strictly causal perturbations'', \emph{Automatica}, vol. 49, no. 10, pp. 2939--2952, 2013.



\bibitem{4.7shs}
A. R. Teel, J. P. Hespanha, and A. Subbaraman, ``Equivalent characterizations
of input-to-state stability for stochastic discrete-time systems'', \emph{IEEE Trans. Automatic Control},
vol. 59, no. 2, pp. 516--522, 2014.


\bibitem{3.3shs}
A. R. Teel, ``A Matrosov theorem for adversarialMarkov decision processes'', \emph{IEEE Trans. Autom. Control},
vol. 58, no. 8, pp. 2142--2148, 2013.


\bibitem{2012teel}
R.~Goebel, R.~G. Sanfelice, A.~R. Teel, \emph{Hybrid Dynamical Systems:
  modeling, stability, and robustness}.\hskip 1em plus 0.5em minus 0.4em\relax
  Princeton University Press, 2012.



\bibitem{q28}
V.~Dolk, P.~Tesi, C.~De~Persis, W.~Heemels, ``Event-triggered control
  systems under denial-of-service attacks,'' \emph{IEEE Trans. Control Netw. Syst.}, vol.~4, no.~1, pp. 93--105, 2017.



\bibitem{D2009}
D.~Ne{\v{s}}i{\'c}, A.~R. Teel, D.~Carnevale, ``Explicit computation of the sampling period in emulation of controllers for nonlinear sampled-data systems,'' \emph{IEEE Trans. Autom. Control}, vol. 54, no. 3, pp. 619--624, 2009.




\bibitem{wangwei2015auto}
W.~Wang, D.~Ne{\v{s}}i{\'c}, and R.~Postoyan, ``Emulation-based stabilization
  of networked control systems implemented on Flexray,'' \emph{Automatica},
  vol.~59, pp. 73--83, 2015.


\bibitem{2014tracking}
R.~Postoyan, N.~Van~de Wouw, D.~Ne{\v{s}}i{\'c}, and W.~M.~H. Heemels,
  ``Tracking control for nonlinear networked control systems,''
  \emph{IEEE Trans. Autom. Control}, vol.~59, no.~6, pp. 1539--1554, 2014.


\bibitem{2011tac}
C. G. Mayhew, R. G. Sanfelice, A. R. Teel, ``Quaternion-based
hybrid control for robust global attitude tracking,'' \emph{IEEE Trans. Autom.
Control}, vol. 56, no. 11, pp. 2555--2566, 2011.

\bibitem{clarke1990}
F. H. Clarke, \emph{Optimization and nonsmooth analysis}, in Classics in
Applied Mathematics, vol. 5. Philadelphia, PA: SIAM, 1990.




%
%
\end{thebibliography}
\end{document}